\documentclass[reqno,12pt,a4paper]{amsart}

\usepackage{mathrsfs}
\usepackage{amssymb}
\usepackage{amsfonts}
\usepackage{latexsym}
\usepackage{amsthm}
\usepackage{graphicx}
\def\lb{\label}

\newcommand{\er}[1]{\textrm{(\ref{#1})}}

\begin{document}

%%%%%%%%%% Some definitions %%%%%%%%%%

%%%%%%%% Equations, theorems %%%%%%%%%
\renewcommand{\theequation}{\arabic{section}.\arabic{equation}}
\theoremstyle{plain}
\newtheorem{theorem}{\bf Theorem}[section]
\newtheorem{lemma}[theorem]{\bf Lemma}
\newtheorem{corollary}[theorem]{\bf Corollary}
\newtheorem{proposition}[theorem]{\bf Proposition}
\newtheorem{definition}[theorem]{\bf Definition}
\newtheorem{remark}[theorem]{\it Remark}
%\theoremstyle{remark}
%\newtheorem{remark}[theorem]{\bf Remark}

%%%%% Alphabet %%%%%
\def\a{\alpha}  \def\cA{{\mathcal A}}     \def\bA{{\bf A}}  \def\mA{{\mathscr A}}
\def\b{\beta}   \def\cB{{\mathcal B}}     \def\bB{{\bf B}}  \def\mB{{\mathscr B}}
\def\g{\gamma}  \def\cC{{\mathcal C}}     \def\bC{{\bf C}}  \def\mC{{\mathscr C}}
\def\G{\Gamma}  \def\cD{{\mathcal D}}     \def\bD{{\bf D}}  \def\mD{{\mathscr D}}
\def\d{\delta}  \def\cE{{\mathcal E}}     \def\bE{{\bf E}}  \def\mE{{\mathscr E}}
\def\D{\Delta}  \def\cF{{\mathcal F}}     \def\bF{{\bf F}}  \def\mF{{\mathscr F}}
\def\c{\chi}    \def\cG{{\mathcal G}}     \def\bG{{\bf G}}  \def\mG{{\mathscr G}}
\def\z{\zeta}   \def\cH{{\mathcal H}}     \def\bH{{\bf H}}  \def\mH{{\mathscr H}}
\def\e{\eta}    \def\cI{{\mathcal I}}     \def\bI{{\bf I}}  \def\mI{{\mathscr I}}
\def\p{\psi}    \def\cJ{{\mathcal J}}     \def\bJ{{\bf J}}  \def\mJ{{\mathscr J}}
\def\vT{\Theta} \def\cK{{\mathcal K}}     \def\bK{{\bf K}}  \def\mK{{\mathscr K}}
\def\k{\kappa}  \def\cL{{\mathcal L}}     \def\bL{{\bf L}}  \def\mL{{\mathscr L}}
\def\l{\lambda} \def\cM{{\mathcal M}}     \def\bM{{\bf M}}  \def\mM{{\mathscr M}}
\def\L{\Lambda} \def\cN{{\mathcal N}}     \def\bN{{\bf N}}  \def\mN{{\mathscr N}}
\def\m{\mu}     \def\cO{{\mathcal O}}     \def\bO{{\bf O}}  \def\mO{{\mathscr O}}
\def\n{\nu}     \def\cP{{\mathcal P}}     \def\bP{{\bf P}}  \def\mP{{\mathscr P}}
\def\r{\rho}    \def\cQ{{\mathcal Q}}     \def\bQ{{\bf Q}}  \def\mQ{{\mathscr Q}}
\def\s{\sigma}  \def\cR{{\mathcal R}}     \def\bR{{\bf R}}  \def\mR{{\mathscr R}}
\def\S{\Sigma}  \def\cS{{\mathcal S}}     \def\bS{{\bf S}}  \def\mS{{\mathscr S}}
\def\t{\tau}    \def\cT{{\mathcal T}}     \def\bT{{\bf T}}  \def\mT{{\mathscr T}}
\def\f{\phi}    \def\cU{{\mathcal U}}     \def\bU{{\bf U}}  \def\mU{{\mathscr U}}
\def\F{\Phi}    \def\cV{{\mathcal V}}     \def\bV{{\bf V}}  \def\mV{{\mathscr V}}
\def\P{\Psi}    \def\cW{{\mathcal W}}     \def\bW{{\bf W}}  \def\mW{{\mathscr W}}
\def\o{\omega}  \def\cX{{\mathcal X}}     \def\bX{{\bf X}}  \def\mX{{\mathscr X}}
\def\x{\xi}     \def\cY{{\mathcal Y}}     \def\bY{{\bf Y}}  \def\mY{{\mathscr Y}}
\def\X{\Xi}     \def\cZ{{\mathcal Z}}     \def\bZ{{\bf Z}}  \def\mZ{{\mathscr Z}}
\def\O{\Omega}

\newcommand{\gA}{\mathfrak{A}}
\newcommand{\gB}{\mathfrak{B}}
\newcommand{\gC}{\mathfrak{C}}
\newcommand{\gD}{\mathfrak{D}}
\newcommand{\gE}{\mathfrak{E}}
\newcommand{\gF}{\mathfrak{F}}
\newcommand{\gG}{\mathfrak{G}}
\newcommand{\gH}{\mathfrak{H}}
\newcommand{\gI}{\mathfrak{I}}
\newcommand{\gJ}{\mathfrak{J}}
\newcommand{\gK}{\mathfrak{K}}
\newcommand{\gL}{\mathfrak{L}}
\newcommand{\gM}{\mathfrak{M}}
\newcommand{\gN}{\mathfrak{N}}
\newcommand{\gO}{\mathfrak{O}}
\newcommand{\gP}{\mathfrak{P}}
\newcommand{\gR}{\mathfrak{R}}
\newcommand{\gS}{\mathfrak{S}}
\newcommand{\gT}{\mathfrak{T}}
\newcommand{\gU}{\mathfrak{U}}
\newcommand{\gV}{\mathfrak{V}}
\newcommand{\gW}{\mathfrak{W}}
\newcommand{\gX}{\mathfrak{X}}
\newcommand{\gY}{\mathfrak{Y}}
\newcommand{\gZ}{\mathfrak{Z}}

\def\ve{\varepsilon}   \def\vt{\vartheta}    \def\vp{\varphi}    \def\vk{\varkappa}

\def\Z{{\mathbb Z}}    \def\R{{\mathbb R}}   \def\C{{\mathbb C}}
\def\T{{\mathbb T}}    \def\N{{\mathbb N}}   \def\dD{{\mathbb D}}
\def\dO{{\mathbb O}}

%%%%% Arrows %%%%%

\def\la{\leftarrow}              \def\ra{\rightarrow}            \def\Ra{\Rightarrow}
\def\ua{\uparrow}                \def\da{\downarrow}
\def\lra{\leftrightarrow}        \def\Lra{\Leftrightarrow}

%%%%% Typography %%%%%

\def\lt{\biggl}                  \def\rt{\biggr}
\def\ol{\overline}               \def\wt{\widetilde}
\def\no{\noindent}

%%%%% Math signs %%%%%

\let\ge\geqslant                 \let\le\leqslant
\def\lan{\langle}                \def\ran{\rangle}
\def\/{\over}                    \def\iy{\infty}
\def\sm{\setminus}               \def\es{\emptyset}
\def\ss{\subset}                 \def\ts{\times}
\def\pa{\partial}                \def\os{\oplus}
\def\om{\ominus}                 \def\ev{\equiv}
\def\iint{\int\!\!\!\int}        \def\iintt{\mathop{\int\!\!\int\!\!\dots\!\!\int}\limits}
\def\el2{\ell^{\,2}}             \def\1{1\!\!1}
\def\sh{\sharp}
\def\wh{\widehat}
%%%%% Math operations %%%%%

\def\where{\mathop{\mathrm{where}}\nolimits}
\def\all{\mathop{\mathrm{all}}\nolimits}
\def\as{\mathop{\mathrm{as}}\nolimits}
\def\Area{\mathop{\mathrm{Area}}\nolimits}
\def\arg{\mathop{\mathrm{arg}}\nolimits}
\def\const{\mathop{\mathrm{const}}\nolimits}
\def\det{\mathop{\mathrm{det}}\nolimits}
\def\diag{\mathop{\mathrm{diag}}\nolimits}
\def\diam{\mathop{\mathrm{diam}}\nolimits}
\def\dim{\mathop{\mathrm{dim}}\nolimits}
\def\dist{\mathop{\mathrm{dist}}\nolimits}
\def\Im{\mathop{\mathrm{Im}}\nolimits}
\def\Iso{\mathop{\mathrm{Iso}}\nolimits}
\def\Ker{\mathop{\mathrm{Ker}}\nolimits}
\def\Lip{\mathop{\mathrm{Lip}}\nolimits}
\def\rank{\mathop{\mathrm{rank}}\limits}
\def\Ran{\mathop{\mathrm{Ran}}\nolimits}
\def\Re{\mathop{\mathrm{Re}}\nolimits}
\def\Res{\mathop{\mathrm{Res}}\nolimits}
\def\res{\mathop{\mathrm{res}}\limits}
\def\sign{\mathop{\mathrm{sign}}\nolimits}
\def\span{\mathop{\mathrm{span}}\nolimits}
\def\supp{\mathop{\mathrm{supp}}\nolimits}
\def\Tr{\mathop{\mathrm{Tr}}\nolimits}
\def\Dom{\mathop{\mathrm{Dom}}\nolimits}
\def\BBox{\hspace{1mm}\vrule height6pt width5.5pt depth0pt \hspace{6pt}}

%%%%%%%%%%%%% specialities %%%%%%%%%%%%%%

\newcommand\nh[2]{\widehat{#1}\vphantom{#1}^{(#2)}}
%{{\mathop{#1}\limits^\wedge}\vphantom{#1}^{(#2)}}
\def\dia{\diamond}

\def\Oplus{\bigoplus\nolimits}

%%%%%%%%%%% End of definitions %%%%%%%%%%

%%%%% OLD OLD OLD

\def\qqq{\qquad}
\def\qq{\quad}
\let\ge\geqslant
\let\le\leqslant
\let\geq\geqslant
\let\leq\leqslant
\newcommand{\ca}{\begin{cases}}
\newcommand{\ac}{\end{cases}}
\newcommand{\ma}{\begin{pmatrix}}
\newcommand{\am}{\end{pmatrix}}
\renewcommand{\[}{\begin{equation}}
\renewcommand{\]}{\end{equation}}
\def\bu{\bullet}

\title[{Even  order periodic  operators on the real line}]
{Even  order periodic  operators on the real line}

\date{\today}

\author[Andrey Badanin]{Andrey Badanin}
\address{Andrey Badanin, Archangelsk St. Technical Univ., Russia, e-mail: an.badanin@gmail.com}

\author[Evgeny Korotyaev]{Evgeny L. Korotyaev}
\address{Evgeny Korotyaev, St.Petersburg State Univ.,
Russia,
e-mail: korotyaev@gmail.com
}


\maketitle
\begin{abstract}
\no We consider $2p\ge 4$ order differential operator on the real
line with a periodic coefficients. The spectrum of this operator is
absolutely continuous and is a union of spectral bands separated by
gaps. We define the Lyapunov function, which is analytic on a
p-sheeted Riemann surface. The Lyapunov function has real or complex
branch points. We prove the following results: (1) The spectrum  at
high energy has multiplicity two. (2) Endpoints of all gaps are
periodic  (or anti-periodic) eigenvalues or real branch points. (3)
The spectrum of operator has an infinite number of open gaps and
there exists only a finite number of non-real branch points for some
specific coefficients (the generic case). (4) The asymptotics of the
periodic, anti-periodic spectrum and branch points are determined at
high energy.

\end{abstract}

\section {Introduction and main results}
\setcounter{equation}{0}

Consider the self-adjoint periodic operator $H$ acting in $L^2(\R)$
and given by
\[
\lb{H}
H=H_0+q,\qqq H_0=(-1)^p{d^{2p}\/dt^{2p}},\qqq
q=\sum_{j=0}^{p-1}{d^{j}\/dt^{j}} q_{j+1} {d^{j}\/dt^{j}},\qqq p\ge 2,
\]
\[
q_j\in L_{real}^1(\T),\qq j\in\N_p=\{1,2,...,p\},\qqq \T=\R/\Z,
\]
where $\Z$ is the set of all integers.
Let $W_j^2(\R),j\in\N=\Z\cap [1,\iy)$, be the Sobolev space of functions
$f,f^{(j)}\in L^2(\R)$. Here and below we use the notation $f'={\pa
f\/\pa t}, f^{(j)}={\pa^jf \/\pa t^j}$. We define the self-adjoint
operator $H$ using the quadratic form
with the form domain $\Dom_{fd}(H)=W_{p}^2(\R)$
(see Proposition \ref{eqf}).

It is well known (see \cite{DS}, Ch.~XIII.7.64) that the spectrum
$\s(H)$ of $H$ for the sufficiently smooth coefficients
$q_j,j\in\N_p$, is absolutely continuous and consists of
non-degenerated intervals $\gS_n,n=1,...,N_G\le\iy$. These intervals
$\gS_n$ and $\gS_{n+1}$ are separated by the gap $\mathfrak{g}_n$
with length $|\mathfrak{g}_n|>0$ and $N_G-1$ is a number of the
gaps. Theorem \ref{T1} extends this result to the  larger case $q_j\in L^1(\T)$.

The typical applications of our operator $H$ are the vibrations of beams, plates and shells:

(1) The
standard Kirchhoff-Love model of the bend of beams and plates
provides the Euler-Bernoulli equation $y''''=\l a y$ (see
\cite{TYW}, Ch.~5.9).

(2) The Vlasov model of the bend of cylinder
shells (see \cite{NCM}, Ch.~I.1.14) gives the equations of vibration
having the form $y^{(8)}+b_1 y=\l b y$.

 Here $y$ is the normal
displacement of the plate (or shell), the functions $a$ (or $b,
b_1$) are defined by the parameters of the plate (or shell): Young's
modulus, Poisson's modulus, rigidity and thickness.

The high order differential operators arise in the inverse problem
method of integration of non-linear evolution equations. There exist
the Lax pairs, where the self-adjoint operator is a high order operator
and
%Consider
%the operators $L=L(t)=\sum_0^n u_i(x,t)\pa_x^i,A=\sum_0^m
%v_i(x,t)\pa_x^i$ with some coefficients $u_i,v_i$, depending on  the
%space variable $x$  and time $t$ (??????? $t$ IS A SPACE VARIABLE BEFORE AND LATER?????????).
%Then
the corresponding non-linear Lax equation
%$\pa_t L=LA-AL$
is integrable by the inverse problem method, see
\cite{DKN}. Many physically interesting equations have this form, see \cite{AC}.

%Now we describe the results for  high order operators $H, p\ge 2$.
Recall that the spectral theory for the high order operators with
decreasing coefficients is well developed, see \cite{BDT}, \cite{Su}
and the references therein. The results for high order periodic
operators are still modest.

We describe our goal. In the case $p=1$ the spectrum of the
Hill operator $-{d^2\/dt^2}+q_1$ is a union of spectral bands, where
all endpoints of the bands are 2-periodic eigenvalues of the
equation $-y''+q_1y=\l y$. In the case $p=2$ the spectrum of the
operator $H$ is also a union of spectral bands, but all endpoints of the
bands are 2-periodic eigenvalues of the equation
$y''''+qy=\l y$ or the branch points of the Lyapunov
function \cite{BK2}. Until now  there are
no any results about the multiplicity of the spectrum at high
energy, number of gaps (is it finite or infinite ?), asymptotics and
type of endpoints of the gaps at high energy etc for the operators $H, p>2$. Our main goal is
to answer some of these questions.

In order to describe our results we consider the equation
\[
\lb{1b}
(-1)^py^{(2p)}+qy=\l y,\qqq q=\sum_{j=0}^{p-1}{d^{j}\/dt^{j}} q_{j+1} {d^{j}\/dt^{j}}
\qqq (t,\l)\in \R\ts\C,
\]
where $\C$ is the complex plane. If all coefficients
$q_j,q_j^{(j-1)}\in L^1(\T)$, then the standard monodromy matrix is
well defined (see \cite{DS}, Ch.~ XIII.7). If some coefficient
$q_j\in L^1(\T), q_j'\notin L^1(\T)$, then the standard monodromy
matrix is not well-defined, since, in general, the derivative of
$y^{(2p-1)}$ is not continuous. In this case we will introduce the
modified symplectic monodromy matrix, see \er{smp}. We think that it
will be convenient even for smooth coefficients $q_j$.
 We rewrite the equation \er{1b} in the vector  form by
\[
\lb{me}
Y'-\cP(\l)Y=\cQ Y,\qqq (t,\l)\in\R\ts\C,
\]
see \cite{Na}, Ch.~ II.4, where the vector-valued function $Y$ is given by
\[
\lb{Y}
Y=\ma y_1\\y_2\\ \vdots\\y_{p+1}\\y_{p+2}\\y_{p+3}\\ \vdots\\y_{2p}\am
=\ma y\\y_1'\\ \vdots\\y_{p}'\\y_{p+1}'+(-1)^pq_py_p\\
y_{p+2}'+(-1)^pq_{p-1}y_{p-1}\\ \vdots\\y_{2p-1}'+(-1)^pq_2y_2\am,
\]
and the $2p\ts 2p$ matrices-valued functions $\cP, \cQ$ are given by
\[
\lb{PQ}
\cP=\ma\dO_{2p-1,1}&\1_{2p-1}\\(-1)^p\l&\dO_{1,2p-1}\am,\qqq
\cQ=(-1)^{p+1}\ma \dO_{p,p}& \dO_{p,p}\\
{\ma
0&...&0& q_p\\
0&...&q_{p-1}&0\\
...&...&...&...\\
q_1&...&0&0
\am}& \dO_{p,p}\am,
\]
$\dO_{m,n}$ is the $m\ts n$ zero matrix, $\1_n$ is the $n\ts n$
identity matrix.
If all $q_j\in L^1(\T)$, then $\cQ\in L^1(\T)$ and there exists a  $2p\ts 2p$ matrix-valued
solution $\cM(t,\l)$  of equation \er{me} with the initial condition
$\cM(0,\l)=\1_{2p}$. In this case the {\it modified monodromy matrix} $\cM(1,\l)$
is well-defined and entire. Its characteristic polynomial $D$ is given by
\[
\lb{1c} D(\t,\l)=\det(\cM(1,\l)-\t \1_{2p}),\qq (\t,\l)\in\C^2.
\]
An eigenvalue
of $\cM(1,\l)$ is called a {\it multiplier}, it is a zero of the
algebraic equation $D(\cdot,\l)=0$. Each $\cM(1,\l), \l\in\C$, has
exactly $2p$ (counted with multiplicities)  multipliers
$\t_j(\l),j\in\N_{2p}$.
Due to \er{smp}, the matrix $\cM$ is symplectic. Then $\t$ is
a multiplier iff $\t^{-1}$ is a multiplier.
The multipliers  have asymptotics
\[
\lb{lom}
\t_j(\l)=e^{z\o_j}(1+O(|z|^{-1}))\qq\text{as}
\qq|\l|\to\iy,\qq\l\in\C_+=\{\l: \Im \l>0\},
\qq\text{all}\qq j\in\N_{2p},
\]
$$
\arg\l\in(-\pi,\pi],\qqq z=\l^{1\/2p}\in
S=\Bigl\{z\in\C:\arg z\in\Bigl(-{\pi\/2p},{\pi\/2p}\Bigr]\Bigr\},
$$
see Lemma \ref{Am}.
Here and below $\o_j$ are the zeros of the
polynomial $\o^{2p}-(-1)^p$ labeling by
\[
\lb{no}
\o_{2j}=\ca\qq e^{-i{\pi\/p}j},\ \ \text{even}\ p \\
e^{i{\pi\/2p}(2j-1)},\ \text{odd} \ p\ac\!\!\!\!\!\!,\qq \o_{2j-1}=-\o_{2(p-j+1)},\qq
\o_j=-\o_{2p-j+1},\qq j\in\N_p.
\]
Note that $\o_{p+1}=-\o_p=i$ and if $p$ is even, then $\o_1=-\o_{2p}=1$.

The coefficients of the polynomial $D(\cdot,\l)$ are entire
functions in $\l$. It is well known (see, e.g.,
\cite{Fo}, Ch.~8) that the roots $\t_j(\l),j\in\N_{2p}$, constitute one or
several branches of $N_F\ge 1$ analytic functions that have only
algebraic singularities in $\C$. Asymptotics \er{lom} show that
$N_F=1$, i.e. $\t_j$ are branches of the unique function $\t$
analytic on the $2p$ sheeted Riemann surface. Moreover, these
asymptotics define the functions $\t_j$ in $\C_+$ for $|\l|$ large
enough. The detailed results about the branches $\t_j$ will be given
in Section 4.

\begin{figure}
\tiny
\unitlength 1mm
\linethickness{0.4pt}
\ifx\plotpoint\undefined\newsavebox{\plotpoint}\fi
\begin{picture}(115.025,68.814)(0,0)
%upper sheet
%contour
\put(26.182,68.123){\line(1,0){77.}}
\put(10.55,52.45){\line(1,0){77.}}
\qbezier(10.55,52.45)(-1.612,52.537)(6.35,60.15)
\qbezier(8.1,62.075)(9.938,66.8)(26.182,68.123)
\qbezier(103.182,68.123)(115.025,60.542)(87.55,52.45)
%cuts
\thicklines
\multiput(18.639,62.064)(-.03642979,-.03362979){47}{\line(-1,0){.03642979}}
\multiput(23.907,62.195)(-.03642979,-.03362979){47}{\line(-1,0){.03642979}}
\multiput(29.289,60.15)(.0336,.04032){50}{\line(0,1){.04032}}
\multiput(36.681,60.374)(.03536842,.03340351){57}{\line(1,0){.03536842}}
\multiput(49.575,62.25)(-.03365385,-.03701923){52}{\line(0,-1){.03701923}}
\put(36.625,60.325){\line(1,0){11.375}}
\put(38.55,62.075){\line(1,0){10.85}}
\put(22.45,60.325){\line(1,0){6.825}}
\put(23.85,62.075){\line(1,0){7.175}}
\put(16.85,60.325){\line(-1,0){10.675}}
\put(18.6,62.075){\line(-1,0){10.675}}
\thinlines
%connection with the middle sheet
\multiput(6.081,60.262)(.032,-1.024){7}{\line(0,-1){1.024}}
\multiput(7.761,62.054)(-.032,-.608){14}{\line(0,-1){.608}}
\multiput(17.081,60.262)(.032,-1.024){7}{\line(0,-1){1.024}}
\multiput(18.761,62.054)(-.032,-.608){14}{\line(0,-1){.608}}
\multiput(22.121,60.262)(.032,-.944){7}{\line(0,-1){.944}}
\multiput(23.801,62.166)(-.0336,-.392){20}{\line(0,-1){.392}}
\multiput(29.401,60.15)(.032,-.976){7}{\line(0,-1){.976}}
\multiput(30.969,62.278)(-.032,-.608){14}{\line(0,-1){.608}}
\multiput(36.569,60.15)(.0336,-.6384){10}{\line(0,-1){.6384}}
\multiput(38.585,62.166)(-.0331852,-.3028148){27}{\line(0,-1){.3028148}}
\multiput(6.55,51.4)(.0328125,-.4703125){32}{\line(0,-1){.4703125}}
\multiput(7.425,51.225)(-.03351064,-.35744681){47}{\line(0,-1){.35744681}}
\multiput(17.55,51.4)(.0328125,-.4703125){32}{\line(0,-1){.4703125}}
\multiput(18.425,51.225)(-.03351064,-.35744681){47}{\line(0,-1){.35744681}}
\multiput(22.45,51.225)(.0328125,-.4648437){32}{\line(0,-1){.4648437}}
\multiput(23.325,51.4)(-.03310811,-.44932432){37}{\line(0,-1){.44932432}}
\multiput(29.8,51.4)(.0328125,-.4648438){32}{\line(0,-1){.4648438}}
\multiput(30.675,51.4)(-.03310811,-.44459459){37}{\line(0,-1){.44459459}}
\multiput(36.975,51.575)(.03333333,-.3625){42}{\line(0,-1){.3625}}
\multiput(37.85,51.4)(-.03310811,-.44932432){37}{\line(0,-1){.44932432}}
\multiput(47.65,59.975)(.031818,-.572727){11}{\line(0,-1){.572727}}
\multiput(49.575,61.725)(-.0333333,-.3333333){21}{\line(0,-1){.3333333}}
\multiput(48,51.75)(.03365385,-.28942308){52}{\line(0,-1){.28942308}}
\multiput(49.05,51.75)(-.03310811,-.46824324){37}{\line(0,-1){.46824324}}
\put(50.876,63.913){\makebox(0,0)[cc]{$r_{1,0}^+$}}
\put(38.151,63.913){\makebox(0,0)[cc]{$r_{1,1}^-$}}
\put(30.625,63.913){\makebox(0,0)[cc]{$r_{1,1}^+$}}
\put(23.636,63.913){\makebox(0,0)[cc]{$r_{1,2}^-$}}
\put(17.035,63.913){\makebox(0,0)[cc]{$r_{1,2}^+$}}
\put(83.326,54.801){\makebox(0,0)[cc]{3rd sheet}}
%middle sheet
%contour
\put(21.4,41.775){\line(1,0){.875}}
\put(31.2,41.775){\line(1,0){5.075}}
\put(49.449,41.775){\line(1,0){53.872}}
\put(10.55,26.375){\line(1,0){77.35}}
\bezier{8}(22.3,41.775)(27.763,41.775)(31.025,41.775)
\bezier{10}(36.3,41.775)(42.763,41.775)(49.025,41.775)
\qbezier(10.725,26.375)(.488,29.525)(6.3,34.775)
\bezier{10}(7.7,36.525)(9.763,38.713)(17.025,40.85)
\qbezier(18.95,41.275)(20,41.6)(21.4,41.775)
\qbezier(87.9,26.375)(101.988,26.988)(105.225,34.25)
\qbezier(103.3,41.95)(107.588,40.2)(106.625,36.35)
\thicklines
%right cuts
\multiput(65.675,35.825)(-.0362069,-.03318966){58}{\line(-1,0){.0362069}}
\multiput(75.939,35.985)(-.03553333,-.03344444){63}{\line(-1,0){.03553333}}
\multiput(82.656,36.116)(-.03344444,-.03344444){63}{\line(0,-1){.03344444}}
\multiput(91.744,35.985)(-.03591636,-.03352364){55}{\line(-1,0){.03591636}}
\multiput(96.75,36.248)(-.03349322,-.03348136){59}{\line(-1,0){.03349322}}
\put(63.75,34.25){\line(1,0){10.15}}
\put(65.85,36.175){\line(1,0){10.325}}
\put(80.55,34.25){\line(1,0){9.1}}
\put(82.825,36.175){\line(1,0){8.925}}
\put(95.075,34.25){\line(1,0){10.325}}
\put(96.65,36.175){\line(1,0){9.975}}
%left cuts
\multiput(18.376,36.034)(-.03644468,-.03362979){47}{\line(-1,0){.03644468}}
\multiput(23.644,36.165)(-.03642979,-.03362979){47}{\line(-1,0){.03642979}}
\multiput(30.757,36.297)(-.03358627,-.03615294){51}{\line(0,-1){.03615294}}
\multiput(38.528,36.165)(-.03832182,-.03352364){55}{\line(-1,0){.03832182}}
\multiput(48,34.425)(.03602941,.03345588){55}{\line(1,0){.03602941}}
\put(16.85,34.425){\line(-1,0){10.725}}
\put(18.6,36.35){\line(-1,0){11.}}
\put(22.45,34.425){\line(1,0){6.65}}
\put(23.675,36.35){\line(1,0){7}}
\put(36.975,34.425){\line(1,0){10.85}}
\put(38.55,36.35){\line(1,0){11.375}}
\thinlines
%connection with the lower sheet
\multiput(66.025,36.175)(-.0328125,-.4375){16}{\line(0,-1){.4375}}
\multiput(63.75,34.25)(.029167,-1.079167){6}{\line(0,-1){1.079167}}
\multiput(63.925,25.325)(.03310811,-.32635135){37}{\line(0,-1){.32635135}}
\multiput(65.15,25.675)(-.03318966,-.25646552){58}{\line(0,-1){.25646552}}
\multiput(73.977,33.942)(.0329412,-.4018824){17}{\line(0,-1){.4018824}}
\multiput(74.649,25.878)(.0332973,-.36021622){37}{\line(0,-1){.36021622}}
\multiput(76.105,35.958)(-.0336,-.4256){20}{\line(0,-1){.4256}}
\multiput(75.321,25.878)(-.03309091,-.34363636){44}{\line(0,-1){.34363636}}
\multiput(82.489,35.958)(-.0326667,-.3546667){24}{\line(0,-1){.3546667}}
\multiput(81.481,25.766)(-.0331852,-.5641481){27}{\line(0,-1){.5641481}}
\multiput(80.585,34.054)(.032,-.912){7}{\line(0,-1){.912}}
\multiput(80.809,25.654)(.03309091,-.29527273){44}{\line(0,-1){.29527273}}
\multiput(89.881,34.054)(.0329412,-.3425882){17}{\line(0,-1){.3425882}}
\multiput(90.553,26.55)(.0332973,-.37232432){37}{\line(0,-1){.37232432}}
\multiput(91.785,36.07)(-.032,-.512){14}{\line(0,-1){.512}}
\multiput(91.225,26.662)(-.03309091,-.36145455){44}{\line(0,-1){.36145455}}
\multiput(94.697,34.054)(.032,-.56){7}{\line(0,-1){.56}}
\multiput(95.369,27.558)(.03294118,-.43152941){34}{\line(0,-1){.43152941}}
\multiput(96.601,36.294)(-.0329412,-.3491765){17}{\line(0,-1){.3491765}}
\multiput(96.041,28.006)(-.03309091,-.392){44}{\line(0,-1){.392}}
\multiput(105.,34.054)(.032,-.56){38}{\line(0,-1){.56}}
\multiput(106.501,36.294)(-.0229412,-.3491765){73}{\line(0,-1){.3491765}}
\put(67.668,37.84){\makebox(0,0)[cc]{$r_{2,0}^+$}}
\put(76.127,37.84){\makebox(0,0)[cc]{$r_{2,1}^-$}}
\put(82.847,37.84){\makebox(0,0)[cc]{$r_{2,1}^+$}}
\put(91.449,37.84){\makebox(0,0)[cc]{$r_{2,2}^-$}}
\put(97.363,37.84){\makebox(0,0)[cc]{$r_{2,2}^+$}}
\put(21.831,28.683){\makebox(0,0)[cc]{2nd sheet}}
%lower sheet
%contour
\put(26.3,18.15){\line(1,0){37.25}}
\put(76.153,18.15){\line(1,0){4.064}}
\put(91.489,18.15){\line(1,0){3.384}}
\put(9.85,2.575){\line(1,0){77.7}}
\bezier{10}(63.625,18.15)(69.412,18.15)(76.05,18.15)
\bezier{8}(81.625,18.15)(86.412,18.15)(91.05,18.15)
\qbezier(9.675,2.575)(-2.662,11.15)(25.95,18.325)
\qbezier(87.55,2.575)(100.563,4.413)(104.725,10.8)
\qbezier(106.125,16.5)(107.412,14.45)(106.05,12.7)
\bezier{10}(106.125,16.5)(103.412,18.50)(94.805,18.15)
%cuts
\thicklines
\multiput(63.4,11.15)(.0362069,.03318966){58}{\line(1,0){.0362069}}
\multiput(75.807,12.709)(-.03554444,-.03345556){57}{\line(-1,0){.03554444}}
\multiput(82.625,12.84)(-.03344444,-.03345556){63}{\line(0,-1){.03345556}}
\multiput(91.813,12.609)(-.03591636,-.03353636){55}{\line(-1,0){.03591636}}
\multiput(96.618,12.672)(-.03348136,-.03348136){59}{\line(-1,0){.03348136}}
\put(63.05,10.8){\line(1,0){11.025}}
\put(65.325,12.725){\line(1,0){10.325}}
\put(80.55,10.8){\line(1,0){9.1}}
\put(82.3,12.725){\line(1,0){9.625}}
\put(94.55,10.8){\line(1,0){10.325}}
\put(96.65,12.725){\line(1,0){9.275}}
\thinlines
\put(21.35,5.277){\makebox(0,0)[cc]{1st sheet}}
\end{picture}
\caption{\footnotesize The Riemann surface of the Lyapunov function
for $p=3$ for the case of small coefficients. Identities \er{t+1}
give $ \D_{2}(r_{1,n}^{\pm})=\D_{3}(r_{1,n}^{\pm})$ and $
\D_{1}(r_{2,n}^{\pm})=\D_{2}(r_{2,n}^{\pm}). $
Then the second and third sheets of the surface are
attached along the cuts
$(r_{1,n-1}^+,r_{1,n}^-),n\in\N$.
We attach the upper (lower) edge of
each cut on the second sheet to the lower (upper) edge of the same cut on the
third sheet.
Similarly,
first and second sheets of the surface are attached along the cuts
$(r_{2,n-1}^+,r_{2,n}^-),n\in\N$. Thus,
whenever we cross the cut, we pass from one sheet to another.}
\lb{RS3}
\end{figure}
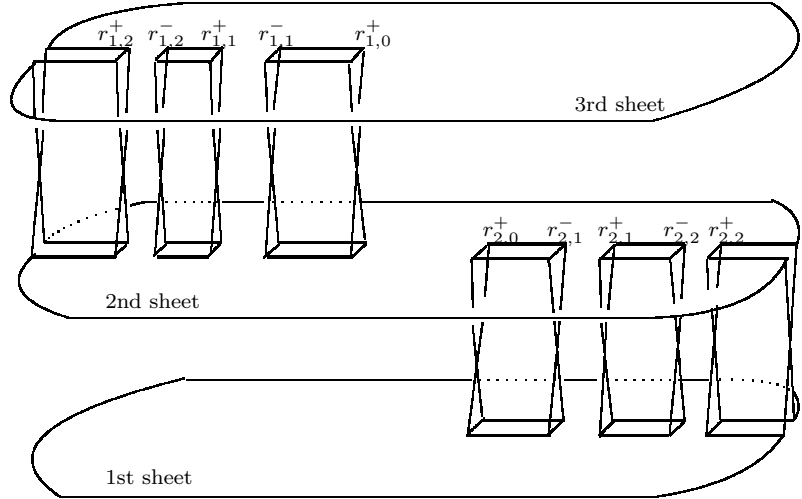
We formulate our first preliminary results.

\begin{theorem}
\lb{T1}
The monodromy matrix $\cM=\cM(1,\cdot)$ is symplectic,
i.e. it satisfies the identity
\[
\lb{smp}
\cM^\top\cJ \cM =\cJ,\qq\text{where}\ \
\cJ=\ma\dO_{p,p}&J_p\\(-1)^pJ_p&\dO_{p,p}\am,\ \
J_p=\ma 0&...&0&0&1\\
0&...&0&-1&0\\
0&...&1&0&0\\
...&...&...&...&...\\
(-1)^{p+1}&...&0&0&0
\am
\]
%$\cM(1,\l)$,
and $\cJ^\top=-\cJ$. Furthermore, there
exists an analytic function $\D$ on the connected $p$-sheeted
Riemann surface $\mR$ having the following properties:

 i) All branches of $\D$ have the form
$\D_j={1\/2}(\t_j+\t_j^{-1}), j\in\N_p$, and satisfy:
\[
\lb{T1-2}
{D(\t,\l)\/(2\t)^p}=\F(\n,\l)=\prod_{j=1}^{p}(\n-\D_j(\l)),\qq
\n={\t+\t^{-1}\/2},\qqq (\t,\l)\in\C^2,
\qq \t\ne 0,
\]
\[
\lb{aD}
\D_j(\l)=\cosh z\o_j+O\lt({e^{|\Re z\o_j|}\/|z|}\rt)\qq\text{as}\qq
|\l|\to\iy,\qq\l\in\C_+.
\]
%(OTHER LABELING???????)

 ii) If $\D_j(\l)\in(-1,1)$ for some $(j,\l)\in\N_p\ts\R$,
and $\l$ is not a branch point of $\D_j$,
then $\D_j'(\l)\ne 0$.

 iii) The spectrum $\s(H)$ of the operator $H$ satisfies
\[
\lb{sd}
\s(H)=\s_{ac}(H)
=\{\l\in\R:\D_j(\l)\in[-1,1]\ \text{for\ some}\ j\in\N_p\}.
\]
\end{theorem}

{\bf Remark.} 1) If $\l\in \s(H)$, then some branch $\D_j(\l)$ is real and the corresponding
multiplier $\t_j(\l)$ is complex and $|\t_j(\l)|=1$. It is more convenient to study the real function
$\D_j(\l)$ on the spectrum $\s(H)$, than the complex multiplier
$\t_j(\l)$ on $\s(H)$.

 2) The surface $\mR$ is connected.
For the first and second order operators with the matrix-valued potentials
the corresponding surface may be disconnected (see \cite{CK},
\cite{K1}, \cite{K2}).

 3)  The proof of i), ii) repeats essentially the argument from \cite{CK}, \cite{K1}, \cite{K2}.

 4) The monodromy matrix for the second order operators has asymptotics in terms of $\cos $
and  $\sin $ bounded  on the real line.
The monodromy matrix for high order operators has asymptotics in terms of $\cosh $
and  $\sinh $, see \er{2if},  unbounded  on the real line.

The zeros of $D(1,\cdot)$  (or $D(-1,\cdot)$) are periodic (or
antiperiodic) eigenvalues of the equation $(-1)^py^{(2p)}+qy=\l y$,
where  $y$ are $1$-periodic (or $1$-antiperiodic) functions. Denote
by $\l_0^+,\l_{2n}^\pm,n\ge 1$, the periodic eigenvalues and  by
$\l_{2n-1}^\pm,n\ge 1$, the antiperiodic eigenvalues labeling by
(counted with multiplicity)
$$
\l_{0}^+\le \l_{2}^-\le \l_{2}^+\le \l_{4}^-\le\l_{4}^+\le \l_{6}^-
\le...,\qq \l_{1}^-\le \l_{1}^+\le \l_{3}^-\le
\l_{3}^+\le\l_{5}^-\le \l_{5}^+ \le....
$$

For the polynomial  $\F$ given by \er{T1-2}
we introduce the discriminant $\r(\l),\l\in\C$, by
\[
\lb{T1-4}
\r=\prod_{1\le j<k\le p}(\D_j-\D_k)^2.
\]
A zero of $\r$ is a {\bf ramification} point (or simply  a {\bf ramification}) of the Lyapunov
function $\D$.

 {\bf Remark.}
1) Ramification is a geometric term used for 'branching out',
in the way that the square root function, for complex numbers,
can be seen to have two branches differing in sign.
We also use it from the opposite perspective (branches coming together)
as when a covering map degenerates at a point of a space,
with some collapsing together of the fibers of the mapping.

 2) Recall that all endpoints of gaps of the spectrum  of the Hill operator (i.e., $p=1$) are periodic or
anti-periodic eigenvalues. The situation is more complicated for the
high order periodic operators and the periodic operators with the
matrix potentials. In these cases the endpoints of gaps are periodic
or anti-periodic eigenvalues, or ramifications (zeros of the
function $\r$).
%These zeros are continuous functions of coefficients
%of the equation.
The numerical analysis for the fourth order operators and the second
order periodic operators with the $2\ts 2$ matrix potential shows
that ramifications can be non-real for some values of the
coefficients and can become real and to create the gap for some
other values of the coefficients. This behavior is similar to the
behavior of the resonances in the scattering problem for the
Schr\"odinger operator (see, e.g., \cite{K5}, \cite{Z}). In fact, this was a
reason for us to use the term {\it resonance} for the zero of the
function $\r$ for the periodic operators in our previous papers
\cite{BK1}, \cite{BK2}, \cite{BBK}, \cite{CK}, ... But now we will
use the term {\it ramification} for such points because the term {\it
resonance} is overloaded and is used in different other senses.

We shortly describe the unperturbed operator
$H^0=(-1)^p{d^{2p}\/dt^{2p}}$, see more in Section 2.
The unperturbed  multipliers $\t_j^0$, the Lyapunov function $\D^0$ with
all branches $\D_j^0$ are given by
$$
\t_j^0(\l)=e^{\o_j z},\qq \D^0(\l)=\cos \l^{1\/2p},\qq \D_j^0(\l)=\cosh z\o_j,\qq
(j,\l)\in\N_{2p}\ts\C_+.
$$
The unperturbed  discriminant $\r^0$ has the form
\[
\lb{asr0}
\r^0=\!\!\prod_{1\le j<\ell\le p}\!\!
(\cosh z\o_j-\cosh z\o_\ell)^2.
\]
The 2-periodic eigenvalues
$\l_n^{0,\pm}=(\pi n)^{2p},n\ge 1$, have multiplicity 2 and the
periodic eigenvalue $\l_0^{0,+}=0$ has multiplicity 1.
The function $\r^0$ is entire and has the zeros
%$r_{k,0}^{0}=0$ and
$r_{k,n}^0,k\in\N_{p-1},n\ge 0$, given by
\[
\lb{r0}
r_{k,n}^0=(-1)^k\Bigl({\pi n\/c_k}\Bigr)^{2p},\qqq
c_k=\cos{\pi k\/2p},\qqq 1>c_1>c_2>...>c_{p-1}>0.
\]
The zero $\l=0$ of the function $\r^0$
has the multiplicity $p-1$ and each another
zero has the multiplicity $2$. The spectrum $\s(H^0)$ has the
multiplicity 2.
The Riemann surface $\mR^0$ of the Lyapunov function $\D^0$
for the operator $H^0$ coincides with the
Riemann surface of the function $\l^{1\/p}$ with the unique branch
point at $\l=0$.

We determine the sharp asymptotics of the
ramifications.

\begin{theorem}
\lb{rho}
i) The function $\r$ is entire,  real on $\R$ and satisfies
\[
\lb{aro}
\r(\l)=\r^0(\l)(1+O(|z|^{-1}))\qq \text{as}\ \ |\l|\to\iy,\ \
|\l-r_{k,n}^0|>1,\ \ \ \forall \ (k,n)\in\N_{p-1}\ts\N.
%\l\in\C\sm\cup_{(k,n)\in\N_{p-1}\ts\N}U_{k,n}.
\]

 ii) The function $\r$ has the zeros $r_{k,0}^+,r_{k,n}^\pm,(k,n)\in\N_{p-1}\ts\N$,
which satisfy
\[
\lb{rnn} r_{k,n}^{\pm}=(-1)^k\lt({\pi n\/c_k}\rt)^{2p}\rt(1
+{c_k^2\/(\pi n)^2}\rt[(-1)^{p+1}\hat q_{p,0} \pm c_{k}|\hat
q_{p,n}|+O\Bigl({1\/n}\Bigr)\rt]\rt)
\]
as $n\to\iy,$ where
\[
\lb{zh}
\hat q_{p,n}=
%h_n=
\int_0^1 q_{p}(t)e^{-i2\pi nt}dt,\qq n\ge 0.
\]

\end{theorem}

 {\bf Remark.} 1) Asymptotics \er{aro} show that $\r\ne 0$, since
$\r^0\ne 0$. Note that for the second (and first) order operators
with the $p\ts p$ matrix-valued potential the
corresponding function may be equal to $0$ (see \cite{CK},
\cite{K1}, \cite{K2}).

 2) The Riemann surface $\mR$,
roughly speaking, is close to $\mR^0$  at high energy.
In general, the points $r_{k,n}^\pm$
are simple branch points of $\D$ (square root type) for $n$ large enough.
The surface $\mR$ for $p=3$ is shown by Fig. \ref{RS3}.

We describe the structure of the bands and the gaps  at high energy.

\begin{theorem}   \lb{T2}
i) The branch $\D_p$ is real analytic function on the interval $( \l_{n_0}^+, \iy)$
for some $n_0\in\N$ and
 $ \l_{n_0}^+< \l_{n_0+1}^-\le
\l_{n_0+1}^+<\l_{n_0+2}^-\le\l_{n_0+2}^+<...$.
 Moreover, if $n\ge n_0$, then each interval
$[\l_{n}^+,\l_{n+1}^-]$ is a spectral band with the
spectrum of the multiplicity $2$, and each interval
$(\l_{n}^-,\l_{n}^+)\ne \es$ is a gap and $\D_p^2(\l_n^\pm)=1$.
There are no other bands of $H$ to the right of $\l_{n_0}^+$.

 ii) The periodic and antiperiodic eigenvalues $\l_n^\pm$ satisfy:
\[
\lb{eln} \l_n^{\pm}=(\pi n)^{2p}\rt(1+{1\/(\pi n)^{2}}
\lt[(-1)^{p+1}\hat q_{p,0} \pm |\hat q_{p,n}|+{O(1)\/n}\rt]\rt)
\qq\text{as}\qq n\to\iy.
\]

\end{theorem}

\no {\bf Remark.}
1) The spectrum of $H$
has multiplicity $2$ at high energy. The spectrum of
the  Schr\"odinger operator  with the $p\ts p$ matrix-valued potential
and the first order operator with the $2p\ts 2p$ matrix-valued potential
has multiplicity $2p$ at high energy (see \cite{CK}, \cite{K1}).

2) The periodic and antiperiodic eigenvalues accumulate at $+\iy$.
The ramifications accumulate at $\pm\iy$.
But for second (or first) order systems the periodic and antiperiodic eigenvalues
and the ramifications accumulate at
$+\iy$ (or $\pm\iy$), see \cite{CK}, \cite{K2} (or  \cite{K1}).

3) The spectrum of $H$ at high energy is described by the branch $\D_p$  of the Lyapunov function.
The structure of the spectrum at high energy is similar to the structure
of the spectrum of the Hill operator:

(a) the spectra of $H$ and of the Hill operator are similar as
the sets, including multiplicities;

(b) endpoints of gaps are periodic and antiperiodic eigenvalues only;

 (c) the sharp asymptotics of these eigenvalues are expressed in terms of
the Fourier coefficients of the potential.

Recall that there exists
an infinite number of open gaps in the spectrum
of the first and the second order operators
for some specific potentials \cite{CK}, \cite{K1}, \cite{K2}.
Now we describe this situation for our case.

\begin{corollary} \lb{ig}
Let the coefficient $q_p$ satisfy
$|\int_0^1 q_{p}(t)e^{-i2\pi n_jt}dt|\ge {1\/{n_j}^{\a}}$
for some  infinite  sequence of indices $n_j\to\iy$ and some $0<\a<1$.
Then

  i) All gaps $\g_{n_j}=(\l_{n}^-,\l_{n}^+)\ne \es$  are open
  and the gap-length $|\g_{n_j}|\to\iy$ as $j\to\iy$.

 ii) All ramifications $r_{k,n_j}^\pm, k\in\N_{p-1}$ are real and
there exist an infinite
number of the non-empty intervals $(r_{k,n_j}^-,r_{k,n_j}^+)\ss\R$
with the length $|r_{k,n_j}^+-r_{k,n_j}^-|\to\iy$ as $j\to\iy$.
\end{corollary}

Note that if $|\int_0^1 q_{p}(t)e^{-i2\pi nt}dt|\ge {1\/{n}^{\a}}$
as $n\to\iy$ and for some $0<\a<1$ (generic periodic coefficients).
Then there exists only a finite number of non-real ramifications
(branch points of the Lyapunov function) and all high energy gaps
are open.

A great number of  papers is devoted to the inverse spectral
theory for the Hill operator: Dubrovin \cite{D}, Garnett and Trubowitz \cite{GT},
Its and Matveev \cite{IM}, Kappeler \cite{Kap}, Kargaev and Korotyaev \cite{KK},
Korotyaev \cite{K3}, Marchenko
and Ostrovski \cite{MO}, Novikov \cite{No} etc.
Note that Korotyaev \cite{K4}
extended the results of \cite{MO}, \cite{GT}, \cite{Kap},\cite{KK},
\cite{K3} for the case $-y''+qy$ to the case of periodic distributions,
i.e. $-y''+q'y$ on $L^2(\R)$, where periodic $q\in L_{loc}^2(\R)$.

We describe now the results for vector differential equations.
Recently the inverse problem for  vector-valued Sturm-Liouville
operators on the unit interval with Dirichlet boundary conditions,
including characterization, was solved by Chelkak, Korotyaev
\cite{CK1},  \cite{CK2}. The periodic case is more complicated and a
lot of papers are devoted only to the direct problem of periodic systems: Carlson
\cite{Ca1}, \cite{Ca2},  Gelfand and Lidskii \cite{GL}, Gesztesy and
coauthors \cite{CHGL}, Korotyaev and coauthors \cite{CK}, \cite{BBK}, \cite{K1}, \cite{K2}, etc.
 We describe results for first and second
order operators with the periodic $p\ts p$ matrix-valued potential
from \cite{CK}, \cite{K1}, \cite{K2}:

1) the properties of the Lyapunov function, defined on the Riemann surface, are described,

2) the conformal mapping with real part given by the
integrated density of states and imaginary part given by the
Lyapunov exponent is constructed and the main properties are obtained,

3) trace formulas (similar to the case of the Hill operators) are determined,

4) an estimate of gap lengths in terms of  potentials is obtained,

5) sharp asymptotics of periodic eigenvalues and ramifications are determined.

Note that the discrete periodic systems  were studied in
\cite{KKu1}, \cite{KKu2}. The results for first and second order
operators are important for us, since we plan to repeat one for even
order periodic  operators. In fact this is the motivation of our
paper. Note that the case of  even order periodic  operators is more
complicated than the case of  first and second order operators,
since in the first case  only  one fundamental solution is bounded
 on the real line
and all other fundamental solutions are unbounded on the real line.

We describe the fourth order operators $H=\pa^4+\pa q_2 \pa +q_1$.
The results for decreasing coefficients are more developed, see
\cite{AP}, \cite{GM}, \cite{HLO}, \cite{LO}. We mention the paper
\cite{CPS}, \cite{McL} about the inverse problem for fourth order
operators on the unit interval. Now we describe the periodic case.
The authors \cite{BK2} obtained the following results for the
operator $H=\pa^4+\pa q_2 \pa +q_1$ (the case $q_2=0$ see in
\cite{BK1}):

 (1) The properties of the Lyapunov function  on the 2-sheeted Riemann
 surface are described.
The asymptotics of the spectral gaps and ramifications
are determined at high energy.

(2) If $q_2=0,q_1\to 0$ or $q_1=0,q_2\to 0$, then there exists
a small non-empty spectral band with the spectrum of multiplicity 4.
The beginner of this band is the ramification, which coincides with the top of the spectrum.
The spectrum in all other bands has multiplicity 2.

(3) There exist both real and non-real ramifications for some specific
potentials.

The spectral properties of the periodic Euler-Bernoulli equation
$(ay'')''=\l by$ were studied by
Papanicolaou \cite{P1}, \cite{P2}, \cite{PK}
(jointly with Kravarritis). It was shown that the spectrum
is a union of non-overlapping bands of multiplicity 2,
similar to the case of the scalar Hill operator.
The beginning of the spectrum is both a simple periodic eigenvalue
and a branch point of the Lyapunov function.
All other ramifications are negative.

Consider the  operator $H$ with $p\ge 2$.
The old well known results see in the book \cite{Na}.
Tkachenko \cite{Tk} obtained the eigenfunction expansion formula for
the operator $H$. Mikhailets and Molyboga \cite{MM1}, \cite{MM2} determined
asymptotics of  eigenvalues for the operator $(-1)^p\pa^{2p}+q$ on
the circle $\T=\R/\Z$, where $q$ is a distribution. Galunov and
Oleinik \cite{GO} considered the operator
$(-1)^p\pa^{2p}+\d_{per}$ on the real line, where $\d_{per}$ is a
periodic $\d$-function.

It is important that
for $p=1$ the spectral analysis of the operator on the circle
(the periodic and antiperiodic spectrum) is equivalent to one
of the operator $H$ on the real line. The main tool is  the analysis
of the entire Lyapunov function.
The situation for
$p\ge 2$ is much more complicated (see \cite{BK1}, \cite{BK2}).
In this case the Lyapunov function $\D$ has the complicated $p$ sheeted  Riemann surface.

%Maksudov and Veliev  \cite{MV} ??????

In the present paper we extend some of results from \cite{BK1}, \cite{BK2}
about the case $p=2$ to the case $p\ge 2$.
We construct the Riemann surface for the Lyapunov
function of $H$ and describe this surface for large $|\l|$.
Moreover, we determine asymptotics of the ramifications
and periodic and antiperiodic eigenvalues at high energy.

The plan of the paper is as follows.
In Sect. 2 we describe the multipliers for the unperturbed operator.
In Sect. 3  we describe the basic properties of the monodromy matrix $\cM$.
In order to determine the asymptotics of the monodromy matrix $\cM$
at high energy we use so-called Jost type solutions with "good"  asymptotics at high energy.
In Sect. 4 we obtain the main properties of the multipliers, the Lyapunov function and the function $\r$
and prove Theorem \ref{T1}. Moreover,  we consider some simple examples.
In Sect. 5 we prove our main Theorems \ref{rho} and \ref{T2}.
In the proof using  the mix of arguments both for the fourth order operator \cite{BK1}, \cite{BK2}
and for the systems  \cite{CK},  \cite{K1}, \cite{K2}, we determine the asymptotics
of the ramifications and periodic eigenvalues analyzing  directly
the determinant $D$ of the monodromy matrix in the neighborhoods of ramifications
(see Lemma \ref{aD1} and the proof of Theorem \ref{rho}, \ref{T2}).
In the end of Sect. 5 we prove the simple Corollary \ref{ig} from Theorem \ref{T2}.
Some technical proofs are placed in Appendix.

\section {Properties of the unperturbed operator}
\setcounter{equation}{0}

The numbers $\o_j$, given by \er{no}, satisfy
\begin{multline}
\lb{omo}
p\ \text{odd}:\qq
\Re\o_{2p}=\Re\o_{2p-1}<...<\Re\o_4=\Re\o_3<\Re\o_2=\Re\o_1,\\
\Im\o_{2j-1}<0,\qq \o_{2j}=\ol\o_{2j-1},\qq\text{all}\qq j\in\N_p,
\end{multline}
\begin{multline}
\lb{om} p\ \text{even}:\
-1=\o_{2p}<\Re\o_{2p-1}=\Re\o_{2p-2}<...<\Re\o_5=\Re\o_4<\Re\o_3=\Re\o_2<
\o_1=1,\\
\Im\o_{2j}<0,\qq
\o_{2j+1}=\ol\o_{2j},
\qq\text{all}\qq j\in\N_{p-1},
\end{multline}
see Fig.\ref{Z}.
\begin{figure}
\tiny
\unitlength=1.00mm
\special{em:linewidth 0.4pt}
\linethickness{0.4pt}
\begin{picture}(150.66,70.33)
%left part
%net lines
\put(1.66,34.90){\line(1,0){68.67}}
\put(32.50,2.33){\line(0,1){68.00}}
\put(4.00,6.33){\line(1,1){60.00}}
\put(61.00,6.33){\line(-1,1){60.00}}
%labels
\put(57.33,34.2){\linethickness{5pt}\line(0,1){1.33}}
\put(57.33,32.33){\makebox(0,0)[cc]{$\o_1$}}
\put(32.50,13.00){\linethickness{5pt}\line(0,1){1.33}}
\put(29.67,12.33){\makebox(0,0)[cc]{$\o_2$}}
\put(32.50,56.00){\linethickness{5pt}\line(0,1){1.33}}
\put(35.67,57.67){\makebox(0,0)[cc]{$\o_3$}}
\put(10.33,34.2){\linethickness{5pt}\line(0,1){1.33}}
\put(12.67,37.00){\makebox(0,0)[cc]{$\o_4$}}
%shade
%sector 0
\put(37.50,34.90){\line(0,1){5.00}}
\put(42.50,34.90){\line(0,1){10.00}}
\put(47.50,34.90){\line(0,1){15.00}}
\put(52.50,34.90){\line(0,1){20.00}}
\put(57.50,34.90){\line(0,1){25.00}}
\put(62.50,34.90){\line(0,1){30.00}}
\put(67.00,50.00){\makebox(0,0)[cc]{$z\o_1$}}
%sector 1
\put(32.50,29.90){\line(1,0){5.00}}
\put(32.50,24.90){\line(1,0){10.00}}
\put(32.50,19.90){\line(1,0){15.00}}
\put(32.50,14.90){\line(1,0){20.00}}
\put(32.50,9.90){\line(1,0){25.00}}
\put(47.00,5.00){\makebox(0,0)[cc]{$z\o_2$}}
%sector 2
\put(32.50,39.90){\line(-1,0){5.00}}
\put(32.50,44.90){\line(-1,0){10.00}}
\put(32.50,49.90){\line(-1,0){15.00}}
\put(32.50,54.90){\line(-1,0){20.00}}
\put(32.50,59.90){\line(-1,0){25.00}}
\put(32.50,64.90){\line(-1,0){30.00}}
\put(17.00,68.00){\makebox(0,0)[cc]{$z\o_3$}}
%sector 3
\put(27.50,34.90){\line(0,-1){5.00}}
\put(22.50,34.90){\line(0,-1){10.00}}
\put(17.50,34.90){\line(0,-1){15.00}}
\put(12.50,34.90){\line(0,-1){20.00}}
\put(7.50,34.90){\line(0,-1){25.00}}
\put(2.00,20.00){\makebox(0,0)[cc]{$z\o_4$}}
%right part
%net lines
\put(81.99,34.90){\line(1,0){68.67}}
\put(112.66,2.33){\line(0,1){68.00}}
\put(82.66,16.83){\line(5,3){60.00}}
\put(82.66,52.83){\line(5,-3){60.00}}
\put(95.66,6.33){\line(3,5){37.00}}
\put(129.66,6.33){\line(-3,5){35.67}}
\put(94.67,34.33){\line(0,0){0.00}}
\put(126.33,13.00){\line(0,0){0.00}}
%labels
\put(133.67,21.67){\linethickness{5pt}\line(1,0){1.33}}
\put(130.67,22.00){\makebox(0,0)[cc]{$\o_1$}}
\put(133.33,47.50){\linethickness{5pt}\line(1,0){1.33}}
\put(134.33,45.00){\makebox(0,0)[cc]{$\o_2$}}
\put(112.00,7.00){\linethickness{5pt}\line(1,0){1.33}}
\put(110.00,6.67){\makebox(0,0)[cc]{$\o_3$}}
\put(112.00,62.00){\linethickness{5pt}\line(1,0){1.33}}
\put(116.00,62.00){\makebox(0,0)[cc]{$\o_4$}}
\put(91.67,22.50){\linethickness{5pt}\line(1,0){1.33}}
\put(91.33,24.67){\makebox(0,0)[cc]{$\o_5$}}
\put(91.67,47.00){\linethickness{5pt}\line(1,0){1.33}}
\put(93.00,49.00){\makebox(0,0)[cc]{$\o_6$}}
%shade
\put(119.67,35.00){\line(-5,-6){2.33}}
\put(124.00,34.67){\line(-2,-3){3.00}}
\put(128.67,35.00){\line(-3,-5){4.33}}
\put(133.00,35.00){\line(-3,-5){5.67}}
\put(136.67,35.00){\line(-3,-5){6.33}}
\put(140.00,34.67){\line(-3,-5){7.00}}
\put(144.00,35.00){\line(-1,-2){7.33}}
\put(148.00,35.00){\line(-1,-2){8.33}}
\put(147.33,25.00){\makebox(0,0)[cc]{$z\o_1$}}
\put(117.33,42.33){\line(2,-1){4.00}}
\put(119.67,45.67){\line(5,-2){6.33}}
\put(122.67,50.67){\line(2,-1){9.00}}
\put(124.67,54.67){\line(2,-1){11.67}}
\put(127.67,59.33){\line(5,-3){12.67}}
\put(130.00,63.33){\line(2,-1){11.67}}
\put(137.33,64.00){\makebox(0,0)[cc]{$z\o_2$}}
\put(113.00,28.33){\line(1,0){3.67}}
\put(112.67,23.67){\line(1,0){6.67}}
\put(113.00,19.00){\line(1,0){8.67}}
\put(113.00,14.67){\line(1,0){11.67}}
\put(112.67,10.33){\line(1,0){15.00}}
\put(121.00,5.67){\makebox(0,0)[cc]{$z\o_3$}}
\put(108.67,41.00){\line(5,3){4.00}}
\put(106.33,45.00){\line(2,1){6.33}}
\put(104.00,49.33){\line(2,1){8.67}}
\put(101.33,53.67){\line(2,1){11.00}}
\put(98.33,58.33){\line(2,1){14.33}}
\put(95.67,63.00){\line(5,2){6.33}}
\put(105.00,68.67){\makebox(0,0)[cc]{$z\o_4$}}
\put(106.33,31.00){\line(3,-2){3.00}}
\put(102.67,29.00){\line(5,-2){5.33}}
\put(99.67,26.67){\line(5,-2){6.67}}
\put(95.67,24.33){\line(3,-1){9.00}}
\put(91.00,21.67){\line(3,-1){11.00}}
\put(87.00,19.00){\line(3,-1){13.67}}
\put(83.33,16.67){\line(3,-1){15.00}}
\put(88.67,10.67){\makebox(0,0)[cc]{$z\o_5$}}
\put(105.33,39.00){\line(0,-1){4.00}}
\put(101.33,41.33){\line(0,-1){6.33}}
\put(97.67,43.33){\line(0,-1){8.33}}
\put(94.00,46.00){\line(0,-1){11.00}}
\put(90.33,48.33){\line(0,-1){13.00}}
\put(86.33,50.33){\line(0,-1){15.33}}
\put(81.67,44.67){\makebox(0,0)[cc]{$z\o_6$}}
\put(55,70){\makebox(0,0)[cc]{$p=2$}}
\put(145,70){\makebox(0,0)[cc]{$p=3$}}
\end{picture}
\caption{\footnotesize
The numbers $\o_k$ for $p=2$ and $p=3$.
The domains $\{z\o_k,0\le\arg z\le{\pi\/2p}\}$ are shaded.}
\lb{Z}
\end{figure}
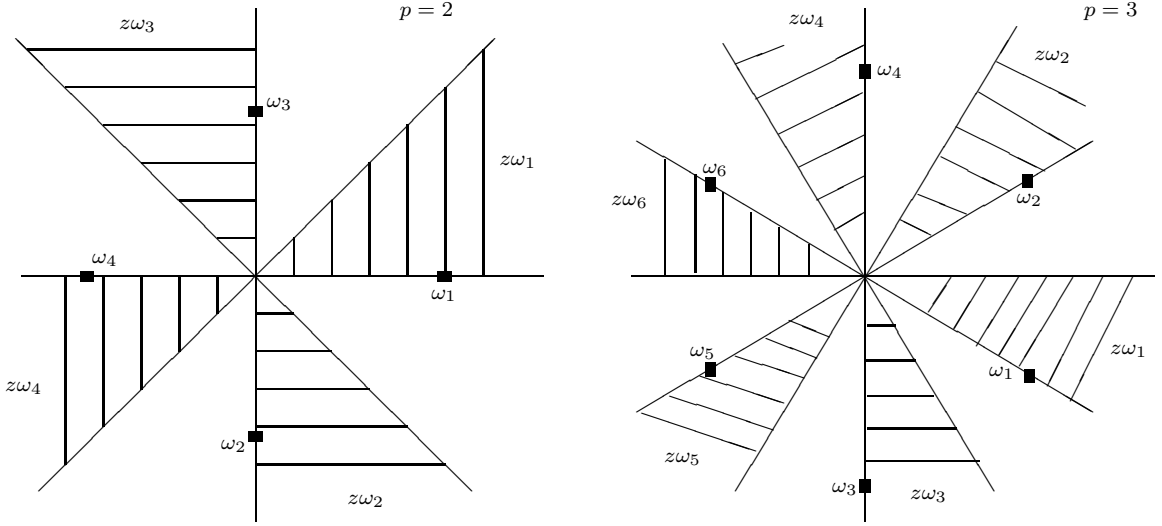
%%%%%%%%%
Moreover,
\[
\lb{ewo}
\o_{p+j+1}=\ve_j\ol\e_j,\qqq \o_{p+j}=\ol\ve_j\ol\e_j,
\qqq\o_{p+j}-\o_{p+j+1}=(-1)^{j+1}2ic_{j}\ol\e_j,
\]
for all $j=-p+1,-p+2,...,p-1$,
where
\[
\lb{ete}
c_j=\cos{\pi j\/2p},\qqq
\ve_j=\ca\qq ie^{i{\pi j\/2p}}\ ,\ j\ \text{even}\\
-ie^{-i{\pi j\/2p}},\  j\ \text{odd}\ac,\qqq
\e_{j}=\ca\ \ 1,\  j\  \text{even}\\
e^{i\pi \/2p},\  j\  \text{odd}\ac.
\]

We introduce the step functions
$\O_j(\l),(j,\l)\in\N_{2p}\ts\C$, which are constant in each half-plane
$\C_\pm=\{\l\in\C:\pm\Im\l>0\}$ and given by
\[
\lb{Oo}
\O_j(\l)=\ca\ \ \ \ \o_j,\ \ \ \ \ \Im\l\ge 0\\
\ol\o_j=\o_\ell,\ \ \Im\l<0\ac,
\]
where
\[
\lb{dEl}
\text{odd}\ p:\ \
\ell=\ca j+1,\ \ j\ \text{odd}\\
j-1,\ \  j\  \text{even}\ac\!\!\!,\qq
\text{even}\ p:\ \ \ell=\ca j+1,\ \  j\  \text{even},\ j\ne 2p\\
j-1,\ \ j\ \text{odd},\ \ j\ne 1\\
\qq j,\qqq j=1,2p\ac\!\!\!.
\]

\begin{lemma}
The functions $\O_j(\l),\l\in\C, j\in \N_{2p}$,
satisfy
\[
\lb{mO}
\Re(z\O_{2p}(\l))\le...\le\Re(z\O_4(\l))\le\Re(z\O_3(\l))\le\Re(z\O_2(\l))\le\Re(z\O_1(\l)),
\]
\[
\lb{mO1}
\Re(z\O_{j}(\l)-z\O_{j+2}(\l))>a|z|,\qq\l\ne 0,\qq\text{all}\qq
j\in\N_{2p-2},
\]
where
\[
\lb{sS}
a=2c_{p-1}\sin{\pi\/4p}>0,\qq
z=\l^{1/2p}\in
S=\Bigl\{z\in\C:\arg z\in\Bigl(-{\pi\/2p},{\pi\/2p}\Bigr]\Bigr\}.
\]
%where $a=2c_{p-1}\sin{\pi\/4p}>0$.

\end{lemma}

\no {\bf Proof.} Assume that \er{mO}, \er{mO1} hold for $\Im\l\ge 0$.
Then identities \er{Oo} give these estimates for $\Im\l<0$.

We will prove \er{mO}, \er{mO1} for $\Im\l\ge 0$,
i.e. $0\le\arg z\le{\pi\/2p}$.
Identities \er{ewo} yield
\[
\lb{oO1}
\Re(z\O_{p+j}(\l)-z\O_{p+j+1}(\l))=
\Re z(\o_{p+j}-\o_{p+j+1})=2c_j(-1)^{j}\Im(z\ol\e_j)
\]
for all $j=-p+1,-p+2,...,p-1$.
Identities
\[
\lb{lo1}
(-1)^j\Im(z\ol\e_j)=|z|
\ca\qq\sin\arg z\ \ ,\ \ \qq\text{if}\ j\ \text{is even}\\
\sin({\pi\/2p}-\arg z)\ ,\ \text{if}\ j\ \text{is odd}\ac\ge 0
\]
yield estimates \er{mO}.
Furthermore, identities \er{oO1} and estimates \er{r0} imply
$$
\Re(z\O_{p+j}(\l)-z\O_{p+j+2}(\l))
=\Re z(\o_{p+j}-\o_{p+j+1})+\Re z(\o_{p+j+1}-\o_{p+j+2})
$$
$$
=(-1)^{j}2c_j\Im(z\ol\e_j)+(-1)^{j+1}2c_{j+1}\Im(z\ol\e_{j+1})
\ge 2c_{j+1}\bigl((-1)^{j}\Im(z\ol\e_j)+(-1)^{j+1}\Im(z\ol\e_{j+1})\bigr).
$$
Identity \er{lo1} and estimates \er{r0} imply
$$
\Re(z\O_{j}(\l)-z\O_{j+2}(\l))
>2c_{p-1}
\max\Bigl\{\sin\arg z,\sin\Bigl({\pi\/2p}-\arg z\Bigr)\Bigr\}|z|,
$$
which yields \er{mO1}.
$\BBox$

We define the branches $\t_j^0,j\in\N_{2p}$,
of the multiplier function $\t^0=e^{i\l^{1/2p}}$
for the unperturbed operator $H^0$ in the upper
half-plane by the identities
$$
\t_j^0(\l)=e^{z\o_j},\qq\text{all}\qq\l\in\C_+.
$$
For each $j\in\N_p$
the functions $\t_{p+j}^0,\t_{p-j+1}^0$ are single-valued analytic
functions in the domain $\mD_{p+j}^0$,
where
\[
\lb{cD}
\mD_{p+j}^0=\C\sm\R,\qq\text{all}\qq j\in\N_{p-1},\qqq
\mD_{2p}^0=\ca\C\sm\R_-\ \text{for even}\ p\\
\C\sm\R_+\ \text{for odd}\ p
\ac.
\]
%The multipliers $\t_j^0,j\in\N_{2p}$, are defined

\begin{figure}
\tiny
\unitlength 1mm % = 2.845pt
\linethickness{0.4pt}
\ifx\plotpoint\undefined\newsavebox{\plotpoint}\fi % GNUPLOT compatibility
\begin{picture}(150.66,70.33)(0,0)
%left part
%net lines
\put(1.66,34.9){\line(1,0){68.67}}
\put(32.5,2.33){\line(0,1){68}}
\put(4,6.33){\line(1,1){60}}
\put(61,6.33){\line(-1,1){60}}
%coordinate line
\put(1.66,14.9){\linethickness{0.1pt}\line(1,0){68.67}}
\put(28.5,16){\makebox(0,0)[cc]{$-2\pi i$}}
\put(1.66,24.9){\linethickness{0.1pt}\line(1,0){68.67}}
\put(28.5,26){\makebox(0,0)[cc]{$-\pi i$}}
\put(1.66,44.9){\linethickness{0.1pt}\line(1,0){68.67}}
\put(34.5,46){\makebox(0,0)[cc]{$\pi i$}}
\put(1.66,54.9){\linethickness{0.1pt}\line(1,0){68.67}}
\put(35.0,56){\makebox(0,0)[cc]{$2\pi i$}}
%shade
%sector 0
\put(37.5,34.9){\line(0,1){5}}
\put(42.5,34.9){\line(0,1){10}}
\put(47.5,34.9){\line(0,1){15}}
\put(52.5,34.9){\line(0,1){20}}
\put(57.5,34.9){\line(0,1){25}}
\put(62.5,34.9){\line(0,1){30}}
\put(67,50){\makebox(0,0)[cc]{$S_1^+$}}
\put(67,20){\makebox(0,0)[cc]{$S_1^-$}}
%sector 1
\put(32.5,29.9){\line(1,0){5}}
\put(32.5,24.9){\line(1,0){10}}
\put(32.5,19.9){\line(1,0){15}}
\put(32.5,14.9){\line(1,0){20}}
\put(32.5,9.9){\line(1,0){25}}
\put(47,5){\makebox(0,0)[cc]{$S_2^+$}}
\put(20,5){\makebox(0,0)[cc]{$S_3^-$}}
%sector 2
\put(32.5,39.9){\line(-1,0){5}}
\put(32.5,44.9){\line(-1,0){10}}
\put(32.5,49.9){\line(-1,0){15}}
\put(32.5,54.9){\line(-1,0){20}}
\put(32.5,59.9){\line(-1,0){25}}
\put(32.5,64.9){\line(-1,0){30}}
\put(17,68){\makebox(0,0)[cc]{$S_3^+$}}
\put(47,68){\makebox(0,0)[cc]{$S_2^-$}}
%sector 3
\put(27.5,34.9){\line(0,-1){5}}
\put(22.5,34.9){\line(0,-1){10}}
\put(17.5,34.9){\line(0,-1){15}}
\put(12.5,34.9){\line(0,-1){20}}
\put(7.5,34.9){\line(0,-1){25}}
\put(2,20){\makebox(0,0)[cc]{$S_4^+$}}
\put(2,50){\makebox(0,0)[cc]{$S_4^-$}}
%right part
%net lines
\put(81.99,34.9){\line(1,0){68.67}}
\put(112.66,2.33){\line(0,1){68}}
\put(82.66,16.83){\line(5,3){60}}
\put(82.66,52.83){\line(5,-3){60}}
\put(95.66,6.33){\line(3,5){37}}
\put(129.66,6.33){\line(-3,5){35.67}}
\put(94.67,34.33){\line(0,0){0}}
\put(126.33,13){\line(0,0){0}}
%coordinate line
\put(81.99,14.9){\linethickness{0.1pt}\line(1,0){68.67}}
\put(109,16){\makebox(0,0)[cc]{$-2\pi i$}}
\put(81.99,24.9){\linethickness{0.1pt}\line(1,0){68.67}}
\put(110,26){\makebox(0,0)[cc]{$-\pi i$}}
\put(81.99,44.9){\linethickness{0.1pt}\line(1,0){68.67}}
\put(114.5,46){\makebox(0,0)[cc]{$\pi i$}}
\put(81.99,54.9){\linethickness{0.1pt}\line(1,0){68.67}}
\put(115.50,56){\makebox(0,0)[cc]{$2\pi i$}}
%shade
%\emline(119.67,35)(117.34,32.204)
%\multiput(119.67,35)(-.03328571,-.03994286){70}{\line(0,-1){.03994286}}
%\end
\put(124,34.67){\line(-2,-3){3}}
\put(128.67,35){\line(-3,-5){4.33}}
\put(133,35){\line(-3,-5){5.67}}
\put(136.67,35){\line(-3,-5){6.33}}
\put(140,34.67){\line(-3,-5){7}}
\put(144,35){\line(-1,-2){7.33}}
\put(148,35){\line(-1,-2){8.33}}
\put(147.33,27){\makebox(0,0)[cc]{$S_1^+$}}
\put(147.33,42.67){\makebox(0,0)[cc]{$S_1^-$}}
%\emline(117.33,42.33)(121.33,40.33)
%\multiput(117.33,42.33)(.06666667,-.03333333){60}{\line(1,0){.06666667}}
%\end
\put(119.67,45.67){\line(5,-2){6.33}}
\put(122.67,50.67){\line(2,-1){9}}
\put(124.67,54.67){\line(2,-1){11.67}}
\put(127.67,59.33){\line(5,-3){12.67}}
\put(130,63.33){\line(2,-1){11.67}}
\put(137.33,64){\makebox(0,0)[cc]{$S_2^+$}}
\put(137.33,13){\makebox(0,0)[cc]{$S_2^-$}}
\put(113,28.33){\line(1,0){3.67}}
\put(112.67,23.67){\line(1,0){6.67}}
\put(113,19){\line(1,0){8.67}}
\put(113,14.67){\line(1,0){11.67}}
\put(112.67,10.33){\line(1,0){15}}
\put(121,68.67){\makebox(0,0)[cc]{$S_3^-$}}
\put(121,5.67){\makebox(0,0)[cc]{$S_3^+$}}
%\emline(108.67,41)(112.67,43.4)
%\multiput(108.67,41)(.05555556,.03333333){72}{\line(1,0){.05555556}}
%\end
\put(106.33,45){\line(2,1){6.33}}
\put(104,49.33){\line(2,1){8.67}}
\put(101.33,53.67){\line(2,1){11}}
\put(98.33,58.33){\line(2,1){14.33}}
\put(95.67,63){\line(5,2){6.33}}
\put(105,68.67){\makebox(0,0)[cc]{$S_4^+$}}
\put(105,5.67){\makebox(0,0)[cc]{$S_4^-$}}
%\emline(106.33,31)(109.33,29)
%\multiput(106.33,31)(.05,-.03333333){60}{\line(1,0){.05}}
%\end
\put(102.67,29){\line(5,-2){5.33}}
\put(99.67,26.67){\line(5,-2){6.67}}
\put(95.67,24.33){\line(3,-1){9}}
\put(91,21.67){\line(3,-1){11}}
\put(87,19){\line(3,-1){13.67}}
\put(83.33,16.67){\line(3,-1){15}}
\put(88.67,10.67){\makebox(0,0)[cc]{$S_5^+$}}
\put(88.67,64){\makebox(0,0)[cc]{$S_5^-$}}
\put(105.33,39){\line(0,-1){4}}
\put(101.33,41.33){\line(0,-1){6.33}}
\put(97.67,43.33){\line(0,-1){8.33}}
\put(94,46){\line(0,-1){11}}
\put(90.33,48.33){\line(0,-1){13}}
\put(86.33,50.33){\line(0,-1){15.33}}
\put(81.67,42.67){\makebox(0,0)[cc]{$S_6^+$}}
\put(81.67,27){\makebox(0,0)[cc]{$S_6^-$}}
\put(119,24.75){\vector(-1,-4){.07}}\put(119,44.75){\vector(-1,4){.07}}\bezier{30}(119,44.75)(121.5,36.25)(119,24.75)
\put(129.75,24.5){\vector(-1,-4){.07}}\put(129.75,44.75){\vector(-1,3){.07}}\bezier{30}(129.75,44.75)(132.75,36.375)(129.75,24.5)
\put(124.75,14.5){\vector(-1,-4){.07}}\put(125,54.75){\vector(-1,3){.07}}\bezier{50}(125,54.75)(130.125,39.375)(124.75,14.5)
\put(112.75,24.75){\vector(-1,-3){.07}}\put(112.75,45){\vector(-1,3){.07}}\bezier{30}(112.75,45)(116.75,35.625)(112.75,24.75)
\put(112.5,14.5){\vector(-1,-2){.07}}\put(112.75,54.75){\vector(-1,2){.07}}\bezier{50}(112.75,54.75)(122.375,37.375)(112.5,14.5)
\put(106.75,24.75){\vector(1,-4){.07}}\put(106.75,44.5){\vector(1,3){.07}}\bezier{30}(106.75,44.5)(103.75,34.875)(106.75,24.75)
\put(100.75,14.75){\vector(1,-4){.07}}\put(100.5,54.75){\vector(1,4){.07}}\bezier{50}(100.5,54.75)(96.375,31.25)(100.75,14.75)
\put(95.5,24.5){\vector(1,-2){.07}}\put(96.25,44.5){\vector(2,3){.07}}\bezier{30}(96.25,44.5)(90.625,35.5)(95.5,24.5)
\put(32.75,24.75){\vector(-1,-4){.07}}\put(32.5,44.75){\vector(-1,3){.07}}\bezier{30}(32.5,44.75)(35.375,35.75)(32.75,24.75)
\put(32.25,14.75){\vector(-1,-3){.07}}\put(32.5,54.75){\vector(-1,2){.07}}\bezier{50}(32.5,54.75)(42.125,37.25)(32.25,14.75)
\put(42.75,24.5){\vector(-1,-4){.07}}\put(42.25,44.75){\vector(-1,3){.07}}\bezier{30}(42.25,44.75)(45,36.875)(42.75,24.5)
\put(52.5,15){\vector(-1,-4){.07}}\put(52.5,54.75){\vector(-1,4){.07}}\bezier{50}(52.5,54.75)(57,34.625)(52.5,15)
\put(22.5,25){\vector(1,-4){.07}}\put(22.75,44.5){\vector(1,4){.07}}\bezier{30}(22.75,44.5)(19.875,34.75)(22.5,25)
\put(12.25,15){\vector(1,-4){.07}}\put(12.5,54.5){\vector(1,3){.07}}\bezier{50}(12.5,54.5)(6.625,38.25)(12.25,15)
\put(65,70){\makebox(0,0)[cc]{$p=2$}}
\put(145,70){\makebox(0,0)[cc]{$p=3$}}
\end{picture}
\caption{\footnotesize
The $\z$-plane for the cases $p=2$ and $p=3$, where $\z=iw^{1\/2p},
w\in\mL^0$.
Each sector $S_j^\pm$ is the image of the $\C_\pm$
half-plane on the $j$-th sheet
of the surface $\mL^0$.
The values of the function $\t^0(w)=e^\z$
at the points, connected by the lines with arrows,
are equal one with other.
}
\lb{Z1}
\end{figure}
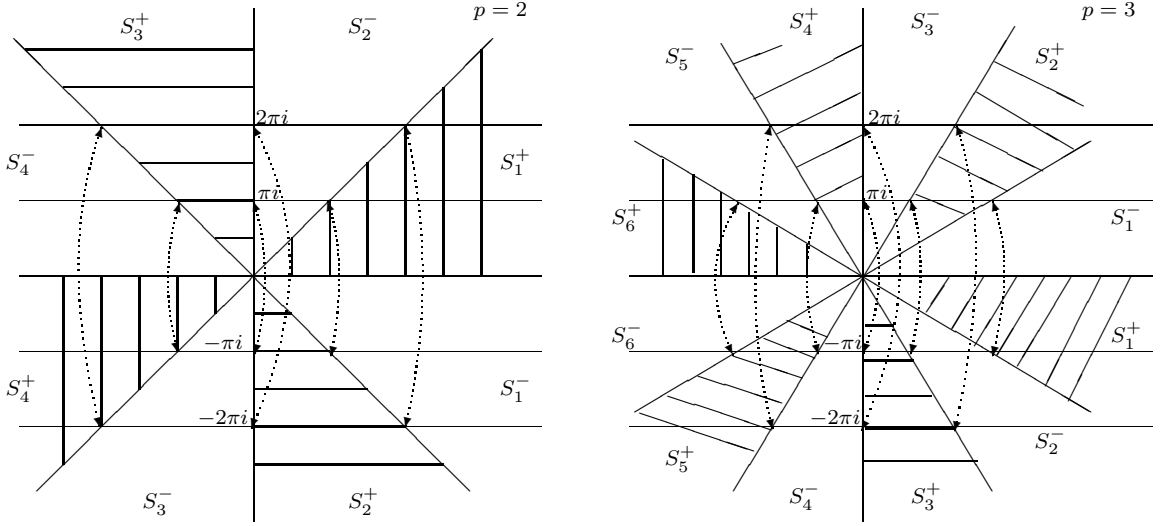
%%%%%%%%%

\begin{lemma}
\lb{dm0}

i) The unperturbed multipliers satisfy the identities
\[
%\lb{sum}
\lb{m=O}
\t_j^0(\ol\l)=\ol{\t_j^0}(\l),\qqq
\t_j^0(\l)=e^{z\O_j(\l)},\qqq
\text{all}\qq
(j,\l)\in\N_{2p}\ts(\C\sm\R),
\]
\[
\lb{mim}
\t_{p-j+1}^0(\l)=(\t_{p+j}^0(\l))^{-1},\qqq\text{all}
\qq(j,\l)\in\N_{p}\ts\mD_{p+j}^0.
\]

 ii) Let $\t_{p+k}^0(\l)=\t_{p+j}^0(\l)$ for some
$\l\in\C\sm\R$, $0\le k<j\le p$. Then
$j=k+1$ and $\l=r_{k,n}^0+i0$ or $\l=r_{k,n}^0-i0$ for some $n\in\N$.
Moreover,
\[
\lb{it0}
\t_{p+k}^0(r_{k,n}^0+i0)\!=\!\t_{p+k}^0(r_{k,n}^0-i0)\!=\!
\t_{p+k+1}^0(r_{k,n}^0+i0)\!=\!\t_{p+k+1}^0(r_{k,n}^0-i0),\ \
k\in\N_{p-1}^0=\{0,1,...,p-1\}.
\]

\end{lemma}

\no {\bf Proof.}
i) The Riemann surface $\mL^0$ of the function $\t^0$
coincides with the
Riemann surface of the function $\l^{1/2p}$ with the unique branch
point at $\l=0$. This surface has $2p$ sheets $\mL_j^0,j\in\N_{2p}$,
corresponding to the branches $\t_j^0$, that is
$\t_j^0(\l)=\t^0(w)|_{w\in\mL_j^0}$,
where $\l\in\C$ is the projection of the point $w\in\mL^0$.
For each $j\in\N_p$ the projection of the sheet $\mL_{p+j}^0$ (and
$\mL_{p-j+1}^0$) is the domain $\mD_{p+j}^0$ and has the cut along
the real axis or semi-axis, see \er{cD}.
The upper (lower) edge of each cut on the sheet
$\mL_{p+j}^0,j\in\N_p$, is attached to the lower (upper) edge of the
corresponding cut on the sheet $\mL_{p+j+1}^0$. Similarly the upper
(lower) edge of each cut on the sheet $\mL_{p-j}^0,j\in\N_p$, is
attached to the lower (upper) edge of the corresponding cut on the
sheet $\mL_{p-j+1}^0$.

The simple parametrization of the surface $\mL^0$ is given by
the analytical mapping $W$ having the form
$w\to\z=iw^{1\/2p}$, where $w\in\mL^0$. We have
$$
W(\mL^0)=\C.
$$
Describe this parametrization in more details.
Introduce the sectors (see Fig. \ref{Z1})
\[
\lb{dS}
S^+=\Bigl\{z\in\C:0<\arg z<{\pi\/2p}\Bigr\},\qq
S_j^+=\o_j S^+,\qq
S_j^-=\{\z\in\C:\ol\z\in S_j^+\},\qq j\in\N_{2p}.
\]
Let $\mL_j^{0,\pm},j\in\N_{2p}$, be the half-plane $\C_\pm$
on the sheet $\mL_j^0$ of the surface $\mL^0$.
Then
\[
\lb{WM}
W(\mL_j^{0,\pm})=S_j^\pm,\qq
W(\mL_j^{0})=S_j,\qq
\qq\text{all}\qq j\in\N_{2p},\qq\text{where}\qq S_j=S_j^+\cup S_j^-.
\]

Each function $\t_j^0,j\in\N_{2p}$,
satisfies the identity $\t_j^0(\l)=\t^0(w)|_{w\in\mL_j^0}$,
where $\l\in\C$ is the projection of the point $w\in\mL^0$.
Let $\z=W(w)\in\C,w\in\mL^0$. Then $\t^0(w)=e^\z$.
Using identities \er{WM} we obtain
$\t_j^0(\l)=e^{\z_j}$, where $\z_j$ is given by
the conditions $\z_j\in S_j, \l=(-1)^p\z_j^{2p}$.
Then $\ol\l=(-1)^p{\ol\z_j}^{2p}$ (see \er{dS} and Fig. \ref{Z1}) and
the identities
$
\t_j^0(\ol\l)=e^{\ol\z_j}=\ol{e^{\z_j}}=\ol{\t_j^0}(\l)
$
for all $(j,\l)\in\N_{2p}\ts(\C\sm\R)$ give
the first identity in \er{m=O}.
Identities \er{Oo} imply the second identity in \er{m=O}.
%Identity \er{sum} gives \er{m=O}.
Identities \er{no} yield \er{mim}.

 ii) Let $\t_{p+k}^0(\l)=\t_{p+j}^0(\l)$ for some
$\l\in\C\sm\R,0\le k<j\le p$.
Then
\[
\lb{um1}
e^{\z_{p+k}}=e^{\z_{p+j}},
\qq\z_{p+k}^{2p}=\z_{p+j}^{2p}=(-1)^p\l,\qq
\z_{p+k}\in S_{p+k},\qq\z_{p+j}\in S_{p+j}.
\]
The first identities in \er{um1} yield $\z_{p+k}-\z_{p+j}=i2\pi n$
for some $n=\pm 1,\pm2,...$
Then we have $\Re\z_{p+k}=\Re\z_{p+j}$.
The second identities in \er{um1} imply $|\z_{p+k}|=|\z_{p+j}|=|z|$,
which yields $\z_{p+k}=\ol{\z_{p+j}}$.
Moreover, $\Im\z_{p+k}-\Im\z_{p+j}=2\pi n$.
Identities \er{dS} and the condition $k<j$ (see also Fig. \ref{Z1})
give:
a) $j=k+1$;
b) $\z_{p+k}={\pi n\/c_k}e^{\pm i{\pi k\/2p}}$,
i.e. $\l=\z_{p+k}^{2p}=(-1)^k({\pi n\/c_k})^{2p}=r_{k,n}^0$;
and
c) identities \er{it0}.
$\BBox$

Consider the operator
$H^\mu=(-1)^p{d^{2p}\/dt^{2p}}+\mu {d^{2p-2}\/dt^{2p-2}},\mu\in\R$.
The equation
\[
\lb{ieso}
(-1)^p y^{(2p)}+\mu y^{(2p-2)}=\l y,\qq(t,\l)\in\R\ts\C,
\]
has the solutions $e^{\o^\mu_j(\l)z t},j\in\N_{2p}$,
where $\o^\mu_j(\l),j\in\N_{2p}$, are the solutions of the equation
\[
\lb{mee}
(\o^\mu)^{2p}+(-1)^{p}(\o^\mu)^{2p-2}\mu z^{-2}-(-1)^{p}=0.
\]
The functions $\o^\mu_j,j\in\N_{2p}$, constitute branches
of the analytic function $\o^\mu$ having only algebraic singularities.
For $\l\in\L_R$ for some $R>0$ large enough, we define these branches
by the asymptotics $\o^\mu_j(\l)=\o_j+o(1)$ as $|\l|\to\iy$,
here and below
$$
\L_{R}=\{\l\in\C:|\l|>R^{2p}\},\qq R>0,\qqq \L_{R}^{\pm}=\L_{R}\cap \C_\pm.
$$
The multipliers have the form
$\t_j^\mu(\l)=e^{z\o_j^\mu(\l)}$ for all $(j,\l)\in\N_{2p}\ts\L_R^+$
for some $R>0$ large enough.
Then
\[
\lb{alb}
\t_j^\mu(\l)=e^{z\O_j^\mu(\l)},\qq\text{all}\qq
(j,\l)\in\N_{2p}\ts\L_R,\qqq\text{where}\qq
\O^\mu_j(\l)=\ca\o^\mu_j(\l),\ \ \Im\l\ge 0\\
\o_\ell^\mu(\l),\ \ \Im\l<0\ac,
\]
where $\ell$ is given by identities \er{dEl}.
Each function $\O^\mu_j(\l),j\in\N_{2p},\m\in\R$, is analytic in
$\l\in\L_R^\pm$
%.Each function $\O^\mu_j,j\in\N_{2p}$, is
and piecewise-continuous in $\l\in\C$,
but the set $\{\O^\mu_j(\l),j\in\N_{2p}\},\m\in\R$, is continuous in $\l\in\C$.
The branches of the Lyapunov function are given by
$\D_j^\mu(\l)=\cosh z\O^\mu_j(\l),(j,\l)\in\N_{p}\ts\L_R$.

\begin{lemma}
\lb{upe}
i) Each function $\o^\mu_j,j\in\N_{2p}$, satisfies the asymptotics
\[
\lb{abe}
\o^\mu_j(\l)=\o_j-{(-1)^p \mu\/2p\o_jz^2}+O(|z|^{-4})
\qq\text{as}\qq|\l|\to\iy,
\]
\[
\lb{b-b0}
\o^\mu_j(\l+\ve)=\o^\mu_j(\l)+O(|z|^{-4})\qq\text{as}\qq
|\l|\to\iy,\qq \ve=O(|z|^{2p-2}).
\]

 ii) The periodic and antiperiodic eigenvalues for equation
$(-1)^p y^{(2p)}+\mu y^{(2p-2)}=\l y$
satisfy:
\[
\lb{les}
\l_0^{\mu,+}=0,\qq\l_n^{\mu,-}=\l_n^{\mu,+}
=(\pi n)^{2p}-(-1)^{p}\mu (\pi n)^{2p-2},
\qq\text{all}\qq n\ge 1.
\]

 iii) Each function $\O^\mu_j,j\in\N_{2p}$,
%is analytic in $\C_\pm\cap\{\l\in\C:|\l|>R\}$ for some $R>0$ and
satisfies the asymptotics
\[
\lb{asb}
\O^\mu_j(\l)=\O_j(\l)+O(|z|^{-2})\qq\text{as}\qq |\l|\to\iy,
\]

\end{lemma}

\no {\bf Proof.} i) Substituting $\o^\mu_j=\o_j+\d,\d=o(1)$, into the identity
$(\o^\mu_j)^{2p}+(-1)^{p}(\o^\mu_j)^{2p-2}\mu z^{-2}-(-1)^{p}=0$ we obtain
\[
\lb{am1}
(\o_j+\d)^{2p}+(-1)^{p}(\o_j+\d)^{2p-2}\mu z^{-2}=(-1)^{p}.
\]
This identity gives $2p\o_j^{2p-1}\d+O(\d^2)+O(|z|^{-2})=0$, which yields
$\d=O(|z|^{-2})$. Using \er{am1} again we obtain
$2p\o_j^{2p-1}\d+(-1)^{p}\o_j^{2p-2}\mu z^{-2}=O(|z|^{-4})$.
This asymptotics gives
$\d=(-1)^{p+1}(2p\o_j)^{-1}\mu z^{-2}+O(|z|^{-4})$, which yields \er{abe}.
Asymptotics \er{abe} yield
\[
\lb{b-b1}
\o^\mu_j(\l+\ve)-\o^\mu_j(\l)
={(-1)^p\mu\/2p\o_j}\lt({1\/z^2}-{1\/\z^2}\rt)+O(|z|^{-4})
\qq\text{as}\qq |\l|\to\iy,
\]
where $\z=(\l+\ve)^{1\/2p}$.
Using the asymptotics $\z=z+O(|z|^{-1})$, we obtain \er{b-b0}.

 ii) We will prove \er{les} for the periodic eigenvalues.
The proof for the antiperiodic eigenvalues is similar.
The periodic eigenvalues
$\l_{0}^{\mu,+},\l_{2n}^{\mu,\pm},n\ge 1$,
are zeros of the entire function
$D_+^\mu
%=2^{-p}\det(\cM^\mu(1,\cdot)-\1_{2p})
=\prod_{j=1}^p(\D_j^\mu-1)$.
We have
\[
\lb{Des}
D_+^\mu=\prod_{j=1}^p(\cosh \o^\mu_j z-1)
=2^p\prod_{j=1}^p\sinh^2{\o^\mu_j z\/2}
={\l\/2^p}\prod_{j=1}^p(\o^\mu_j)^2\prod_{n=1}^\iy
\lt(1+{(\o^\mu_j)^2 z^2\/(2n\pi)^2}\rt)^2.
\]
Using the simple identity
$$
\o^{2p}+(-1)^{p}\o^{2p-2}\mu z^{-2}-(-1)^{p}
=\prod_{j=1}^p(\o^2-(\o^\mu_j)^2),\qq\text{all}\qq \o\in\C,
$$
we obtain $\prod_{j=1}^p(\o^\mu_j)^2=-1$ (put $\o=0$) and
$$
\prod_{j=1}^p\lt(1+{(\o^\mu_j)^2 z^2\/(2n\pi)^2}\rt)
={(-1)^p\l\/(2n\pi)^{2p}}\prod_{j=1}^p\lt(\lt({2in\pi\/z}\rt)^2
-(\o^\mu_j)^2\rt)
=1-{(-1)^p \mu \/(2n\pi)^{2}}-{\l\/(2n\pi)^{2p}}.
$$
Substituting these identities into \er{Des} we obtain
$$
D_+^{\mu}=-{\l\/2^p}\prod_{n=1}^\iy
\lt(1-{(-1)^p \mu \/(2n\pi)^{2}}-{\l\/(2n\pi)^{2p}}\rt)^2,
$$
which yields \er{les} for the periodic eigenvalues.

 iii) Asymptotics \er{abe} and definition \er{alb} of $\O_j^\mu$
yield \er{asb}.
$\BBox$

\no {\bf Remark.} The periodic eigenvalue $\l=0$ for equation \er{ieso} is simple
and other periodic and antiperiodic eigenvalues
have multiplicities $2$.

\section {Fundamental solutions}
\setcounter{equation}{0}

In this section we consider the operator $H=H_0+q,\  H_0=(-1)^p{d^{2p}\/dt^{2p}}$ and recall that
$$
q=\sum_{j=0}^{p-1}{d^{j}\/dt^{j}} q_{j+1} {d^{j}\/dt^{j}},\qqq
q_j\in L_{real}^1(\T),\qq j\in\N_p=\{1,2,...,p\},\qq\ p\ge 2,
$$
The form domain of the
self-adjoint operator $H_0$ is the set $\Dom_{fd}(H_0)=W_{p}^2(\R)$.
The quadratic form $(qy,y)$ is defined by
$
(qy,y)=\sum_{j=0}^{p-1}(-1)^j(q_{j+1}y^{(j)},y^{(j)})
$,
$y\in W^2_p(\R)$.
Let $q_j=\hat q_{j,0}+a_j'$, where  $\hat q_{j,0}=\int_0^1q_j(t)dt$ and
$a_j\in W_1^1(\T), \int_0^1a_j(t)dt=0,
j\in\N_p$. The integration by parts gives the form
\[
\lb{qfq}
(qy,y)=\sum_{j=0}^{p-1}(-1)^j\lt(\hat q_{j+1,0}\|y^{(j)}\|^2
-2\Re (a_{j+1}y^{(j+1)},y^{(j)})\rt),
\]
correctly defined on the form domain $y\in W^2_p(\R)$.

For each $q_{j}, j\in\N_p$, we introduce the sequence $(q_{j,n})_{n=1}^\iy$
of the smooth functions
$q_{j,n}\in W_{j-1}^1(\T)$, such that
\[
\lb{ben}
\int_0^1(q_j-q_{j,n})dt=0\qq\text{ all}\qq n\in\N,\qqq
\b_{n}=\sup_{j\in\N_p}\int_0^1|q_j-q_{j,n}|dt\to 0\qq\text{ as}\qq n\to\iy.
\]
Let
\[
\lb{Hn}
H_n=H_0+q_{(n)},\qq n\in\N,\qq\text{where}\qq
q_{(n)}=\sum_{j=0}^{p-1}{d^{j}\/dt^{j}} q_{j+1,n} {d^{j}\/dt^{j}}.
\]

\begin{proposition}
\lb{eqf}
i) The quadratic form $(qy,y)$ satisfies
\[
\lb{VE}
|(qy,y)|\le{1\/2}\|y^{(p)}\|^2+C\|y\|^2,\qqq \text{all} \qq y\in W_{p}^2(\R)
\]
for some constant $C>0$, where $\|y\|^2=(y,y)$ is the scalar product in $L^2(\R)$.

 ii) There exists a unique self-adjoint operator  $H=H_0+q$
with the form domain $\Dom_{fd}(H)=W_{p}^2(\R)$ and
\[
\lb{qf4}
(Hy,y_1)=(H_0y,y_1)+(qy,y_1),\qq\text{all}\qq
y,y_1\in W_{p}^2(\R).
\]

 iii) Let $y\in W_p^2(\R)$.
Then there exists the sequence $(\ve_n)_1^\iy$ such that
$\ve_n>0$ for all $n\in\N$, $\ve_n\to 0$ as $n\to\iy$,
and
\[
\lb{qf7}
|((H-H_n)y,y)|\le \ve_n(|(Hy,y)|+\|y\|^2),\qq\text{all}\qq n\in\N.
\]
\end{proposition}

\no {\bf Proof.}
i) We get
\[
\lb{qf1}
2|(a_{j}y^{(j)},y^{(j-1)})|\le 2b\|y^{(j)}\|\|y^{(j-1)}\|
\le\ve\|y^{(j)}\|^2+{b^2\/\ve}\|y^{(j-1)}\|^2, \qq b=\sup_{(t,j)\in \T\ts \N_p} |a_{j}(t)|
\]
for any $\ve >0$, where $(y,y)=\|y\|^2=\int_\R |y|^2dt$.
Thus \er{qfq} yields
\[
\lb{qf3}
|(qy,y)|\le \ve\|y^{(p)}\|^2+ \wt C\sum _0^{p-1}\|y^{(j)}\|^2, \qqq \wt C=q_0+\ve +{b^2\/\ve},\qqq
\hat q_0=\sum _0^{p-1}|\hat q_{j,0}|.
\]
Using the simple estimate
$
\sum _0^{p-1}\|y^{(j)}\|^2\le\ \ve^2\|y^{(p)}\|^2+C_\ve\|y\|^2
$
for some $C_\ve>0$, we obtain
\[
\lb{qf33}
|(qy,y)|\le \ve(1+\ve \wt C)\|y^{(p)}\|^2+ \wt C C_\ve\|y\|^2,
\]
which yields  \er{VE} with ${1\/2}=\ve(1+\ve \wt C)$ and $C=\wt C C_\ve$
for $\ve$ small enough.

ii) The operator $q$ on the domain $\Dom_{fd}(H_0)$ is given by \er{qfq}
and satisfies the estimate \er{VE}.
Using the KLMN theorem (see \cite{RS1}) we obtain that
there exists a unique self-adjoint operator  $H=H_0+q$
with the form domain $W_{p}^2(\R)$ and identity \er{qf4}
holds true.

 iii)
Repeating the previous arguments and using $\|y^{(j)}\|^2\le\|y^{(p)}\|^2+\|y\|^2$
we obtain
$$
|((q_{j}-q_{j,n})y^{(j-1)},y^{(j-1)})|\le 2\b_{n}\|y^{(j)}\|\|y^{(j-1)}\|
\le\b_{n}(\|y^{(j)}\|^2+\|y^{(j-1)}\|^2)
$$
$$
\le
2\b_{n}(\|y^{(p)}\|^2+\|y\|^2)=2\b_{n}((H_0y,y)+\|y\|^2),
$$
where $\b_n$ is given by \er{ben}.
This yields
\[
\lb{qf2}
|((H-H_n)y,y)|\le
\sum_0^{p-1}|((q_{j+1}-q_{j+1,n})y^{(j)},y^{(j)})|\le
2p\b_n\bigl((H_0y,y)+\|y\|^2\bigr).
%=\a_n|(H_0y,y)|,
\]
Using estimates \er{VE} we obtain
$$
(H_0y,y)\le|(Hy,y)|+|(qy,y)|\le|(Hy,y)|+{1\/2}(H_0y,y)+C\|y\|^2,
$$
which yields $|(H_0y,y)|\le2|(Hy,y)|+2C\|y\|^2$. Substituting
this estimate into \er{qf2} we obtain \er{qf7},
where $\ve_n=4p\b_n (1+C)$.
% and denoting $\ve_n=2\a_n,t_n=()$.
\BBox

%\no {\bf Remark.}
%Any domain of essential self-adjointness for $H_0$ is a form core for $H$.

Below a vector $h=(h_n)_1^N\in\C^N$ has the norm $|h|=\sum_1^N|h_n|$,
while an $N\ts N$ matrix $\cA=(\cA_{ij})_{i,j=1}^N$
has the operator norm given by $|\cA|=\sup_{|h|=1}|\cA h|=\max_{1\le j\le N}\sum_{i=1}^N|\cA_{ij}|$.
Always below we denote $n\ts n$ diagonal matrices by
$\diag(a_j)_{j=1}^n=(a_j\d_{jk})_{j,k=1}^{n}$.
The following Lemma (proof in Appendix) describes the basic properties
of the monodromy matrix $\cM(1,\l)$.

\begin{lemma}
\lb{T21}
The matrix-valued function $\cM(1,\l)$ is entire and satisfies:
\[
\lb{2if}
|\cM(1,\l)|\le 2pe^{z_0+\vk},\ \ \text{all}\ \ \l\in\C,\qqq
|\cZ^{-1}(\l)\cM(1,\l)\cZ(\l)|\le 2pe^{z_0+\vk},\ \
\text{all}\ \ |\l|\ge 1,
\]
where
$$
\cZ=\diag(z^{j-1})_1^{2p},\qq
\vk=\max_{j\in\N_p}\int_0^1|q_j(t)|dt,\qq z_0=\max_{j\in\N_p}|\Re(z\o_j)|.
$$
Moreover, $\cM(1,\l)$ is a continuous function of $(q_j)_1^p\in L^1(\T)^p$.

\end{lemma}

\begin{lemma}
\lb{Spec}
Let each $q_j\in W_{j-1}^1(\T)$, $j\in\N_{p}$ and let $\l\in\C$.
Then the spectrum of the $2p\ts 2p$ matrix $(\vp_j^{(k-1)}(1,\l))_{k,j=1}^{2p}$
coincides with the spectrum of the matrix $\cM(1,\l)$, counted with multiplicity.
Moreover, in this case
\[
\lb{st}
\s(H)=\{\l\in\R:|\t_j(\l)|=1\ \text{for some}\ j\in\N_p\}.
\]
\end{lemma}

\no {\bf Proof.}
Introduce the vector-valued function $\wt Y=(y^{(j-1)})_{j=1}^{2p}$.
Identity \er{Y} shows that
%\[
%\lb{cs1}
%Y=\cS\wt Y,\qq\text{where}\qq
%\cS=\ma\1_{p+1}&\dO_{p+1,p-1}\\\ma\dO_{p-1,1}&
%{\ma
%0&...&0& q_p\\
%0&...&q_{p-1}&0\\
%...&...&...&...\\
%q_2&...&0&0
%\am}
%&\dO_{p-1,1}\am&\1_{p-1}\am.
%\]
\[
\lb{cs1}
Y=\cS\wt Y,\qq\text{where}\qq
\cS=\ma\1_{p+1}&\dO_{p+1,p-1}\\
\wt S&\1_{p-1}\am,
\qq\wt S=\ma
0&0&...&0& q_p&0\\
0&0&...&q_{p-1}&0&0\\
...&...&...&...\\
0&q_2&...&0&0&0
\am.
\]
Let $\wt\cM(t,\l)=(\vp_j^{(k-1)}(t,\l))_{k,j=1}^{2p},
(t,\l)\in\R\ts\C$.
Identity \er{cs1} shows that each matrix-valued function
$\cS\wt\cM(\cdot,\l)$, $\l\in\C$,
satisfies equation \er{me}, which yields
$\cM(t,\l)=\cS(t)\wt\cM(t,\l)\cS^{-1}(0)$.
Using $\cS(1)=\cS(0)$ we obtain
$\cM(1,\l)=\cS(0)\wt\cM(1,\l)\cS^{-1}(0)$.
Thus the matrices
$\cM(1,\l)$ and $\wt\cM(1,\l)$ are similar and
the spectra of these matrices coincide one with
other, counted with multiplicity.
Identity \er{st} follows (see, e.g., \cite{DS}, Th.~ XIII.7.64).
$\BBox$

We will introduce the Jost type fundamental
matrix solution $\cT(t,\l)$ of equation \er{me2}.
This solution will be described below.
Recall that the monodromy matrix has the form $\cM(1,\l)$,
where the matrix-valued function $\cM(t,\l)$ satisfies the matrix
equation \er{me}.
Rewrite equation \er{me} in the form
\[
\lb{me1}
\cM'-\cP^\mu(\l)\cM=\cQ^\mu(t)\cM,\qqq (t,\l)\in\R\ts\C,
\]
where the $2p\ts 2p$ matrices $\cP^\mu$ and $\cQ^\mu$ are given by
\[
\lb{mu}
\cP^\mu=\cP-(-1)^p\mu\cE,\qq \cQ^\mu=\cQ+(-1)^p\mu\cE,
\qq \mu=\int_0^1q_p(t)dt\in\R,
\]
the matrices $\cP,\cQ$ are defined by \er{PQ},
and the matrix $\cE$ is given by
\[
\lb{mK}
\cE=(\cE_{jk})_{j,k=1}^{2p},\qq \cE_{p+1,p}=1,\qq
\cE_{jk}=0,\qq\text{all}\qq (j,k)\ne(p+1,p).
\]
Each matrix $\cP^\mu(\l)$, $\l\in\L_R=\{\l\in\C:|\l|>R^{2p}\}$ for some $R>0$ large enough,
has eigenvalues $z\O^\mu_j(\l),j\in\N_{2p}$, all these eigenvalues are simple
and the corresponding eigenvectors
%$U_j(\l)$
are given by
$$
U_j=
\ma1\\ z\O^\mu_j\\(z\O^\mu_j)^2\\ \vdots\\
(z\O^\mu_j)^p\\(z\O^\mu_j)^{p+1}+(-1)^p\mu(z\O^\mu_j)^{p-1}\\
\vdots\\(z\O^\mu_j)^{2p-1}+(-1)^p\mu(z\O^\mu_j)^{2p-3}\am,\qqq j\in\N_{2p}.
$$
%Since all eigenvalues of $\cP^\mu(\l)$ are simple,
Then the matrix $\cP^\mu$ is similar to the diagonal matrix
$$
z\cB^\mu=\cU^{-1}\cP^\mu\cU,\qqq\text{where}\qqq\cB^\mu=\diag(\O^\mu_j)_{1}^{2p},
$$
and the $2p\ts 2p$ matrix $\cU$ has the form
$$
\cU=\ma
U_1& U_2 & ... & U_{2p}
\am.
$$
We rewrite equation \er{me1} in the form
\[
\lb{me2}
\wt\cM'-z\cB^\mu(\l)\wt\cM=\wt\cQ(t,\l)\wt\cM,\qqq (t,\l)\in\R\ts\L_R,
\]
where
$$
\wt\cM=\cU^{-1}\cM\cU,
\qqq\wt\cQ=(\wt\cQ_{ij})_{i,j=1}^{2p}=\cU^{-1}\cQ^\mu\cU.
$$
The following Lemma, proved in Appendix, shows that the matrix $\wt\cQ$
is decreasing at large $|\l|$.

\begin{lemma}
\lb{dcQ}
The matrix-valued function
$\wt\cQ$ satisfies the following asymptotics:
\[
\lb{wQ}
\wt\cQ(t,\l)={(-1)^{p+1}\/z}\Bigl((q_p(t)-\mu)\cL(\l)+b(t)O(|z|^{-1})\Bigr),\qq
\cL=(\cL_{jk})_{j,k=1}^{2p},\qq
\cL_{jk}={\ol\O_{j}^{p}\O_{k}^{p-1}\/2p},
\]
uniformly on $t\in[0,1]$ as $|\l|\to\iy$,
where $b(t)=\max_{j\ne p}|q_j(t)|$.

\end{lemma}

%{\bf Remark.}
Consider equation \er{me2}.
Assume that this equation has the
$2p\ts 2p$ matrix-valued solution $\cT(t,\l)$ for all $t\in[0,1]$
and some $\l\in\L_R$.
%, i.e. $\cT'-z\cB^\mu\cT=\wt\cQ\cT$.
Then $\wt\cM(t,\l)=\cT(t,\l)\cT(0,\l)^{-1}$ and
\[
\lb{dE}
\cM(t,\l)=\cU(\l)\wt\cM(t,\l)\cU^{-1}(\l)
=\cU(\l)\cT(t,\l)\cT(0,\l)^{-1}\cU^{-1}(\l),\qq (t,\l)\in[0,1]\ts\L_R.
\]
%It seems to be reasonable to rewrite equation \er{me2}
%in the form $\wt\cT'=e^{-z\cB^\mu t}\wt\cQ e^{z\cB^\mu t}\wt\cT$, where
%$\wt\cT=e^{-z\cB^\mu t}\cT$.
%However, this form is not available, since
%it is difficult to analyze the asymptotics  of the matrix valued function
%$e^{-z\cB^\mu t}\wt\cQ e^{z\cB^\mu t},t\in[0,1]$, as
%$|\l|\to\iy$.
In order to analyze equation \er{me2} by the Birkhoff method (see \cite{Na}),
we write the solution $\cT$ in the form
\[
\lb{FT}
\cT=\cG e^{z\cB^\mu t},
\]
where $\cG(t,\l),(t,\l)\in [0,1]\ts\L_R$, is some matrix-valued function.
Substituting \er{FT} into \er{me2} we obtain
\[
\lb{ecG}
\cG'+z(\cG\cB^\mu-\cB^\mu\cG)=\wt\cQ\cG.
\]
This equation is equivalent to the integral equation
\[
\lb{me5}
\cG=\1_{2p}+K\wt\cQ\cG,
\]
where $K$ is an integral operator given by
\[
\lb{dcL}
(K\cA)_{ij}(t,\l)=\ca
\ \ \int_0^t e^{z(t-s)(\O^\mu_i(\l)-\O^\mu_j(\l))}\cA_{ij}(s)ds,
\ \ \ \text{if}\ i>j\\
-\int_t^1 e^{z(t-s)(\O^\mu_i(\l)-\O^\mu_j(\l))}\cA_{ij}(s)ds,
\ \ \text{if}\ i\le j\ac\!\!\!\!
=\int_0^1e_{ij}(t-s,\l)\cA_{ij}(s)ds
\]
for the matrix-valued function $\cA$, where $(i,j,t,\l)\in\N_{2p}^2\ts[0,1]\ts\L_R$
and
\[
\lb{e}
e_{i j}(t,\l)=\ca\ \ \,  e^{zt(\O^\mu_i(\l)-\O^\mu_j(\l))}\chi(t)\ \ ,\  \text{if}\ i>j\\
-e^{zt(\O^\mu_i(\l)-\O^\mu_j(\l))}\chi(-t),\ \text{if}\ i\le j
\ac\!\!\!\!,\qqq
\chi(t)=\ca 0,\ \text{if}\ t<0\\ 1,\ \text{if}\ t\ge 0\ac\!\!\!\!.
\]
In fact, differentiating \er{me5}
%substituting \er{dcL}, \er{e} into \er{cY}
we obtain
\begin{multline}
\cG_{ij}'(t,\l)=
(K\wt\cQ\cG)_{ij}'(t,\l)={d\/dt}\ca
\ \ \int_0^t e^{z(t-s)(\O^\mu_i(\l)-\O^\mu_j(\l))}(\wt\cQ\cG)_{ij}(s)ds,
\ \ \text{if}\ i>j\\
-\int_t^1 e^{z(t-s)(\O^\mu_i(\l)-\O^\mu_j(\l))}(\wt\cQ\cG)_{ij}(s)ds,
\ \ \text{if}\ i\le j\ac \\
=(\wt\cQ\cG)_{ij}(t)+z(\O^\mu_i(\l)-\O^\mu_j(\l))(K\wt\cQ\cG)_{ij}(t,\l),
\qq\text{all}\qq(i,j,t,\l)\in\N_{2p}^2\ts[0,1]\ts\L_R.
\end{multline}
Using the identities
$$
(\O^\mu_i-\O^\mu_j)(K\wt\cQ\cG)_{ij}
=(\cB^\mu(K\wt\cQ\cG)-(K\wt\cQ\cG)\cB^\mu)_{ij}
=(\cB^\mu\cG-\cG\cB^\mu)_{ij}
$$
we obtain
$
\cG'=\wt\cQ\cG+z(\cB^\mu\cG-\cG\cB^\mu).
$
Thus $\cG$ satisfies \er{ecG}, which yields the equivalence
of equations \er{ecG} and \er{me5}.

%Using \er{ecA}, \er{cY}  we rewrite equation \er{ecG} in the
%form of the integral equation

In Lemma \ref{l32}, proved in Appendix, we describe
 the Jost type fundamental
matrix solution $\cT$  of equation \er{me2}. In fact we will prove that  equation \er{me5}
has a unique solution for  $|\l|$
large enough. For the $2p\ts 2p$ matrix-valued function $\cA\in L^\iy(0,1)$ we introduce the norm
$$\|\cA\|_\iy=\sup_{t\in[0,1]}|\cA(t)|.$$

\begin{lemma} \lb{l32}
i) Let $\cA\in L^\iy(0,1)$ be
a $2p\ts 2p$ matrix-valued function.
Then for each $\l\in\L_R$ for some $R>0$ large enough
the operator $K$ satisfies the estimate
\[
\lb{KK}
\|(K\wt\cQ\cA)(\cdot,\l)\|_\iy\le {\x\/|z|}\|\cA\|_\iy,\qq
\text{where}\qq \x=\max\Bigl\{4\int_0^1|q_p(t)-\mu|dt;1\Bigr\}.
\]

 ii) For each $\l\in\L_R$ the integral equation \er{me5}
has the unique solution $\cG(t,\l)$.
Each matrix-valued function $\cG(t,\cdot),t\in[0,1]$,
is analytic in $\L_{R}^\pm$ and satisfies the estimates
\[
\lb{emF}
\|\cG(\cdot,\l)\|_\iy\le 2,\qq
\|\cG(\cdot,\l)-\1_{2p}\|_\iy\le {2\x\/|z|},\qq
\|\cG(\cdot,\l)-\1_{2p}-\cG_1(t,\l)\|_\iy\le {2\x^2\/|z|^2},
\]
for all $\l\in\L_{2R}$, where $\cG_1=K\wt\cQ$,
\[
\lb{asG}
(\cG_1(t,\l))_{ij}={(-1)^{p+1}\ol\O_i^p\O_j^{p-1}\/2pz}\int_0^1
e_{ij}(t-s,\l)(q_p(s)-\mu)ds+O(|z|^{-2}),\qq i,j\in\N_{2p},
\]
as $|\l|\to\iy$, uniformly on $t\in[0,1]$.

 iii) For each $t\in[0,1]$ the function $\cT(t,\cdot)$ is analytic in
$\L_{R}^\pm$ and satisfies
\[
\lb{amT}
\cT(t,\l)=e^{z\cB^\mu(\l)t}(\1_{2p}+O(|z|^{-1}))\qq\text{as}\qq |\l|\to\iy,
\]
uniformly on $t\in[0,1]$.
\end{lemma}

Now we will prove the main result of this
Section. In this Lemma \ref{PHI} we obtain the representation and asymptotics
of the monodromy matrix $\cM$, see \er{MY}-\er{pbn}.
%, where
%the properties of the matrix $\cF e^{z\cB^\mu}$ and
%the asymptotics of entries of the matrix $\cF$ will be
%important below.

\begin{lemma}
\lb{PHI}
i) The monodromy matrix $\cM(1,\cdot)$ satisfies the identity
\[
\lb{MY}
\cM(1,\cdot)=\cU\cG(0,\cdot)\cF e^{z\cB^\mu}(\cU\cG(0,\cdot))^{-1},\qq\text{where}\qq
\cF=\cG^{-1}(0,\cdot)\cG(1,\cdot).
\]

 ii) The matrix-valued function
$\cF=(\cF_{ij})_{i,j=1}^{2p}$ is analytic in
$\L_{R}^{\pm}$ and satisfies the asymptotics
\[
\lb{phi}
\cF_{ij}(\l)=\d_{ij}+{(-1)^{p+1}\/2pz}\ol\O_i^p\O_j^{p-1}
\int_0^1\x_{ij}(t,\l)(q_p(t)-\mu)dt+O(|z|^{-2}),
\]
\[
\lb{pns}
\cF_{ij}(\l)=O(|z|^{-1})\qq \ \text{for} \ \ i\ne j,
\qqq \cF_{jj}(\l)=1+O(|z|^{-2})
\]
as $|\l|\to\iy$, where
\[
\lb{xij}
\x_{i j}(t,\l)=\ca  e^{z(1-t)(\O_i(\l)-\O_j(\l))},\  \text{if}\ i>j\\
\qq e^{-zt(\O_i(\l)-\O_j(\l))},\ \text{if}\ i\le j
\ac\!\!\!\!.
\]

 iii) The functions $\cF_{p+k,p+k+1}$ and $\cF_{p+k+1,p+k}$
for all $k\in\N_{p-1}^0=\{0,1,...,p-1\}$
satisfy
\[
\lb{pbn}
\cF_{p+k,p+k+1}(\l)
=f_{k,n}+O(n^{-2}),\qq
\cF_{p+k+1,p+k}(\l)=\ol f_{k,n}+O(n^{-2})
\]
as  $n\to\iy$, $\l=(-1)^k({\pi n\/c_k})^{2p}+O(n^{2p-2}),\Im\l\ge 0$,
where
\[
\lb{hkn}
f_{k,n}
={i(-1)^{k+1}c_{k}\/2p\pi n}
\ca\qq\ e^{i{\pi k\/2p}}\hat q_{p,n},\ \ k\ \text{odd}\\
-e^{-i{\pi k\/2p}}\ol{\hat q}_{p,n},\ \ k\ \text{even}\ac.
%\qq\text{for}\qq
\]
%and $f_{k,n}(\l)=\ol f_{k,n}(\ol\l)$ for $\Im\l<0$.
\end{lemma}

\no {\bf Proof.}
i) Identities \er{dE}, \er{FT} yield
$\cM(t,\l)=\cU(\l)\cG(t,\l)e^{z\cB^\mu t}\cG^{-1}(0,\l)\cU^{-1}(\l)$,
which implies \er{MY}.

 ii) Estimates \er{emF} yield $\cG(t,\l)=\1_{2p}+\cG_1(t,\l)+O(|z|^{-2})$
as $|\l|\to\iy$, uniformly on $t\in[0,1]$, and
$$
\cF(\l)=\cG^{-1}(0,\l)\cG(1,\l)=\1_{2p}+\cG_1(1,\l)-\cG_1(0,\l)+O(|z|^{-2})
\qq\text{as}\qq|\l|\to\iy.
$$
Substituting \er{asG} into this asymptotics we obtain
$$
\cF_{ij}(\l)=\d_{ij}+{(-1)^{p+1}\/2pz}\ol\O_i^p\O_j^{p-1}
\int_0^1(e_{ij}(1-t,\l)-e_{ij}(-t,\l))(q_p(t)-\mu)dt+O(|z|^{-2})
$$
as $|\l|\to\iy$. Substituting \er{e} into the last asymptotics and using
\er{asb} we obtain \er{phi},
which yields the first asymptotics in \er{pns}.
The second asymptotics in \er{pns} follows from the identities
$\x_{jj}=1$, see \er{xij}, and the identity $\int_0^1q_p(t)dt=\mu$,
see \er{mu}.

 iii) Identity \er{Oo} gives $\O_j=\o_j$ for all $j\in\N_{2p}$.
%The proof for $\Im\l<0$ is similar.
Identities \er{ewo}, $\ve_k^{-1}=\ol\ve_k$ and $\ve_k^{2p}=(-1)^{p+k}$
(see \er{ete})
yield for $k\in\N_{p-1}^0$:
\[
\lb{oo1}
\ol\o_{s}^p\o_{s+1}^{p-1}=\ve_k^{2p-1}\e_k=(-1)^{s}\ol\ve_k\e_k,\qq
\ol\o_{s+1}^p\o_{s}^{p-1}=\ol\ve_k^{2p-1}\e_k=(-1)^{s}\ve_k\e_k,
\qq s=p+k.
\]
Substituting identities \er{oo1} into asymptotics \er{phi}
and using \er{xij} we obtain
\begin{multline}
\lb{pp1}
\cF_{s,s+1}(\l)={(-1)^{k+1}\ol\ve_k\e_k\/2pz}
\int_0^1e^{-zt(\o_{s}-\o_{s+1})}(q_p(t)-\mu)dt+O(|z|^{-2}),\\
\cF_{s+1,s}(\l)={(-1)^{k+1}\ve_k\e_k\/2pz}
\int_0^1e^{z(1-t)(\o_{s+1}-\o_{s})}(q_p(t)-\mu)dt+O(|z|^{-2})
\end{multline}
as $|\l|\to\iy,k\in\N_{p-1}^0$.
Let $\l=(-1)^k({\pi n\/c_{k}})^{2p}+O(n^{2p-2})$ as
$n\to\iy$.
Then $z=\l^{1\/2p}={\e_k\pi n\/c_{k}}+O(n^{-1})$
and using \er{ewo} we have
$z(\o_{s}-\o_{s+1})=i(-1)^{k+1}2\pi n+O(n^{-1})$.
Substituting this asymptotics
into \er{pp1} we obtain
\begin{multline}
\lb{pp1n}
\cF_{s,s+1}(\l)={(-1)^{k+1}\ol\ve_kc_{k}\/2p\pi n}
\int_0^1e^{i(-1)^{k}2\pi nt}(q_p(t)-\mu)dt+O(n^{-2}),\\
\cF_{s+1,s}(\l)={(-1)^{k+1}\ve_kc_{k}\/2p\pi n}
\int_0^1e^{i(-1)^{k+1}2\pi nt}(q_p(t)-\mu)dt+O(n^{-2})
\end{multline}
as $n\to\iy$. Substituting $\ve_k$ from \er{ete}
into \er{pp1n}, we get \er{pbn}.
$\BBox$

\section {Properties of the multipliers}
\setcounter{equation}{0}

Define the single-valued branches of the multiplier $\t$ at high
energy.
Here we use the results of Lemma \ref{nre1} which will be proved later.
The zeros
of the function $\r$ for high energy are close to the real axis.
Then the functions $\t_j,j\in\N_{2p}$, are analytic
in the domain $\L_R\cap\{\l\in\C:\d<\arg\l<\pi-\d\}$ for some $R>0$
large enough and for any $\d>0$ small enough.
Asymptotics \er{lom} define the branches $\t_j,j\in\N_{2p}$,
of the function $\t$ in this domain.

Moreover, the points $r_{k,n}^\pm\in\L_R,k\in\{j-1,j\},j\in\N_{p-1},n\ge n_0$
for some (large) $n_0\ge 1$, are ramification points of the functions $\t_{p+j}$ and
$\t_{p-j+1}$ and these functions have no any
other singularities in $\L_R$ (see discussion after the proof of Lemma \ref{nre}).
Here $r_{k,n}^\pm$ satisfy:

 1) $|r_{k,n}^\pm-r_{k,n}^0|<1$ for all
$k\in\N_{p-1}^0=\{0,...,p-1\},n\ge n_0$,

 2) all $r_{0,n}^\pm,n\ge n_0$, are positive
numbers (anti-periodic and periodic eigenvalues),

 3) all $r_{k,n}^\pm,k\in\N_p,n\ge n_0$, are real or non-real numbers
(ramification points of the Lyapunov function),
$\pm\Im r_{k,n}^\pm\ge 0$ and if $\Im r_{k,n}^+>0$ for some $k\in\N_p,n\ge n_0$,
then $r_{k,n}^-=\ol r_{k,n}^+$.

 Then each function $\t_{p+j},\t_{p-j+1},j\in\N_{p}$,
is a single-valued analytic
function in the domain $\mD_{p+j}$, where
$$
\mD_{p+j}=\L_R\sm\cup_{n\ge n_0}(\G_{j-1,n}\cup\G_{j,n})
\qq\text{all}\qq j\in\N_{p-1},\qq
\mD_{2p}=\L_R\sm\cup_{n\ge n_0}\G_{p-1,n},
$$
each $\G_{j,n},j\in\N_{p-1}^0,n\ge n_0$,
is the segment $[r_{j,n-1}^+,r_{j,n}^-]$ of the line.

Note that for each $j\in\N_p$ the cuts $\G_{j,n}^0=[r_{j,n-1}^0,r_{j,n}^0],n\ge 1$,
on the real axis for the multiplier function $\t^0$ of the operator $H^0$
merge and constitute the cuts along the negative (for odd $j$)
or positive (for even $j$) semi-axis, see \er{cD}.

%Moreover, we will prove in Lemma \ref{nre} that each point
%$r_{k,n}^\pm$ for $k\in\N_{p-1}^0,n\ge n_0$, is a zero of
%the functions $\t_{p+k}-\t_{p+k+1}$ and $\t_{p-k}-\t_{p-k+1}$ and all other functions
%$\t_j-\t_{\ell},1\le j<\ell\le 2p,(j,\ell)\not\in\{(p+k,p+k+1),(p-k,p-k+1)\}$
%have no a zero at this point.
%This result determines the Riemann surface $\mL$ of the function $\t$,
%and then the Riemann surface $\mR$ of the Lyapunov function $\D$,
%at high energy (see Fig. \ref{RS3}).

The following result is a simple consequence of
the standard perturbation theory.

\begin{lemma}
\lb{Am}
i) Let $\l\in\L_R$ for some $R>0$ large enough.
Then for each $j\in\N_{2p}$ the monodromy matrix $\cM(1,\l)$
has the eigenvalue $\t_j$
satisfying the asymptotics
\[
\lb{awt}
\t_j(\l)=e^{z\O_j}(1+O(|z|^{-1}))\qq\text{as}\qq |\l|\to\iy,
\]
which yields \er{lom}.
Moreover, if $\t_j(\l)=\t_k(\l)$ for some $1\le j<k\le 2p$ and some $\l\in\L_R$,
then $k=j+1$. No more than two multipliers $\t_j, j\in\N_{2p}$, can coincide one with other
at the point $\l$.

ii)
The multipliers satisfy the identities
\[
\lb{mn1}
\t_{p-j+1}(\l)=\t_{p+j}^{-1}(\l),\qq \text{all}
\qq (j,\l)\in\N_{p}\ts\mD_{p+j}.
\]

\end{lemma}

\no {\bf Proof.}
i) Recall the following Gershgorin's result
from the matrix theory, see \cite{HJ}:
\vspace{2mm}

 {\it Let $A=(A_{ij})_{i,j=1}^{m}$ be a complex $m\ts m$
matrix and let $R_i=\sum_{j=1,j\ne i}^{m}|A_{ij}|$
for $i\in\N_{m}$.
Then every eigenvalue of $A$ lies within at least one of the
discs $\{\t\in\C:|\t-A_{ii}|<R_i\},i\in\N_{m}$.
If the union of $k$ discs is disjoint from the union of
the other $n-k$ discs then
the former union contains exactly $k$ and the latter $n-k$
eigenvalues of $A$.
}
\vspace{2mm}

 Identity \er{MY} implies that
the matrices $\cM(1,\cdot)$ and $\cX=e^{z\cB^\mu}\cF$
have the same eigenvalues (the multipliers).
Asymptotics \er{asb}, \er{pns} give
$$
\cX=e^{z\cB^0}\lt(\1_{2p}+{\cW(z)\/|z|}\rt),\qq
\text{where}\qq
\cB^0=\diag(\O_j)_{1}^{2p},
$$
and the matrix-valued function $\cW(z)=(w_{ij}(z))_{i,j=1}^{2p}$
is uniformly bounded on $|z|>R$ for some $R>0$.
By Gershgorin's theorem every multiplier $\t$
lies within at least one of the discs
$$
\mK_j=\lt\{\t\in\C:|\t-e^{z\O_j}|<|e^{z\O_j}|{w_j\/|z|}\rt\},
\qq j\in\N_{2p},
\qq\text{where}\qq w_j=\max_{|z|>R}\sum_{k=1}^{2p}|w_{jk}(z)|.
$$
Firstly, let the disc $\mK_j$ for some $j\in\N_{2p}$
be disjoint from the other discs
$\mK_k,k\ne j$. Then, by Gershgorin's theorem,
the disc $\mK_j$ contains exactly one multiplier
$\t_j(z)$, which satisfies the estimate
$|\t_j(z)-e^{z\O_j}|<|e^{z\O_j}|{w_j\/|z|}$.
This estimate gives \er{awt} for this case.

Secondly, consider all $k,j,1\le j<k\le 2p$, such that
$\mK_j\cap \mK_k\ne\es$.
Then the distance between the centers of these discs is less
than the sum of their radii:
$
|e^{z\O_j}-e^{z\O_k}|<{1\/|z|}(|e^{z\O_j}|w_j+|e^{z\O_k}|w_k).
$
Then
\[
\lb{e00}
|e^{z(\O_k-\O_j)}-1|<{1\/|z|}(w_j+w_k),
\]
where we used the estimate
$|e^{z(\O_k-\O_j)}|=e^{\Re z(\O_k-\O_j)}\le 1$
(see \er{mO}). If $k\ge j+2$, then estimates \er{mO1}
together with \er{mO} yields
$\Re z(\O_k-\O_j)<-a|z|,a>0$, and then
$e^{z(\O_k-\O_j)}\to 0$ as $|z|\to\iy$.
For $|z|$ large enough we have a contradiction with \er{e00}.
Hence $k=j+1$.
Moreover, the similar arguments show that
only two domains $\mK_j,\mK_{j+1}$
can intersect each other and they are disjoint from
other domains $\mK_m,m\ne j,j+1$. In fact, let $\mK_j\cap \mK_{j+1}\ne\es$
and $\mK_{j+1}\cap \mK_{j+2}\ne\es$.
Then we have two estimates
$$
|e^{z\O_j}-e^{z\O_{j+1}}|<{1\/|z|}(|e^{z\O_j}|w_j+|e^{z\O_{j+1}}|w_{j+1}),
\ \
|e^{z\O_{j+1}}-e^{z\O_{j+2}}|
<{1\/|z|}(|e^{z\O_{j+1}}|w_{j+1}+|e^{z\O_{j+2}}|w_{j+2}),
$$
which yield
%$$
%|e^{z(\O_{j+1}-\O_j)}-1|<{1\/|z|}(w_j+w_{j+1}),\qq
%|e^{z(\O_{j+2}-\O_{j+1})}-1|<{1\/|z|}(w_{j+1}+w_{j+2}).
%$$
$$
|e^{z\O_j}-e^{z\O_{j+2}}|\le
|e^{z\O_j}-e^{z\O_{j+1}}|+|e^{z\O_{j+1}}-e^{z\O_{j+2}}|<
{1\/|z|}\bigl(|e^{z\O_j}|w_j+2|e^{z\O_{j+1}}|w_{j+1}
+|e^{z\O_{j+2}}|w_{j+2}\bigr).
$$
Then
$$
|1-e^{z(\O_{j+2}-\O_j)}|<{1\/|z|}(w_j+2w_{j+1}+w_{j+2}),
$$
which is in contradiction with the estimate
$\Re z(\O_{j+2}-\O_j)<-a|z|,a>0$.
Thus only two domains $\mK_j,\mK_{j+1}$
can intersect each other and they are disjoint from
other domains $\mK_m,m\ne j,j+1$.

Let $\mK_j\cap \mK_{j+1}\ne\es$ for some $j\in\N_{2p-1}$.
By Gershgorin's theorem, the domain $\mK_j\cup \mK_{j+1}$ contains exactly
two multipliers $\t_j(z),\t_{j+1}(z)$ and
$
|\t_{j+1}-e^{z\O_{j+1}}|
<{1\/|z|}(2|e^{z\O_j}|w_j+|e^{z\O_{j+1}}|w_{j+1}),
$
which implies
$$
|\t_{j+1}e^{-z\O_{j+1}}-1|<{1\/|z|}(2w_j+|e^{z(\O_{j}-\O_{j+1})}|w_{j+1}).
%\le{1\/|z|}(w_j+2w_{j+1}).
$$
Estimate \er{e00} yields that $e^{z(\O_{j}-\O_{j+1})}=1+O(|z|^{-1})$
and then $\t_{j+1}$ satisfies asymptotics \er{awt}.
The similar arguments show that
$\t_{j}$ also satisfies \er{awt}.
In fact,
$
|\t_{j}-e^{z\O_{j}}|
<{1\/|z|}(|e^{z\O_j}|w_j+2|e^{z\O_{j+1}}|w_{j+1}),
$
which implies
$$
|\t_{j}e^{-z\O_{j}}-1|<{1\/|z|}(2w_{j+1}+|e^{z(\O_{j+1}-\O_{j})}|w_{j+1}).
$$
Since $e^{\Re z(\O_{j+1}-\O_{j})}\le 1$, we obtain
asymptotics \er{awt} for $\t_{j}$.

ii) Asymptotics \er{awt} and identities \er{mim} provide \er{mn1}.
$\BBox$

\no{\bf Proof of Theorem \ref{T1}.}
The proof of identity \er{smp} is standard.
We rewrite equation \er{me} in the form
$\cJ\cM'=\cH\cM$, where $\cH=\cJ(\cP+\cQ)$,
$\cJ\cQ$ is a diagonal $2p\ts 2p$ matrix and
$$
\cJ\cP=\ma(-1)^p\l&\dO_{2p-1,p}\\
\dO_{1,2p-1}& \wt J\am,\qq
\wt J=\ma\dO_{p-1,p-1}&\dO_{p-1,1}&-J_{p-1}\\
\dO_{1,p-1}&1&\dO_{1,p-1}\\
-J_{p-1}&\dO_{p-1,1}&\dO_{p-1,p-1}\am.
$$
Then the matrix $\cH$ is symmetric: $\cH^\top=\cH$.
Using the identity $\cJ^\top=-\cJ$ we obtain $-(\cM^\top)'\cJ=\cM^\top\cH$,
which yields
$$
(\cM^\top\cJ\cM)'=(\cM^\top)'\cJ\cM+\cM^\top\cJ\cM'=-\cM^\top\cH\cM+\cM^\top\cH\cM=0.
$$
Then $\cM^\top\cJ\cM=\const$ and using $\cM(0)=\1_{2p}$ we obtain \er{smp}.

 i) We will use the arguments from \cite{CK}. Identity \er{smp} yields
\[
\lb{rD}
D(\t,\cdot)=\t^{2p} D(\t^{-1},\cdot),\ \ \ \t\ne 0,
\]
and
$ D(\t,\l)=\sum_0^{2p}\vk_k(\l)\t^{2p-k}$,
where the functions $\vk_k$ are given by
$$
\vk_0=1,\ \ \ \vk_1=-2pT_1,\ \  \vk_2=-{2p\/2}(T_2+T_1\vk_1),\ \
... ,
\vk_k=-{2p\/k}\sum_0^{k-1}T_{k-j}\vk_j,...,\ \ T_k={\Tr \cM^k(1,\cdot)\/2p},
$$
(see \cite{RS2}, p.331-333).
By Lemma \ref{T21}, the coefficients $\vk_k(\l)$ are entire in $\l\in\C$.
Using the identity \er{rD}
we obtain $\vk_{2p-j}=\vk_j,j\in\N_{p}^0$, which yields
$
 D(\t,\cdot)=(\t^{2p}+1)+\vk_1(\t^{2p-1}+\t)+
...+\vk_{p-1}(\t^{p+1}+\t^{p-1})+\vk_{p}\t^p.
$
Then
\[
\lb{14}
{D(\t,\l)\/(2\t)^p}
=\n ^p+f_1(\l)\n^{p-1}+...+f_p(\l),
\qqq \n={\t+\t^{-1}\/2},
\]
where $f_1,..,f_p$ are some linear combinations
of $\vk_0,..,\vk_p$. In
particular, all coefficients $f_1(\l),...,$ $f_p(\l)$ are entire functions.
The function $\Phi(\n,\l)={D(\t,\l)\/(2\t)^p}$ is a polynomial of $\n$ of degree $p$.
Each zero $\D_j,j\in\N_p$, of this function satisfies
$
\D_j={1\/2}(\t_j+\t_j^{-1}), j\in\N_p,
$
where $\t_j,\t_j^{-1}$ are multipliers, and identity \er{T1-2}
holds.
Asymptotics \er{lom} yields \er{aD}.
Recall that
$\t_j$ are branches of the function
$\t$ analytic on the $2p$ sheeted Riemann surface (see Sect.1).
Then $\D_j(\l),j\in\N_p$
constitute $p$ branches of one analytic function $\D(\l)$ on
the connected $p$-sheeted Riemann surface $\mR$.

 ii) Proof  repeats the standard arguments (see \cite{CK}).

 iii)
Let $\cM_n, n\ge 1$, be the monodromy matrix
for the operator $H_n$ given by \er{Hn}.
Let $\t_{j,n},j\in\N_p$, be the multipliers of $H_n$.
Identity \er{st} gives
\[
\lb{sts}
\s(H_n)=\{\l\in\R:|\t_{j,n}(\l)|=1\ \text{for some}\ j\in\N_p\}.
\]
Lemma \ref{T21} provides $\cM_n\to\cM(1,\cdot)$ as $n\to\iy$
uniformly on any compact in $\C$.
Then $\t_{j,n}\to\t_j$
as $n\to\iy$ uniformly on any compact in $\C$ for all $j\in\N_p$.
Identity \er{sts} implies
\[
\lb{qf8}
\s(H_n)\to\{\l\in\R:|\t_j(\l)|=1\ \text{for some}\ j\in\N_p\}
\qq\text{as}\qq n\to\iy.
\]
Assume that
\[
\lb{qf9}
\s(H_n)\to\s(H)\qq\text{as}\qq n\to\iy.
\]
Then using \er{qf8} we obtain
$\s(H)=\{\l\in\R:|\t_j(\l)|=1\ \text{for some}\ j\in\N_p\}$,
which yields \er{sd}.

Now we will prove \er{qf9}. We need the following result
(see \cite{Ka}, Th.~VI.5.13, Cor.~V.4.2):

 {\it
Let $A\ge 0$ be an operator in a Hilbert space $\mH$
and let $Q_n$ be a symmetric operator in $\mH$ with the form domain $\Dom_{fd} Q_n\ss\Dom_{fd} A$.
Assume that
$$
|(Q_n u,u)|\le \ve_n\bigl(\|u\|^2+(Au,u)\bigr),\qq u\in\Dom_{fd} Q_n,
\qq\text{where}\qq\ve_n>0,\qq\ve_n\to 0.
$$
Then the the Friedrichs extension $A_n$ of the operator
$A+Q_n$ is selfadjoint for
sufficiently large $n$ and $A_n\to A$ in the uniform resolvent sense.
If $\s(A)$ has a gap at $\a$, then
$\s(A_n)$ has a gap at $\a$ for sufficiently large $n$.
}

 Estimate \er{qf7} shows that
$H_n\to H$ in the uniform resolvent sense
(and then in the strong resolvent sense) as $n\to\iy$,
and $\lim_{n\to\iy}\s(H_n)\ss\s(H)$.
Then relation
\er{qf9} is obtained from the following
result (see \cite{Ka}, Th.~VIII.1.14):

 {\it Let $H,H_n,n\ge 1$, be selfadjoint operators in a Hilbert space $\mH$
and let $H_n\to H$ in the strong resolvent sense.
Then every open set containing a point of $\s(H)$
contains at least a point of $\s(H_n)$ for sufficiently large $n$.
}
$\BBox$

Consider two simple examples.

\no {\bf Example 1.} Consider the operator
$H=(-1)^p{d^{2p}\/dt^{2p}}+\sum_{j=0}^{p-1}q_{j+1}{d^{2j}\/dt^{2j}}$
with the constant coefficients $q_j$.
Equation \er{1b} has the solutions $e^{\pm i\z_j(\l)t}$,
where $\z_j(\l)=\sqrt {w_j(\l)},j\in\N_p$, and $w_j$
are values of the algebraic function $w(\l)$ which is a solution of the equations
$P(w)-\l=0$, $P(w)=w^{p}+\sum_{j=0}^{p-1}(-1)^{j} q_{j+1} w^{j}$.
Then the multipliers have the form $e^{\pm i\z_j(\l)}$ and
the Lyapunov function is given by $\D_j(\l)=\cos\z_j(\l)$.
If the polynomial $P(w)-\l$ for some $\l\in\R$ has $n$ ($1\le n\le p$)
positive simple zeros,
then the spectrum $\s(H)$ in some interval $(\l-\ve,\l+\ve),\ve>0$,
has multiplicity $2n$.
Let $P(w)=T_p(w-1)$, where
$T_p(w)=\sum_0^{[p/2]}{p\choose 2n}(w^2-1)^nw^{p-2n}$
is the Chebyshev polynomial. The properties of these polynomials
(see \cite{AS})
provide that the spectrum is given by
$\s(H)=[-1,+\iy)$ and has multiplicity $2p$ (maximal multiplicity)
on the interval $(-1,1)$ and the multiplicity $2$
(minimal multiplicity) on $(1,+\iy)$.

\no {\bf Example 2.}
Consider the operator $H=\wt H^p$, where $\wt H=-{d^2\/dt^2}+q$
is the Hill operator
with the 1-periodic function $q\in L^2(\T)$.
The branches of the Lyapunov function are given by
$\D_j(\l)=\wt\D(-z^2\o_j^2),j\in\N_p$, where
$\wt\D$ is the (entire) Lyapunov function of $\wt H$.
Recall that the spectrum $\s(\wt H)$ is semi-bounded below and
consists of bands separated by gaps.
Let $\s_j=\{\l\in\R:\wt\D(-z^2\o_j^2)\in[-1,1]\},j\in\N_p,z=\l^{1\/2p}\in S$.
The spectrum of $H$ satisfies the identity
$\s(H)=\cup_1^p\s_j=\s_1\cup\s_p.$
We have
$$
\s_1=\ca
\qqq\{\l\in[0,\iy): -z^2\in\s_-\}, \qq p\ \text{even}\\
\{\l\in(-\iy,0]:
-e^{-i{\pi\/p}}z^2\in\s_-\},\ \ p\ \text{odd}\ac,\qq
\s_p=\{\l\in[0,\iy): z^2 \in\s(\wt H)\cap[0,\iy)\},
$$
where $\s_-=\s(\wt H)\cap(-\iy,0]$.
If $p$ is odd, then the spectrum $\s(H)$ has multiplicity 2.
If $p$ is even, then the spectrum has multiplicity $4$
in the set $\s_1\cap\s_p$ and multiplicity $2$ in the other
intervals.

Recall that all functions
$\r,D_\pm=2^{-p}D(\pm 1,\cdot)=\prod_{j=1}^{p}(\D_j\mp 1)$ are entire.
The zeros of $\r$ are ramifications of the Lyapunov function,
the zeros of $D_\pm$ are periodic and
antiperiodic eigenvalues.
We introduce the contours $C_n(r)=\{\l:|z-\pi n|=\pi r\},r>0,n\ge 0$.

\begin{lemma}
\lb{Dpm}
Let $N\in\N$ be large enough. Then  the function $D_+$ (and $D_-$) has
exactly $2N+1$ (and $2N$) zeros in the domain  $\{|z|<2\pi(N+{1\/2})\}$ (and in $\{|z|<2\pi N\}$),
counted with multiplicity. Moreover, for each $n>N$ the function $D_+$ (and $D_-$) has
exactly two zeros in the disk $\{|z-2\pi n|<{\pi\/2}\}$ (and in $\{|z-\pi(2n+1)|<{\pi\/2}\}$),
counted with multiplicity. There are no other zeros.
\end{lemma}

\no {\bf Proof.} We consider the function $D_+$.
The proof for $D_-$ is similar.
The function $D_+$ for the operator $H^0$ is given by
$$
D_+^0
=\prod_{j=1}^{p}(\cosh z\o_j- 1)
=-{\l\/2^p}\prod_{n=1}^\iy\lt(1-{\l\/(2n\pi)^{2p}}\rt)^2.
$$
Assume that for each $j\in\N_p$ and for some $R>0$ large enough
\[
\lb{s10}
|\cosh z\o_j- 1|=2\Bigl|\sinh{z\o_j\/2}\Bigr|^2
>{1\/8}e^{|\Re z\o_j|},\qq\text{all}\qq
\l\in\L_R\sm\cup_{n\in\N}\{|z-2\pi n|\le{\pi\/2}\}.
\]
Then asymptotics \er{aD} and estimates \er{s10} yield
\[
\lb{3i}
{D_+(\l)\/D_+^0(\l)}
=\prod_{j=1}^{p}{\D_j(\l)-1\/\cosh z\o_j- 1}
=\prod_{j=1}^{p}{\cosh z\o_j-1+O(|z|^{-1}e^{|\Re z\o_j|})\/\cosh z\o_j-1}
=1+O(|z|^{-1})
\]
as $\l\in\C\sm\cup_{n\in\N}\{|z-2\pi n|\le{\pi\/2}\},
|\l|\to\iy$.
Let $N\ge 1$ be large enough and let $N'>N$ be another integer.
Let $\l$ belong to the
contours $C_0(2N+{1\/2}),C_0(2N'+{1\/2}),C_{2 n}({1\/2}),|n|>N$.
Asymptotics \er{3i} yields
$$
|D_+(\l)-D_+^0(\l)|=|D_+^0(\l)|\lt|{D_+(\l)\/D_+^0(\l)}-1\rt|
=|D_+^0(\l)|O(|z|^{-1})<|D_+^0(\l)|
$$
on all contours.
Hence, by Rouch\'e's theorem, $D_+$ has as many zeros,
as $D_+^0$ in each of the
bounded domains and the remaining unbounded domain. Since
$D_+^0$ has exactly one simple zero at $\l=0$
and exactly one zero of multiplicity two
at $(2\pi n)^{2p},n\ge 1$,
and since $N'>N$ can be chosen arbitrarily large,
the statement for $D_+$ follows.

We will prove \er{s10}.
Using the simple estimate $e^{|\Im z|}<4|\sin z|$
as $|z-\pi n|>{\pi\/4}$
for all $n\in\Z$ (see \cite{PT}, Lemma 2.1) we obtain that
estimates \er{s10} hold in the domain
$\C\sm\cup_{n\in\Z}\{|z\o_j+i2\pi n|\le{\pi\/2}\}$.
For each $j\ne p$
the estimates $|z\o_j+i2\pi n|>{\pi\/2}$
hold for all $n\in\Z$ and $|\l|$ large enough.
Moreover, the estimates $|z\o_p+i2\pi n|=|z-2\pi n|>{\pi\/2}$ hold
for all $n\le 0$ and $|\l|$ large enough.
Thus,
%only the conditions $|z-2\pi n|>{\pi\/2}$ for all $n\in\N$
%is essential in \er{s10} and
estimates \er{s10}
hold in $\L_R\sm\cup_{n\in\N}\{|z-2\pi n|\le{\pi\/2}\}$.
%with $|\l|>0$ large enough.
$\BBox$

Now we will describe the zeros of the function $\r$.
Identifying the sides of the sector
$S=\{z\in\C:\arg z\in(-{\pi\/2p},{\pi\/2p}]\}$ (i.e.
we identify each point $xe^{i{\pi\/2p}},x\in\R_+$,
on the $z$-plane with the point $xe^{-i{\pi\/2p}}$)
we obtain the cone $S_{con}$.
For each $(k,n)\in\N_{p-1}\ts\N^0,\N^0=\N\cup\{0\}$,
we introduce the domain $\mU_{k,n}$  given by
\[
\lb{dD}
\mU_{k,n}=\Bigl\{\l=z^{2p},z\in S_{con}:\Bigl|z-\pi n{\e_k\/c_{k}}\Bigr|<\beta\Bigr\},
\qq  \beta>0\qq
\text{is small enough}.
\]
Each domain $\mU_{k,n}$ is a neighborhood of the zero
$r_{k,n}^0=(\pi n{\e_k\/c_{k}})^{2p}$ of the function $\r^0$, see \er{r0},
where $\e_k$ is given by \er{ete}.

We have $\mU_{1,0}=...=\mU_{p-1,0}$
and $\mU_{1,0}\cap \mU_{k,n}=\es$ for all $(k,n)\in\N_{p-1}\ts\N$.
If $p=2$ or $3$, then
the domains $\mU_{k,n},\mU_{k',n'},(k,n),(k',n')\in\N_{p-1}\ts\N,(k',n')\ne(k,n)$,
are separated, that is $\mU_{k,n}\cap \mU_{k',n'}=\es$.
The situation is more complicated for $p\ge 4$
(see Fig. \ref{clas}).
In this case we have $\mU_{2k,n}\cap \mU_{2k'-1,n'}=\es$
for all $2k,2k'-1\in\N_{p-1},n,n'\in\N$.
However, the domains $\mU_{2k,n},\mU_{2k',n'}$ can have non-empty intersection
for some $2k,2k'\in\N_{p-1},n,n'\in\N$ such that $(k',n')\ne(k,n)$.
The similar statement for the domains $\mU_{2k-1,n},\mU_{2k'-1,n'}$ holds.

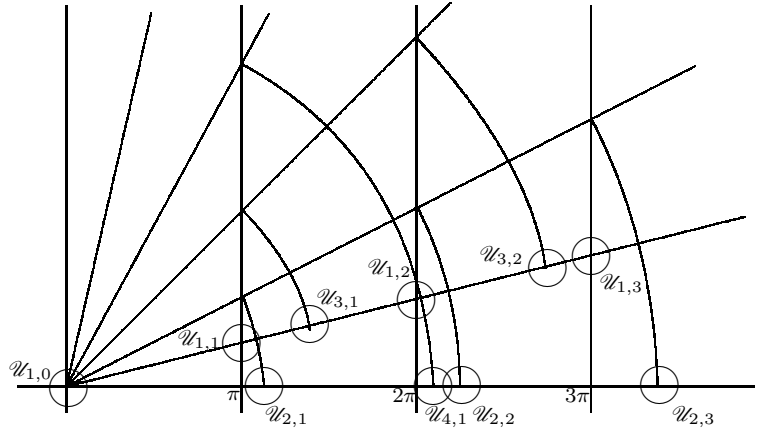
\begin{figure}
\tiny
\unitlength 1mm % = 2.845pt
\linethickness{0.4pt}
\ifx\plotpoint\undefined\newsavebox{\plotpoint}\fi % GNUPLOT compatibility
\begin{picture}(103.25,55.5)(0,0)
\put(12.75,57.25){\line(0,-1){54}}
\put(35.75,57.25){\line(0,-1){54}}
\put(58.75,57.25){\line(0,-1){54}}
\put(81.75,57.25){\line(0,-1){54}}
\put(6.25,6.75){\line(1,0){97}}
%\emline(12.75,6.75)(87.25,81.75)
\multiput(12.75,6.75)(.033725667723,.033952014486){1500}{\line(0,1){.033952014486}}
%\end
%\emline(12.5,6.5)(95.25,49)
\multiput(12.75,6.75)(.06567460317,.03373015873){1260}{\line(1,0){.06567460317}}
%\end
%\emline(13.25,6.75)(30.5,83.25)
\multiput(12.75,6.75)(.0336914063,.1494140625){332}{\line(0,1){.1494140625}}
%\end
%\emline(12.75,6.5)(102,29)
\multiput(12.75,6.75)(.13380809595,.03373313343){667}{\line(1,0){.13380809595}}
%\end
%\emline(12.75,6.5)(54.25,83.5)
\multiput(12.75,6.75)(.0337398374,.06260162602){790}{\line(0,1){.06260162602}}
%\end
\qbezier(36,18.5)(38.125,12.875)(38.75,6.75)
\qbezier(59,30.25)(64.25,19.875)(64.5,7)
\qbezier(81.75,42.25)(89.875,26.5)(90.5,6.75)
\qbezier(35.75,49.5)(59.125,36.75)(61,7)
\put(38.75,6.75){\circle{5.00}}
\put(61,6.75){\circle{5.00}}
\put(64.75,7){\circle{5.00}}
\put(90.75,7){\circle{5.00}}
\qbezier(36,30)(43.625,21.875)(44.75,14.25)
\qbezier(58.75,53)(73.75,36.75)(75.75,22.5)
\put(35.75,12.5){\circle{5.00}}
\put(44.75,15){\circle{5.00}}
\put(76,22.5){\circle{5.00}}
%\put(12.05,5.65){\makebox(0,0)[cc]{$0$}}
\put(34.75,5.55){\makebox(0,0)[cc]{$\pi$}}
\put(57.25,5.55){\makebox(0,0)[cc]{$2\pi$}}
\put(80.0,5.55){\makebox(0,0)[cc]{$3\pi$}}
\put(13,6.5){\circle{5.00}}
\put(41.75,2.75){\makebox(0,0)[cc]{$\mU_{2,1}$}}
\put(69,2.75){\makebox(0,0)[cc]{$\mU_{2,2}$}}
\put(95.25,2.75){\makebox(0,0)[cc]{$\mU_{2,3}$}}
\put(62.75,2.75){\makebox(0,0)[cc]{$\mU_{4,1}$}}
\put(30.25,12.8){\makebox(0,0)[cc]{$\mU_{1,1}$}}
\put(48.5,18.25){\makebox(0,0)[cc]{$\mU_{3,1}$}}
\put(8,8.75){\makebox(0,0)[cc]{$\mU_{1,0}$}}
\put(58.75,18.25){\circle{5.00}}
\put(81.75,24){\circle{5.00}}
\put(70.0,24){\makebox(0,0)[cc]{$\mU_{3,2}$}}
\put(55.25,22.25){\makebox(0,0)[cc]{$\mU_{1,2}$}}
\put(85.75,20.25){\makebox(0,0)[cc]{$\mU_{1,3}$}}
\end{picture}
\caption{\footnotesize
The domains $\mU_{k,n}$ in the $z$-plane for $p=6$.}
\lb{clas}
\end{figure}
%%%%%%%%%

We introduce the cluster decomposition
of the set of indices $(k,n)\in\N_{p-1}\ts\N^0$,
having the form $\cup_{j=-\iy}^\iy \mC_j=\N_{p-1}\ts\N^0$,
where

{\it  i) The indices $(k,n),(\ell,m)$ belong to the cluster $\mC_j$
iff the domains $\mU_{k,n}$ and $\mU_{\ell,m}$ are connected one with
other by a chain of the pairwise intersecting domains $\mU_{k_s,n_s}$.

 ii) $\mC_0=\{(k,0),k\in\N_{p-1}\}$.

 iii) If $r_{k,n}^0<r_{k',n'}^0$ and $(k,n)\in \mC_j,(k',n')\in \mC_{j'}$,
then $j\le j'$.
}

 Then the clusters have the form
\[
\lb{fc}
\mC_{j}=\{(2k_m,n_m),m\in\N_{\ell_j}\},\qq
\mC_{-j}=\{(2k_m-1,n_m),m\in\N_{\ell_{-j}}\}
\qq\text{for all}\qq j\in\N
\]
where the number  $\ell_{\pm j}$ of elements of the cluster
$\mC_{\pm j}$ satisfies the estimate $\ell_{\pm j}\le{p\/2}$,
$n_m\in\N$, $1\le k_1<...<k_{\ell_{\pm j}}\le {p\/2}$,
$1\le n_{\ell_{\pm j}}<...<n_1$.
Introduce the domains
\[
\lb{dwU}
\mV_j=\cup_{(k,n)\in \mC_j}\mU_{k,n},\qq j\in\Z.
\]
These domains satisfy the following relations:
\[
\lb{lc}
\mV_j\ss\{\l\in\C:|\Im z|<\beta\},\qqq
\mV_{-j}\ss\{\l\in\C:|\Im ze^{-i{\pi\/2p}}|<\beta\},
\]
for all $j\ge 0$.
Moreover, if $\l_j\in\mV_j$, $\l_{j'}\in\mV_{j'}$ and $j<j'$,
then $\Re\l_j<\Re\l_{j'}$,
and each domain $\mV_{j-1}$ is separated from $\mV_j,j\in\Z$,
by the line $\{\l:\Re z=R_j\}$ for some
$R_j\in\R$, $...<R_{-1}<R_0<R_1<R_2<...$.

\begin{lemma}
\lb{nre1}
The function $\r$ has as many
zeros, counted with multiplicity, as the function $\r^0$, in each domain
$\{\l\in\C:R_{-N}<\Re z<R_N,|\Im z|<N\}$ and in each domain $\mV_j$,
$|j|>N$ for $N\in\N$ large enough. There are no other zeros.
\end{lemma}

\no {\bf Proof.} Recall that
$$
\r^0=\prod_{1\le j<\ell\le p}(\cosh z\o_j-\cosh z\o_\ell)^2
=-{(-1)^{p(p+1)\/2}p^p\l^{p-1}\/2^{(p-1)p}}\prod_{n=1}^\iy\prod_{k=1}^{p-1}
\lt(1-{\l\/r_{k,n}^0}\rt)^2.
$$
Assume that for each $1\le j<\ell\le p$, and for some $c>0$
\[
\lb{s12}
|\cosh z\o_j-\cosh z\o_\ell|>ce^{\max\{|\Re z\o_j|,|\Re z\o_\ell|\}}\qq
\text{as}\qq\bigl|z-\pi n{\e_k\/c_{k}}\bigr|>\beta,
\qq \text{all}\qq (k,n)\in\N_{p-1}\ts\N.
 \]
Asymptotics \er{aD} and estimates \er{s12} yield
\begin{multline}
\lb{4i}
{\r(\l)\/\r^0(\l)}
=\prod_{1\le j<\ell\le p}\lt({\D_j(\l)-\D_\ell(\l)\/\cosh z\o_j-\cosh z\o_\ell}\rt)^2\\
=\prod_{1\le j<\ell\le p}\lt({\cosh z\o_j-\cosh z\o_\ell+O(|z|^{-1}e^{|\Re z\o_j|})
+O(|z|^{-1}e^{|\Re z\o_\ell|})\/\cosh z\o_j-\cosh z\o_\ell}\rt)^2
=1+O(|z|^{-1})
\end{multline}
as $|\l|\to\iy$.
Let $N\in\N$ be large enough and let $N'>N$ be another integer.
Let $\l$ belong to the contours $C_0(R_N),C_0(R_{N'}),
\pa \mV_n,|n|>N$.
Asymptotics \er{4i} on all contours yields
$$
|\r(\l)-\r^0(\l)|=\r^0(\l)\lt|{\r(\l)\/\r^0(\l)}-1\rt|=|\r^0(\l)|O(|z|^{-1})<|\r^0(\l)|.
$$
Hence, by Rouch\'e's theorem, $\r$ has as many zeros,
as $\r^0$ in each of the
bounded domains and the remaining unbounded domain.
Since $N_1>N$ can be chosen arbitrarily large,
the statement follows.

We have to prove estimates \er{s12}.
Let $|z-\pi n{\e_k\/c_{k}}|>\beta$
for all $(k,n)\in\N_{p-1}\ts\N$. Then $|2z\ol\e_kc_k-2\pi n|>2\beta c_{k}$.
Identities \er{ewo} give $|z(\o_{p+k}-\o_{p+k+1})\pm i2\pi n|>2\beta c_{k}$
and a fortiori $|z(\o_j\pm\o_\ell)+i2\pi n|>2\beta c_{k}$
for all $1\le j<\ell\le p$.
Using the standard estimates we obtain
$$
|\cosh z\o_j-\cosh z\o_\ell|
=2\lt|\sinh{z(\o_j-\o_\ell)\/2}\rt|\lt|\sinh{z(\o_j+\o_\ell)\/2}\rt|
>ce^{{1\/2}(|\Re z(\o_j-\o_\ell)|+|\Re z(\o_j+\o_\ell)|)},
$$
for some $c>0$, which yields \er{s12}.
$\BBox$

\begin{lemma}
\lb{nre}
i) Let $\t_{p+k}(\l)=\t_{p+j}(\l)$ for some
$0\le k<j\le p,\l\in\L_R$, where $R>0$ is large enough.
Then $j=k+1$ and $\l\in\mU_{k,n}$ for some (large) $n\in\N$.
Moreover, in this case
%if $\Im\l\ne 0$, then
$\ol\l\in\mU_{k,n}$ and $\ol\l$ is also the zero
of the function $\t_{p+k}-\t_{p+k+1}$.

 ii) Let $r_{k,n}^{\mu,\pm},k\in\N_{p-1},n\ge 0$,
be ramifications of the Lyapunov function for the operator $H^\mu$.
Then
\[
\lb{res}
r_{k,n}^{\mu,\pm}=(-1)^k\Bigl({\pi n\/c_k}\Bigr)^{2p}\lt(1+
(-1)^{p+1} {\mu c_k^2\/(\pi n)^2}\bigl(1+O(n^{-2})\bigr)\rt)\qq
\text{as}\qq
n\to\iy,\qq\text{all}\qq k\in\N_{p-1}.
\]
Moreover, all  ramifications $r_{k,n}^{\mu,\pm}$
are real for $n$ large enough.
\end{lemma}

\no {\bf Proof.} i)
Let $\t_{p+k}(\l)=\t_{p+j}(\l)$ for some
$0\le k<j\le p,\l\in\L_R$.
Then, due to Lemma \ref{Am}, $j=k+1$.
Asymptotics \er{lom} gives
\[
\lb{pr1}
e^{z(\O_{p+k}-\O_{p+k+1})}=1+O(|z|^{-1})\qq\text{as}\qq |\l|\to\iy,
\qq\text{where}\qq z=\l^{1\/2p}.
\]
Substituting identity \er{ewo} into \er{pr1} we obtain
$z=\pi n{\e_k\/c_{k}}+O(n^{-1})$ as $n\to\iy$.
Then $\l\in\mU_{k,n}$ for $n\in\N$ large enough.

The domain $\mU_{k,n}$ is symmetric with respect to the real axis,
then $\ol\l\in\mU_{k,n}$. The function $\r$ is real on $\R$
(see Theorem \ref{rho} i), then $\ol\l$ is the ramification.
Note that $\l$
is a zero of the function $\t_{p+k}-\t_{p+k+1}$.
Then asymptotics \er{awt}
and the first identity in \er{m=O}
%Identities \er{mnO}
show that $\ol\l$
is also a zero of the function $\t_{p+k}-\t_{p+k+1}$.

 ii) Let $\l=r_{k,2n}^{\mu,\pm}$ for some $(k,n)\in\N_{p-1}\ts\N$.
By Lemma \ref{Dpm} i),
$z=(r_{k,2n}^{\mu,\pm})^{1\/2p}=z_0+\ve$, where $z_0=(r_{k,2n}^0)^{1\/2p}$
and $|\ve|<p\beta$ for all $n\in\N$ large enough.
Then
$1=\t_{s}^\mu(\l)(\t_{s+1}^\mu)^{-1}(\l)=e^{z(\O_{s}^\mu(\l)-\O_{s+1}^\mu(\l))}$,
where $s=p+k$ and we used identities \er{alb}.
Consider the case $\Im\l\ge 0$, the proof for the other case is similar.
Then $z$ satisfies the identity
\[
\lb{s4}
z={2\pi n i\/\o^\mu_{s}(\l)-\o^\mu_{s+1}(\l)}.
\]
Asymptotics \er{abe} yields
\[
\lb{s5}
z={2\pi n i\/\o_{s}-\o_{s+1}}+O(n^{-2})
={(-1)^{k+1}\pi n \e_k\/c_{k}}+O(n^{-2})
\qq\text{as}\qq n\to\iy,
\]
where we used \er{ewo}.
Asymptotics \er{abe} and \er{s5} give
\begin{multline}
\lb{s6}
\o^\mu_{s}(\l)-\o^\mu_{s+1}(\l)
=\o_{s}-\o_{s+1}
-{(-1)^p \mu \/2pz^2}\lt({1\/\o_{s}}-{1\/\o_{s+1}}\rt)+O(n^{-4})\\
=(-1)^{k+1}{2ic_{k}\/\e_k}
\lt(1+{(-1)^p \mu c_{k}^2\/2p(\pi n)^2}\rt)+O(n^{-4}),
\end{multline}
where we used \er{ewo} and the simple identity
$\o_{s}\o_{s+1}=\e_k^{-2}$.
Substituting asymptotics \er{s6} into identity \er{s4} we obtain
$$
z={(-1)^{k+1}\pi n \e_k\/c_{k}}
\lt(1-{(-1)^p \mu c_{k}^2\/2p(\pi n)^2}\rt)+O(n^{-3}),
$$
which yields \er{res}.

Assume that $r_{k,n}^{\mu,\pm}$ are non-real for some
$n\in\N$ large enough.
Then $r_{k,n}^{\mu,-}=\ol{r_{k,n}^{\mu,+}}$,
which is in contradiction with
asymptotics \er{res}. Hence $r_{k,n}^{\mu,\pm}\in\R$.
$\BBox$

We introduce the labeling of
the ramifications at high energy:

\vspace{1mm}
\no {\it
For each $k\in\N_{p-1}, n\ge n_0$ for some $n_0\in\N$ large enough,
$r_{k,n}^\pm$ are zeros of the function $\t_{p+k}-\t_{p+k+1}$
and $r_{k,n}^\pm\in\mU_{j,n}$. Moreover, we assume
that $\Im r_{k,n}^+\ge 0$ and

\no if $\Im r_{k,n}^+>0$,
then $r_{k,n}^-=\ol{r_{k,n}^+}$,

\no if $\Im r_{k,n}^+=0$,
then $(-1)^kr_{k,n}^-<(-1)^kr_{k,n}^+$.
}
\vspace{1mm}

\begin{corollary}
The following identities hold true
\[
\lb{t+1}
\t_{p+k}(r_{k,n}^\pm)=\t_{p+k+1}(r_{k,n}^\pm),\qq
\t_{p-k}(r_{k,n}^\pm)=\t_{p-k+1}(r_{k,n}^\pm),\qq
\D_{p-k}(r_{k,n}^\pm)=\D_{p-k+1}(r_{k,n}^\pm),
\]
for all $k\in\N_{p-1}, n\ge n_0$
for some $n_0\in\N$ large enough.
\end{corollary}

\no {\bf Proof.} The results of Lemmas \ref{nre1} and \ref{nre} i)
yield the first identities in \er{t+1}. Identities \er{mn1} give
the second identities in \er{t+1}. The definition of the functions
$\D_j$, see Theorem \ref{T1} i), implies the third identities in \er{t+1}.
$\BBox$

\no {\bf Remark}.
Identities \er{t+1} define the order of attachment
of the sheets of the Riemann surface $\mR$ at high energy
(see Fig. \ref{RS3}).

\section {Asymptotics}
%\section {Proof of Theorems \ref{rho} and \ref{T2}}
\setcounter{equation}{0}

Now we will determine the rough asymptotics of the periodic and
antiperiodic eigenvalues and ramifications of the Lyapunov function.

\begin{lemma} The periodic and antiperiodic eigenvalues $\l_n^\pm$
and the ramifications $r_{k,n}^{\pm}$ satisfy:
\[
\lb{rae}
\l_n^\pm=(\pi n)^{2p}+O(n^{2p-2})\qq\text{as}\qq n\to\iy,
\]
\[
\lb{rar}
r_{k,n}^{\pm}=r_{k,n}^{0}+O(n^{2p-2})\qq\text{as}\qq n\to\iy,\qq k\in\N_{p-1},
\qq r_{k,n}^0=(-1)^k\lt({\pi n\/c_k}\rt)^{2p}.
\]
\end{lemma}

\no {\bf Proof.}
Let $\l=\l_{n}^\pm$ for some $n\in\N$.
Lemma \ref{Dpm} gives
$z=(\l_{n}^\pm)^{1\/2p}=\pi n+\d,$
where $|\d|<{\pi\/2}$ for all $n\ge 1$ large enough.
The periodic and antiperiodic eigenvalues are real zeros of
the functions $\t_j^2-1,j\in\N_p$. Asymptotics \er{awt} show
that these functions, with only exception $\t_p^2-1$,
have no any large real zeros.
Then $1=\t_p^2(\l)=e^{2iz}(1+O(n^{-1}))$,
where we used \er{awt}.
Substituting $z=\pi n+\d$ into this identity we obtain
$e^{2i\d}=1+O(n^{-1})$. Then $\d=O(n^{-1})$ and $z=\pi n+O(n^{-1})$,
which yields \er{rae}.

We will prove \er{rar}.
Let $\l=r_{k,2n}^{\pm}$ for some $(k,n)\in\N_{p-1}\ts\N$.
By Lemma \ref{Dpm},
$z=(r_{k,2n}^{\pm})^{1\/2p}=z_0+\ve$, where $z_0=(r_{k,2n}^0)^{1\/2p}$
and $|\ve|<p\beta$
for all $n\in\N$ large enough.
Identities \er{t+1} show that
$1=\t_{s}(\l)\t_{s+1}^{-1}(\l)=e^{z(\O_{s}-\O_{s+1})}(1+O(n^{-1}))$,
where $s=p+k$ and we used asymptotics \er{awt}.
Substituting $z=z_0+\ve$ into this identity and using
$e^{z_0(\O_{s}-\O_{s+1})}=1$
we obtain
$e^{\ve(\O_{s}-\O_{s+1})}=1+O(n^{-1})$.
Then $\ve=O(n^{-1})$ and $z=z_0+O(n^{-1})$, which yields
\er{rar}.
$\BBox$

In order to improve asymptotics \er{rae}, \er{rar} we determine
the asymptotics of the function
$D(\t,\l)=\det(\cM(1,\l)-\t \1_{2p})$ in the neighborhoods
of the unperturbed ramifications at high energy.

\begin{lemma}
\lb{aD1}
Let $k\in\N_{p-1}^0$ and let $\l\in \ol \C_+$ and $\t\in\C$ satisfy
\[
\lb{t1}
\l=r_{k,n}^0+O(n^{2p-2}),\qq \t=\t_{p+k}^0(\l)\bigl(1+O(n^{-1})\bigr)\qq\text{as}\qq
n\to\iy.
\]
Then
\[
\lb{t2}
\t=\t_{p+k+1}^0(\l)\bigl(1+O(n^{-1})\bigr)\qq\text{as}\qq
n\to\iy.
\]
Moreover, the determinant $D(\t,\l)$, given by \er{1c},
satisfies the asymptotics
\[
\lb{asD}
D(\t,\l)=\a(\t,\l)\det\lt(\ma
1-(\t^\mu_{p+k}(\l))^{-1}\t &f_{k,n}\\
\ol f_{k,n}&1-(\t^\mu_{p+k+1}(\l))^{-1}\t
\am+O(n^{-2})\rt)
\]
as $n\to\iy$, where $\a=\t^{p-k-1}\prod_{j=p+k+2}^{2p}(\t^\mu_{j}(\l))^{-1}\ne 0$
and $f_{k,n}$ are given by \er{hkn}.
\end{lemma}

\no {\bf Proof.}
Identities \er{alb} give $\O_j^\mu(\l)=\o_j^\mu$ for all $j\in\N_{2p}$.
We have $z=\l^{1\/2p}=z^0+O(n^{-1})$,
where $z^0=(r_{k,n}^0)^{1\/2p}=\pi n{\e_k\/c_k}$.
Asymptotics \er{t1} and identities \er{ewo} give
\[
\lb{ad2}
\t e^{-z\o_{s+1}}=e^{z(\o_{s}-\o_{s+1})}(1+O(n^{-1}))
=e^{z^0(\o_{s}-\o_{s+1})}(1+O(n^{-1}))=1+O(n^{-1}),\qq s =p+k
\]
as $n\to\iy$.
Asymptotics \er{ad2} yields \er{t2}.

We will prove asymptotics \er{asD}.
Identity \er{MY} yields
\[
\lb{ar1b}
D(\t,\cdot)=\det(\cF e^{z\cB^\mu}-\t \1_{2p})=
\det(\cF-\t e^{-z\cB^\mu})
=\det\ma A_1-\t e^{-zB_1}&A_2\\A_3&A_4-\t e^{-zB_2}\am,
\]
where the matrices $B_1=\diag(\o^\mu_1,...,\o^\mu_{s +1}),
B_2=\diag(\o^\mu_{s +2},...,\o^\mu_{2p})$,
$$
A_1=\ma \cF_{11}&...&\cF_{1,s +1}\\
...&...&...\\
\cF_{s +1,1}&...&\cF_{s +1,s +1}\am,\qq
A_2=\ma \cF_{1,s +2}&...&\cF_{1,2p}\\
...&...&...\\
\cF_{s +1,s +2}&...&\cF_{s +1,2p}\am,\qq
$$
$$
A_3=\ma \cF_{s +2,1}&...&\cF_{s +2,s +1}\\
...&...&...\\
\cF_{2p,1}&...&\cF_{2p,s +2}\am,\qq
A_4=\ma \cF_{s +2,s +2}&...&\cF_{s +2,2p}\\
...&...&...\\
\cF_{2p,s +2}&...&\cF_{2p,2p}\am.
$$
Due to \er{pns},
the matrices $A_1,...,A_4$ are bounded for $|\l|>0$ large enough.
Identity \er{ar1b} yields
$$
D(\t,\cdot)=\a\det\ma A_1-\t e^{-zB_1}&A_2\\
\t^{-1} e^{zB_2}A_3&\t^{-1}e^{zB_2}A_4-\1_{p-k-1}\am,\qq \a=\a(\t,\l).
$$
Estimates \er{mO}, asymptotics \er{asb}
and \er{t1} yield
$|\t|^{-1}e^{zB_2(\l)}=e^{-z\o_{s }}e^{z\o_{s +2}}(1+O(n^{-1}))$
as $n\to\iy.$
Relations \er{mO1} show that $\Re z(\o_{s }-\o_{s +2})>a|z|$,
$a>0$. Then
$|\t|^{-1}e^{zB_2(\l)}=O(e^{-a n}).$
These asymptotics show that the matrix $\t^{-1}e^{zB_2}A_4-\1_{p-k-1}$
is invertible for $n$ large enough.
Using the standard formula (see \cite{Ga}, Ch.2.5) we obtain
\[
\lb{ar8b}
D(\t,\cdot)=\a\det\bigl(A_1-\t e^{-zB_1}
-A_2(\t^{-1}e^{zB_2}A_4-\1_{p-k-1})^{-1}\t^{-1} e^{zB_2}A_3\bigr)
\det\bigl(\t^{-1}e^{zB_2}A_4-\1_{p-k-1}\bigr).
\]
Substituting the asymptotics $|\t|^{-1}e^{zB_2(\l)}=O(e^{-a n})$
into identity \er{ar8b} we obtain
\[
\lb{ar3b}
D(\t,\l)=\a
\bigl(\det(A_1(\l)-\t e^{-zB_1(\l)})+O(e^{-a n})\bigr)
\qq\text{as}\qq n\to\iy,\qq a>0.
\]
Furthermore, we have
\[
\lb{s7}
A_1-\t e^{-zB_1}=\ma A_5-\t e^{-zB_3}&A_6\\A_7&A_0-\t e^{-zB_0}\am,
\]
where $B_3=\diag(e^{-z\o^\mu_{1}},...,e^{-z\o^\mu_{s -1}}),
B_0=\diag(e^{-z\o^\mu_{s }},e^{-z\o^\mu_{s +1}})$,
$$
A_5=\ma \cF_{11}&...&\cF_{1,s -1}\\
...&...&...\\
\cF_{s -1,1}&...&\cF_{s -1,s -1}\am,\qq
A_6=\ma \cF_{1,s }&\cF_{1,s +1}\\
...&...\\
\cF_{s -1,s }&\cF_{s -1,s +1}\am,\qq
$$
$$
A_7=\ma \cF_{s ,1}&...&\cF_{s ,s -1}\\
\cF_{s +1,1}&...&\cF_{s +1,s -1}\am,\qq
A_0=\ma \cF_{s ,s }&\cF_{s ,s +1}\\
\cF_{s +1,s }&\cF_{s +1,s +1}\am.
$$
Estimates \er{mO}  imply
$
\t e^{-zB_3(\l)}=\t e^{-z\o_{s -1}}(1+o(1))
=e^{z\o_{s +1}}e^{-z\o_{s -1}}(1+o(1)),
$
where we used \er{t2}.
Relations \er{mO1} show that
$\Re z(\o_{s-1}-\o_{s+1})>a|z|$.
Then $\t e^{-zB_3(\l)}=O(e^{-an})$.
Asymptotics \er{pns} show that
$\cF_{jj}(\l)=1+O(n^{-2}),j\in\N_{2p}$,
which yields
\[
\lb{s8}
A_5(\l)-\t e^{-zB_3(\l)}=\1_{s -1}+O(n^{-2})\qq\text{as}\qq n\to\iy.
\]
Thus the matrix $A_5(\l)-\t e^{-zB_3(\l)}$ is invertible for large $n$
and \er{s7} gives
\[
\lb{s9}
\det(A_1-\t e^{-zB_1})
=\det(A_5-\t e^{-zB_3})
\det\bigl(A_0-\t e^{-zB_0}-A_7(A_5-\t e^{-zB_3})^{-1}A_6\bigr).
\]
Substituting asymptotics
\er{s8} into identity \er{s9} we get
$$
\det(A_1(\l)-\t e^{-zB_1(\l)})
=\det(A_0(\l)-\t e^{-zB_0(\l)}+O(n^{-2}))\qq\text{as}\qq n\to\iy.
$$
Substituting this asymptotics into
\er{ar3b} we have
$$
D(\t,\l)=\a\det\lt(
\ma\!
\cF_{s ,s }(\l)-\t e^{-z\o^\mu_{s }(\l)}\!\!\!\!\!\!&
\cF_{s ,s +1}(\l)\\
\cF_{s +1,s }(\l)\!\!\!\!\!\!&
\cF_{s +1,s +1}(\l)-\t e^{-z\o^\mu_{s +1}(\l)}
\!\am+O(n^{-2})\rt).
$$
Substituting \er{pns}, \er{pbn}
into the last asymptotics we obtain \er{asD}.
$\BBox$

Below we write $a_n=b_n+\ell^2(n)$
iff the sequence $(a_n-b_n)_{n\ge 1}\in\ell^2$.

\no {\bf Proof of Theorem \ref{rho}.}
i) Repeating the arguments from \cite{CK} we obtain that
$\r$ is entire and real on $\R$.
Asymptotics \er{aD}  yields \er{aro}.

 ii) Let $\l=r_{k,n}^{\pm}$ for some $(k,n)\in\N_{p-1}\ts\N$.
We assume that $\Im\l\ge 0$. Then $\O_j=\o_j$ and $\O_j^\mu=\o_j^\mu$
for all $j\in\N_{2p}$.
Using the identity $r_{k,n}^-=\ol{r_{k,n}^+}$ we obtain
asymptotics for $r_{k,n}^-\in\C_-$.
Let $\l^\mu=r_{k,n}^{\mu,\pm}$, where $\mu=\hat q_{p,0}$,
be the unperturbed ramification.
Asymptotics \er{res}, \er{rar} yield
\[
\lb{ar15}
z=\l^{1\/2p}=z^\mu+\xi\d,\qq
\text{where}\qq z^\mu=(\l^\mu)^{1\/2p},\qq \xi={\e_k\/c_{k}},\qq
\d=O(n^{-1})\qq \text{as}\qq n\to\iy.
\]
Since $\l=r_{k,n}^{\pm}$ is the ramification,
the monodromy matrix $\cM(1,\l)$ has the eigenvalue $\t$
of multiplicity 2, i.e. its characteristic polynomial
$D(\cdot,\l)$ has the zero $\t$
of multiplicity 2. If $n\in\N$ is large enough, then
identities \er{t+1} show that
$\t=\t_{p+k}(\l)=\t_{p+k+1}(\l)$.
Using asymptotics \er{awt} we obtain
$\t=\t^0_{p+k}(\l)(1+O(n^{-1}))$ as $n\to\iy$.
Then we can apply asymptotics \er{asD}.
Note that the function $A(\t)=\det\ma a_1-\t a_2& a_3\\ a_4&a_5-\t a_6\am$,
where $a_j\in\C$ for all $j\in\N_6$, has the zero of multiplicity $2$ iff
$(a_2a_5-a_1a_6)^2+4a_2a_6a_3a_4=0$.
Using asymptotics \er{asD} we deduce that
\begin{multline}
\lb{ar9b}
\Bigl( (\t^\mu_{s}(\l))^{-1}\bigl(1+O(n^{-2})\bigr)
-(\t^\mu_{s+1}(\l))^{-1}\bigl(1+O(n^{-2})\bigr)\Bigr)^2\\
+4\bigl(\t^\mu_{s}(\l)\t^\mu_{s+1}(\l)\bigr)^{-1}
\bigl(f_{k,n}+O(n^{-2})\bigr)\bigl(\ol f_{k,n}+O(n^{-2})\bigr)=0
\qq\text{as}\qq n\to\iy,
\end{multline}
where $s=p+k$ and $\t_j^\mu$ are given by \er{alb}.
Identity \er{ar9b} yields
\[
\lb{ar7}
\Bigl((\t^\mu_{s}(\l))^{-{1\/2}}(\t^\mu_{s+1}(\l))^{1\/2}\bigl(1+O(n^{-2})\bigr)
-(\t^\mu_{s}(\l))^{1\/2}(\t^\mu_{s+1}(\l))^{-{1\/2}}\bigl(1+O(n^{-2})\bigr)\Bigr)^2
+4|f_{k,n}|^2=n^{-3}\ell^2(n)
\]
as $n\to\iy$, where we used $f_{k,n}=\ell^2(n)O(n^{-1})$, see \er{hkn}.

Identities \er{t+1}, applied to the operator $H^\mu$,
yields
$\t^\mu_{s+1}(\l^\mu)=\t^\mu_{s}(\l^\mu)$.
%Since $z=z^\mu+\xi\d$, we have
Asymptotics \er{ar15} imply
\begin{multline}
\lb{tt}
\t^\mu_{s}(\l)(\t^\mu_{s+1}(\l))^{-1}
=\t^\mu_{s}(\l)(\t^\mu_{s+1}(\l))^{-1}
(\t^\mu_{s}(\l^\mu))^{-1}\t^\mu_{s+1}(\l^\mu)
\\
=e^{(z^\mu+\xi\d)(\o^\mu_{s}(\l)-\o^\mu_{s+1}(\l))}
e^{-z^\mu(\o^\mu_{s}(\l^\mu)-\o^\mu_{s+1}(\l^\mu))}
\\
=e^{z^\mu(\o^\mu_{s}(\l)-\o^\mu_{s}(\l^\mu)-\o^\mu_{s+1}(\l)+\o^\mu_{s+1}(\l^\mu))}
e^{\xi\d(\o^\mu_{s}(\l)-\o^\mu_{s+1}(\l))}.
\end{multline}
Asymptotics \er{asb} gives $\o^\mu_j(\l)=\o_j(\l)+O(n^{-2})$. Asymptotics
\er{b-b0} yields $\o^\mu_j(\l)-\o^\mu_j(\l^\mu)=O(n^{-4})$ as $n\to\iy,j\in\N_{2p}$.
Then \er{tt} implies
$$
(\t^\mu_{s}(\l))^{1\/2}(\t^\mu_{s+1}(\l))^{-{1\/2}}
=e^{\xi{\d\/2}(\o_{s}(\l)-\o_{s+1}(\l))}(1+O(n^{-2}))
=1+i(-1)^{k+1}\d+O(n^{-2})
$$
as $n\to\iy$, where we used \er{ewo}.
Substituting these asymptotics into \er{ar7} we obtain
$$
\d=\pm(-1)^k|f_{k,n}|+O(n^{-2})
=\pm{(-1)^kc_{k}|\hat q_{p,n}|\/2p\pi n}+O(n^{-2})\qq\text{as}\qq n\to\iy,
$$
where we used \er{hkn}.
Then
$$
(r_{k,n}^{\pm})^{1\/2p}
=z^\mu\pm {(-1)^k\e_{k}|\hat q_{p,n}|\/2p\pi n}
+O(n^{-2}),
$$
and
\[
\lb{z7}
r_{k,n}^{\pm}
=r_{k,n}^{\mu,\pm}
\pm {(z^\mu)^{2p-1}(-1)^k\e_k|\hat q_{p,n}|\/\pi n}
+O(n^{2p-3})
=r_{k,n}^{\mu,\pm}
\pm \lt({\pi n\/c_{k}}\rt)^{2p-2}c_{k}|\hat q_{p,n}|
+O(n^{2p-3}),
\]
where we used $z^\mu=\xi\pi n+O(n^{-1})$, see \er{res},
and $\e_k^{2p}=(-1)^k$, see \er{ete}.
Substituting asymptotics \er{res} into \er{z7} we obtain \er{rnn}.
$\BBox$

\no {\bf Proof of Theorem \ref{T2}}.
i) Asymptotics \er{aD} shows
that the branches of the Lyapunov function $\D$
on the interval $(K,+\iy)$ for some $K\in\R$ satisfy:

 if $p$ is odd, then there is exactly one real branch $\D_1$
and the other branches are non-real;

 if $p$ is even, then there are two real branches $\D_1$ and
$\D_{{p\/2}+1}$, $\D_{{p\/2}+1}>1$ and the other branches are non-real.

 Moreover, $\D_1(\l)\ne\D_j(\l)$ for all $j=2,...,p,\l\in(K,+\iy)$,
hence $\D_1$ is analytic on the interval $(K,+\iy)$.
Asymptotics \er{aD} for $\D_{1}$ and Theorem \ref{T1} ii) show that
the function $\D_1$ oscillates on $(K,+\iy)$ similar to the Lyapunov
function for the Hill operator. Then using identity \er{sd}
and the standard arguments (see \cite{BK1}) we obtain the needed statement.

 ii)
Let $\l=\l_{n}^\pm$ for some $n\in\N$.
Recall that $\l\in\R$ and satisfies
\[
\lb{D=0}
D_n=D((-1)^n,\l)=\det(\cM(1,\l)-(-1)^n\1_{2p})=0.
\]
Let $\l^\mu=\l_{n}^{\mu,\pm}$ be the unperturbed 2-periodic
eigenvalues.
Asymptotics \er{les}, \er{rae} give
$z=\l^{1\/2p}=z^{\mu}+\d$, where
$z^\mu=(\l^{\mu})^{1\/2p},$
$\d\in\R,\d=O(n^{-1})$ as $n\to\iy$,
$\l^{\mu}$ satisfies \er{les}.
Asymptotics \er{asD} for $k=0$ give
\[
\lb{adp}
D_n=\a_n\det\lt(\ma
1-(-1)^n (\t_p^\mu(\l))^{-1}&f_{0,n}\\
\ol f_{0,n}&1-(-1)^n (\t_{p+1}^\mu(\l))^{-1}
\am+O(n^{-2})\rt)\qq\text{as}\qq n\to\iy,
\]
where
$\a_n=\a((-1)^n,\l_{n}^\pm)\ne 0$,
$f_{0,n}=-i{\hat q_{p,n}\/2p\pi n}$.
Each $\l^{\mu}$ is a periodic or antiperiodic eigenvalue
of the operator $H^\mu$ and the corresponding multipliers
satisfy the identities
$\t_p^\mu(\l^\mu)=\t_{p+1}^\mu(\l^\mu)=(-1)^n$.
Then
\[
\lb{z8}
(-1)^n(\t_j^\mu(\l))^{-1}=(\t_j^\mu(\l))^{-1}\t_j^\mu(\l^\mu)
=e^{-(z^\mu+\d)\o^\mu_j(\l)}e^{z^\mu\o^\mu_j(\l^\mu)}
=e^{-\d\o^\mu_j}e^{-z^\mu(\o^\mu_j(\l)-\o^\mu_j(\l^\mu))}
\]
for $j=p,p+1.$
Asymptotics \er{asb} yield
$e^{-\d\o^\mu_j(\l)}=e^{-\d\o_j}(1+O(n^{-3}))$
as $n\to\iy.$
Asymptotics \er{b-b0} show that
$\o^\mu_j(\l)-\o^\mu_j(\l^\mu)=O(n^{-4})$.
Then \er{z8} gives
$$
(\t_j^\mu(\l))^{-1}=(-1)^ne^{-\d\o_j}(1+O(n^{-3}))=(-1)^n(1-\d\o_j)+O(n^{-2})
\qq\text{as}\qq n\to\iy,\qq
j=p,p+1.
$$
Using the identities $\o_{p}=-\o_{p+1}=-i$ we obtain
$$
(\t_p^\mu(\l))^{-1}=(-1)^n(1+i\d)+O(n^{-2}),\qq
(\t_{p+1}^\mu(\l))^{-1}=(-1)^n(1-i\d)+O(n^{-2})\qq\text{as}\qq n\to\iy.
$$
Substituting these asymptotics
into \er{adp} we obtain
$$
D_n=\a_n\det\lt(\ma
-i\d&f_{0,n}\\
\ol f_{0,n}&i\d\am+O(n^{-2})\rt)
=\a_n\det(\cN
+\d \1_2+O(n^{-2})),\  \cN=\ma 0&if_{0,n}\\
-i\ol f_{0,n}&0\am.
$$
Using identity \er{D=0} we conclude that
$\d$ is an eigenvalue of the matrix $\cN+O(n^{-2})$.
Since the eigenvalues of the matrix
$\cN$ have the form $\pm |f_{0,n}|$, we obtain
$\d=\pm|f_{0,n}|+O(n^{-2}).$
Using the identity $|f_{0,n}|={|\hat q_{p,n}|\/2p\pi n}$ we obtain
$
\l^{1\/2p}=z^\mu\pm {|\hat q_{p,n}|\/2p\pi n}+O(n^{-2}).
$
Then \er{les} gives \er{eln}.
$\BBox$

\no {\bf Proof of Corollary \ref{ig}.}
%We will prove the statement
i)
%. The proof of ii) is similar.
Asymptotics \er{eln} and the estimates
$|\hat q_{p,n_k}|\ge {1\/{n_k}^{\a}}$ give
the asymptotics
$$
|\g_n|=\l_n^+-\l_n^-=2(\pi n)^{2p-2}|\hat q_{p,n}|+O(n^{2p-3})\qq\text{as}
\qq n\to \iy,
$$
and the estimates
$|\g_{n_k}|\ge2\pi^{2p-2} n_k^{2p-2-\a}(1+O({n_k}^{\a-1}))$ as $k\to \iy$,
which yields the statement.

 ii) Assume that $r_{k,n_j}^\pm$ are non-real.
Then $r_{k,n_j}^-=\ol{r_{k,n_j}^+}$, which is in contradiction with
asymptotics \er{rnn}. Hence $r_{k,n_j}^\pm\in\R$.
Moreover, asymptotics \er{rnn} gives
$$
|r_{k,n_j}^+-r_{k,n_j}^-|
=2\Bigl({\pi n\/c_k}\Bigr)^{2p-2}c_{k}|\hat q_{p,n}|+O(n^{2p-3})\qq\text{as}
\qq  n\to \iy,
$$
which yields $|r_{k,n_j}^+-r_{k,n_j}^-|\to\iy$ as $k\to\iy$.
$\BBox$

\section {Appendix}
\setcounter{equation}{0}

\no {\bf Proof of Lemma \ref{T21}.} The standard arguments yield that the
fundamental solution $\cM(t,\l)$ of equation \er{Y},
with the initial condition $\cM(0,\l)=\1_{2p}$, satisfies the integral equation
\[
\lb{iev}
\cM(t,\l)=\cM_0(t,\l)+\int_0^t\cM_0(t-s,\l)Q(s)\cM(s,\l)ds,
\]
where $\cM_0(t,\l)=e^{t\cP(\l)}$ is a solution at $Q=0$.

Describe the properties of the matrix-valued function $\cM_0$.
Each function $\cM_0(t,\cdot)$, $t\in\R$, is entire.
Moreover, $\cM_0=((\vp_j^0)^{(k-1)})_{k,j=1}^{2p}$,
where $\vp_j^0,j\in\N_{2p}$, are the solutions of the equation $y^{(2p)}=\l y$,
satisfying the conditions $(\vp_j^0)^{(k-1)}(0,\l)=\d_{jk}$
for all $k\in\N_{2p}$. These solutions are given by the identities
$$
\vp_1^0={1\/2p}\sum_1^{2p}e^{z\o_n t},\qqq
\vp_{j+1}^0(t,\l)=\int_0^t\vp_{j}^0(s,\l)ds,\qq\text{ all}\qq
j\in\N_{2p-1}.
$$
Then $|(\vp_j^0)^{(k-1)}(t,\l)|\le e^{z_0|t|}$
for all $j,k\in\N_{2p},(t,\l)\in\R\ts\C$,
and
\[
\lb{eu1}
|\cM_0(t,\l)|\le 2pe^{z_0|t|},\qqq\text{all}\qq (t,\l)\in\R\ts\C.
\]
Estimate \er{eu1} will be useful for obtaining the first estimate in \er{2if}.
Now we will prove the other estimate of $\cM_0$ (see below \er{eu}),
which will be effective to obtain the second estimate in \er{2if}.
In fact, direct calculations show that $\cP=z\cZ\cC\cB(\cZ\cC)^{-1}$,
where the diagonal matrix $\cB$ is given by
$\cB=\diag(\o_j)_1^{2p}$,
and the matrix $\cC$ has the form $\cC=(\o_j^{k-1})_{k,j=1}^{2p}$.
Then $\cC^{-1}={1\/2p}\cC^*$,
$\cM_0=\cZ\cC e^{zt\cB}(\cZ\cC)^{-1}$ and
\[
\lb{eu}
|\cZ^{-1}(\l)\cM_0(t,\l)\cZ(\l)|\le 2pe^{z_0|t|},\qq\text{all}\qq
(t,\l)\in\R\ts(\C\sm\{0\}).
\]

The standard iterations in \er{iev} yield
\[
\lb{evj}
\cM(t,\l)=\sum_{n\ge 0}\cM_n(t,\l),\qqq \cM_n(t,\l)=\int_0^t\cM_0(t-s,\l)Q(s)\cM_{n-1}(s,\l)ds.
\]
and  $\cM_n(t,\l)$ is given by
\[
\lb{2ig}
\cM_n(t,\l)=\int\limits_{T}\prod\limits_{k=1}^{n}
\rt(\cM_0(t_{k+1}-t_k,\l)Q(t_k)\rt)\cM_0(t_1,\l)dt_1dt_2...dt_n
\]
the factors are ordering from right to left,
$T=\{0< t_1<...< t_n<t_{n+1}=t\}$.
Substituting estimates \er{eu1} into identities \er{2ig}
we obtain
\[
|\cM_n(t,\l)|\le{2p\/n!}e^{z_0|t|}\lt(2p\int_0^t|Q(s)|ds\rt)^n,
\qqq \all \qq (n,t,\l)\in\N\ts\R\ts\C.
\]
These estimates show that for each fixed
$t\in\R$ the formal series \er{evj} converges
absolutely and uniformly on
bounded subset of $\C$. Each term of this series is an entire
function. Hence the sum is an entire function.
Summing the majorants we get
\[
|\cM(t,\l)|\le 2p e^{z_0|t|+\int_0^t|Q(s)|ds}
\qq \all \qq (t,\l)\in\R\ts\C,
\]
which yields the first estimate in \er{2if}. Moreover,
we deduce that the monodromy matrix $\cM(1,\l)$ is a
continuous function of all $q_j\in L^1(\T),j\in\N_p$.

Substituting estimates \er{eu} into identities \er{2ig}
we obtain
$$
|\cZ^{-1}(\l)\cM_n(t,\l)\cZ(\l)|\le
{2p\/n!}e^{z_0|t|}\lt(2p\int_0^t|\cZ^{-1}(\l)Q(s)\cZ(\l)|ds\rt)^n,
$$
for all $(n,t,\l)\in\N\ts\R\ts(\C\sm\{0\})$. Substituting these estimates into
the series \er{evj} we obtain
$$
|\cZ^{-1}(\l)\cM(t,\l)\cZ(\l)|\le
2p e^{z_0|t|+\int_0^t|\cZ^{-1}(\l)Q(s)\cZ(\l)|ds},
\qq \all \qq (t,\l)\in\R\ts(\C\sm\{0\}),
$$
which yields the second estimate in \er{2if}.
$\BBox$

\no {\bf Proof of Lemma \ref{dcQ}.}
Asymptotics \er{asb} implies
$$
\cU(\l)=\cZ(\l)\cU_0(\1_{2p}+O(|z|^{-2})),\qq  \text{where}\qq
\cZ=\diag(z^{j-1})_{j}^{2p},
\qq \cU_0=(\O_k^{j-1})_{j,k=1}^{2p}.
$$
Then
$$
\wt\cQ(t,\l)=\cU^{-1}(\l)\cQ^\mu(t,\l)\cU(\l)
=\cU_0^{-1}\cZ^{-1}(\l)\cQ^\mu(t,\l)\cZ(\l)\cU_0(\1_{2p}+O(|z|^{-2}))
\qq\text{as}\qq|\l|\to\iy
$$
uniformly on $t\in[0,1]$.
We have
$$
\cZ^{-1}(\l)\cQ^\mu(t,\l)\cZ(\l)
={(-1)^{p+1}\/z}\Bigl((q_p(t)-\mu)\cE+b(t)O(|z|^{-1})\Bigr)
\qq\text{as}\qq|\l|\to\iy,
$$
 uniformly on $t\in[0,1]$,
where $\cE$ is given by \er{mK}. Then we obtain \er{wQ} with
$\cL=\cU_0^{-1}\cE\cU_0$.
Using the identity $\cU_0^{-1}={1\/2p}\cU_0^*$ we get
$\cL={1\/2p}\cU_0^*\cE\cU_0$. Then
$$
\cL_{jk}={1\/2p}\sum_{\ell,n=1}^{2p} \ol\O_{j}^{\ell-1}\cE_{\ell n}\O_{k}^{n-1}
={\ol\O_{j}^{p}\O_{k}^{p-1}\/2p},
$$
which yields the identity for $\cL_{jk}$ in \er{wQ}.
$\BBox$

\no {\bf Proof of Lemma \ref{l32}.}
i) Asymptotics \er{wQ} show that
% each function $\wt\cQ_{kj},k,j\in\N_{2p}$, satisfies
the estimate
\[
\lb{ewq}
\max_{k,j\in\N_{2p}}|\wt\cQ_{kj}(t,\l)|\le
{\max\{|q_p(t)-\mu|,{1\/4}\}\/p|z|},\qq\text{all}\qq
(t,\l)\in\ts[0,1]\ts\L_{R_1}
\]
for some $R_1>0$. Assume that
%each function $e_{kj},k,j\in\N_{2p}$, satisfies
\[
\lb{ee}
\max_{k,j\in\N_{2p}}|e_{kj}(t,\l)|\le 2,\qq\text{all}\qq (t,\l)\in\R\ts\L_{R_2},
\qq\text{for some}\qq R_2>0.
\]
Then substituting estimates \er{ee} into \er{dcL} we obtain
\begin{multline}
\!\!\!\!\!\!
\|(K\wt\cQ\cA)(\cdot,\l)\|_\iy=\max_{t\in[0,1]}|(K\wt\cQ\cA)(t,\l)|
=\!\!\max_{(t,j)\in[0,1]\ts\N_{2p}}\sum_{i=1}^{2p}
\lt|\int_0^1e_{ij}(t-s,\l)\sum_{n=1}^{2p}\wt\cQ_{i n}(s,\l)\cA_{nj}(s)ds\rt|\\
\le 2\max_{1\le j\le 2p}\sum_{i,n=1}^{2p}\int_0^1|\wt\cQ_{i n}(s,\l)||\cA_{nj}(s)|ds
\le 2\max_{1\le j\le 2p}\sum_{i,n=1}^{2p}
\max_{t\in[0,1]}|\cA_{nj}(t)|\int_0^1|\wt\cQ_{i n}(s,\l)|ds.
\end{multline}
for all $\l\in\L_{\max\{R_1,R_2\}}$.
Using estimates \er{ewq} we obtain
$$
\|(K\wt\cQ\cA)(\cdot,\l)\|_\iy\le {\x\/|z|}
\max_{t\in[0,1]}\max_{1\le j\le 2p}\sum_{n=1}^{2p}|\cA_{nj}(t)|,
$$
which yields \er{KK}.

We will prove \er{ee}.
We will consider the case $i>j$. The proof for $i\le j$ is similar.
Identities \er{e} shows that $e_{ij}(t,\l)=0$ for $t<0$.
%Then we have to prove \er{ee} for $t\ge 0$.
Asymptotics \er{asb} shows that
$$
\O^\mu_j=\O_j+{\wt\O_j\/z},\qq j\in\N_{2p},\qq\text{where}\ \
\max_{(j,\l)\in\N_{2p}\ts\{|z|>R_2\}}|\wt\O_j(\l)|<{1\/2}\log 2
$$
for some $R_2>0$.
Estimates \er{mO} give $\Re z(\O_i -\O_j)\le 0$ for $i>j$. Then
\[
\lb{Ese}
|e_{i  j}(t,\l)|=e^{t\Re z(\O^\mu_i (\l)-\O^\mu_j(\l))}\le
e^{t\Re z(\O_i -\O_j)+t\Re(\wt\O_i (\l)-\wt\O_j(\l))}
\le e^{|\wt\O_i (\l)-\wt\O_j(\l)|}\le 2,\qq
\]
for $t\ge 0,i>j,|z|>R_2$, which yields
%The similar estimates for $i\ge j$ hold true.
estimates \er{ee} for $i>j$.

 ii) The standard iterations in \er{me5} give
\[
\lb{me6}
\cG=\sum_{n=0}^\iy \cG_{n},
\qq  \cG_{0}=\1_{2p},\qq \cG_{n}=K\wt\cQ\cG_{n-1}=(K\wt\cQ)^n,\qq\text{all}\qq n\in\N.
\]
Estimate \er{KK} yields
\[
\lb{me7}
\|\cG_{n}(\cdot,\l)\|_\iy=\|(K\wt\cQ)^n(\cdot,\l)\|_\iy\le\lt({\x\/|z|}\rt)^n,\qq
\text{all}\qq (n,\l)\in\N\ts\L_{\max\{R_1,R_2\}}.
\]
These estimates show that the formal series \er{me6} converges
absolutely and uniformly on any
bounded subset of $\L_R,R=\max\{R_1,R_2,\x^{2p}\}$.
Hence it gives the unique solution of equation \er{me5}.
Each term of this series is analytic in $\L_{R}^\pm$. Hence the matrix-valued
function $\cG$ is analytic in $\L_{R}^\pm$.
If $\l\in\L_{2R}$, then substituting estimates \er{me7}
into the series \er{me6} we obtain
$$
\|\cG(\cdot,\l)\|_\iy\le \sum_{n=0}^\iy\lt({\x\/|z|}\rt)^n=
{1\/1-{\x\/|z|}}\le 2,\qq
%$$
%$$
\|\cG(\cdot,\l)-\1_{2p}\|_\iy\le \sum_{n=1}^\iy\lt({\x\/|z|}\rt)^n=
{{\x\/|z|}\/1-{\x\/|z|}}
\le {2\x\/|z|},
$$
$$
\|\cG(\cdot,\l)-\1_{2p}-\cG_1(\cdot,\l)\|_\iy\le \sum_{n=2}^\iy\lt({\x\/|z|}\rt)^n=
{({\x\/|z|})^2\/1-{\x\/|z|}}
\le {2\x^2\/|z|^2},
$$
which yields \er{emF}.
Substituting \er{wQ} into the identity $\cG_1=K\wt\cQ$
we obtain
$$
(\cG_1(t,\l))_{ij}
={(-1)^{p+1}\/z}\int_0^1e_{ij}(t-s,\l)
(q_p(s)-\mu)\cL_{ij}(\l)ds+a_{ij}(t,\l)O(|z|^{-2}),\qq i,j\in\N_{2p},
$$
where $a_{ij}(t,\l)=
(-1)^{p+1}\int_0^1e_{ij}(t-s,\l)
b(s)ds,i,j\in\N_{2p}$.
Estimate \er{Ese} gives $|a_{ij}(t,\l)|\le 2\|b\|$,
which yields \er{asG}.

 iii) The matrix-valued function $\cG(t,\cdot),t\in[0,1]$,
is analytic in $\L_{R}^\pm$, then $\cT(t,\cdot)$ is also analytic.
The second estimate in \er{emF} yields asymptotics \er{amT}.
$\BBox$

\

\

{\bf Acknowledgments.} \footnotesize
Andrey Badanin was partially supported by DAAD grant "Mikhail Lomonosov-2007".
 The various parts of this paper were written at
 Mathematical Institute of the Tsukuba University, Japan (March, 2010) and Ecole Polytechnique,
 France (April -- July, 2010). Evgeny Korotyaev is grateful to the institutes for the hospitality.

\end{document}